\def\spose#1{\hbox to 0pt{#1\hss}}
\def\ltsimm{\mathrel{\spose{\lower 3pt\hbox{$\sim$}}
        \raise 2.0pt\hbox{$<$}}}
\def\gtsimm{\mathrel{\spose{\lower 3pt\hbox{$\sim$}}
        \raise 2.0pt\hbox{$>$}}}

\documentclass[useAMS,usenatbib]{mn2e}
\usepackage{graphicx}
\usepackage{multirow}
\usepackage{amssymb}

\title{Feedback from Winds and Supernovae in Massive Stellar Clusters. I: Hydrodynamics}
\author[H. Rogers and J.M. Pittard]{H. Rogers and J.M. Pittard\\
School of Physics and Astronomy, The University of Leeds, Leeds, LS2 9JT}

\date{Released 2012 Xxxxx XX}

\pagerange{\pageref{firstpage}--\pageref{lastpage}} \pubyear{2012}

\def\LaTeX{L\kern-.36em\raise.3ex\hbox{a}\kern-.15em
    T\kern-.1667em\lower.7ex\hbox{E}\kern-.125emX}

\begin{document}

\label{firstpage}

\maketitle

\begin{abstract}
We use 3D hydrodynamical models to investigate the effects of massive star feedback from winds and supernovae on inhomogeneous molecular material left over from the formation of a massive stellar cluster. We simulate the interaction of the mechanical energy input from a cluster with 3 O-stars into a giant molecular cloud (GMC) clump containing 3240\,M$_{\odot}$ of molecular material within a 4\,pc radius. The cluster wind blows out of the molecular clump along low-density channels, into which denser clump material is entrained. We find that the densest molecular regions are surprisingly resistant to ablation by the cluster wind, in part due to shielding by other dense regions closer to the cluster. Nonetheless, molecular material is gradually removed by the cluster wind
during which mass-loading factors in excess of several 100 are obtained. Because the clump is very porous, $60-75$ per cent of the injected wind energy escapes the simulation domain, with the difference being radiated. After 4.4\,Myr, the massive stars in our simulation begin to explode as supernovae. The highly structured environment into which the SN energy is released allows even weaker coupling to the remaining dense material and practically all of the SN energy reaches the wider environment. The molecular material is almost completely dispersed and destroyed after 6\,Myr. The escape fraction of ionizing radiation is estimated to be about 50 per cent during the first 4\,Myr of the cluster's life. A similar model with a larger and more massive GMC clump reveals the same general picture, though more time is needed for it to be destroyed.
\end{abstract}


\section{Introduction}
Massive stars have a profound effect on their natal environment creating wind-blown shells, cavities and HII regions. Their winds and supernovae (SNe) chemically enrich the interstellar medium (ISM) and also help to sustain turbulence within it. Massive stars embedded within molecular clouds likely inhibit further star formation as their winds and ionizing radiation disperse and destroy the remaining molecular gas, though in some circumstances massive stars may also trigger new star formation \citep{Koenig12} and new cluster formation \citep{Beuther08,Gray11}. The removal of molecular material is also crucial to the question of cluster dissolution \citep{PortegiesZwart10, Pfalzner11,Pelupessy12}.
 
Stellar feedback is also recognized as having significant influence on galactic and extragalactic scales. For instance, without strong stellar feedback, cosmological models predict around 10 times the stellar mass found in real galaxies \citep[e.g.][]{Cole00, Keres09}. Feedback from massive stars can also drive galactic winds from starburst galaxies \citep[e.g.][]{Axon78,Bland88,Heckman00,Adelberger03}, and appears to be responsible for the low star formation efficiency of galaxies with dark matter haloes somewhat less than the halo mass of the Milky Way \citep{Guo10}. Ionizing radiation from massive stars is also important for the ionization of galaxies \citep{Reynolds84} and the reionization of the early universe \citep{Fan06}.

In order to understand the effects of massive stars on galactic scales, however, we must first understand their impact on their local (cluster) environment. 
The extent to which a cloud is affected by stellar feedback is clearly dependent on a number of parameters, including the mass of the cloud and the stellar cluster, the structure of the cloud, the position of the cluster relative to the cloud, and the age of the system.
However, the degree to which stellar feedback processes (stellar winds, SNe, and ionizing radiation) couple to the clumpy, inhomogeneous molecular clouds which initially surround a massive stellar cluster is exceedingly ill-determined, and the dominant feedback process is still to be settled \citep[e.g.][]{Yorke89,Draine91,Matzner02,Wang10,Lopez11,Pellegrini11}. 

That stellar winds play a significant role in stellar cluster feedback is apparent from the fact that many young (pre-supernova) massive star forming regions contain diffuse X-ray emission - the hot gas responsible can only have been created by winds. Observations reveal that the surrounding cold molecular material can sometimes confine this hot gas \citep[e.g.][]{Townsley06}, but around other clusters the cold clouds appear to be shaped and removed by the hot gas. In fact, there are now several lines of evidence that indicate that the hot X-ray emitting gas often escapes or leaks out of the local cluster environment, instead of being bottled up inside a swept-up shell. Firstly, the diffuse X-ray emission in M17 and the Rosette nebula reveals that only a small proportion of the cluster wind energy is radiated in the X-ray, and  \citet{Townsley03} conclude that most of the hot gas must flow without cooling into the wider ISM. This picture is supported by analysis of the Omega nebula, the Arches cluster, and NGC\,3603, all of which contain an amount of hot gas equal to only 1 per cent of the wind material from the O-stars over the age of the cluster. Secondly, direct evidence for outflowing gas comes from observations of M17 and particularly RCW\,49 which show stellar bow shocks around O-stars outside of the central clusters. These indicate large scale gas outflows away from the stellar cluster with velocities of at least a few hundred ${\rm km\,s^{-1}}$ \citep{Povich08}.

Stellar wind feedback into an inhomogeneous environment has been considered by \citet[][]{Tan01} and more recently by \citet{Harper-Clark09}. The latter postulate that the non-uniform surrounding medium causes gaps in the swept-up shell surrounding the wind-blown bubble where some of the high-pressure gas in the bubble interior can leak out. This scenario has received backing from \citet{Lopez11} who conclude that such leakage may be occuring within 30 Doradus. \citet{Lopez11} also conclude that direct stellar radiation pressure dominates the interior dynamics, but this claim has proved far more controversial, and other works argue in favour of the thermal pressure of hot X-ray emitting plasma shaping the large-scale structure and dynamics in 30 Doradus \citep{Pellegrini11,Povich12}. One must recognize that there are significant uncertainties in determining the hot gas pressure in regions like 30 Doradus, where it is unclear whether the X-ray emitting gas should be treated as a single large bubble or as multiple smaller bubbles with distinct identities. 

The possibility that the pressure exerted by stellar radiation may be dynamically important in massive young stellar clusters has received much attention in recent years, with \citet{Krumholz09}, \citet{Fall10} and \citet{Murray10} all arguing that radiation pressure is the dominant feedback mechanism. However, these works disagree on the net momentum coupling between the radiation field and the gas, partially because this depends on the degree of inhomogeneity of the gas and the effect that this has on the radiation field \citep[e.g.][]{Krumholz12b}. 
Complementary work on the ionized gas pressure has shown that ionization feedback into a highly inhomogeneous medium is not very effective at high cluster masses \citep{Dale11}, but becomes more so at lower masses \citep{Dale12}. 

Given these competing processes, our aim in this work is to examine the extent to which the {\em mechanical} energy input from a cluster of massive stars is confined by and shapes the local environment. In this initial investigation we focus solely on feedback due to stellar winds and supernovae, and defer investigations of other forms of feedback (radiation pressure, photoionization etc.) to future works. We conduct our investigation through 3D hydrodynamic simulations of this interaction. In Section~\ref{sec:model_details} we describe the numerical models and initial conditions used in this work. We present and discuss results from the simulations in Section~\ref{sec:results}. Section~\ref{sec:conclusions} summarizes and concludes this work. 

\section{Simulations of Stellar Feedback}
\label{sec:model_details}
We use an MPI-parallelized numerical scheme to solve the Euler equations of hydrodynamics using a lagrangian formulation and a remap onto the original grid.  A piecewise parabolic interpolation and characteristic tracing is used to obtain the time-averaged fluid variables at each zone interface. The code then solves a Riemann problem to determine the time-averaged fluxes, and then solves the equations of hydrodynamics:

\begin{eqnarray}
\label{eq:mass}
\frac{\partial \rho}{\partial t} + \nabla \cdot (\rho {\bf u}) & = & 0,\\  
\frac{\partial \rho {\bf u}}{\partial t} + \nabla \cdot (\rho {\bf u}u + P) & = & 0, \\ 
\frac{\partial \rho \varepsilon}{\partial t} + \nabla \cdot [(\rho \varepsilon + P){\bf u}] & = & n\Gamma - n^2\Lambda, 
\end{eqnarray}

\noindent where $\varepsilon = {\bf u}^{2}/2 + e/\rho$ is the total specific energy, $\rho$ is the mass density, $e$ is the internal energy density, $P$ is the pressure, and $T$ is the temperature. We adopt an ideal gas equation of state, $e = P/(\gamma - 1)$, and solar abundances.

The net heating/cooling rate per unit volume is parameterized as $\dot{e} = n\Gamma - n^{2}\Lambda$, where $n=\rho/m_{\rm H}$, and $\Gamma$ and $\Lambda$ are heating and cooling coefficients which are assumed to depend only on temperature.  In the ISM, $\Gamma$ decreases with increasing density as the starlight, soft X-ray, and cosmic ray flux are attenuated by the high column densities associated with dense clouds. Because the exact form of the attenuation depends on details which remain uncertain (for instance the size and abundance of PAHs), the heating rate at $\rm T \lesssim 10^{4}\;{\rm K}$ is similarly uncertain. In this work we assume that $\Gamma = 10^{-26}{\rm\,erg\,s^{-1}}$ (independent of $\rho$ or $T$).  The low-temperature ($\rm T \lesssim 10^{4}\;{\rm K}$) cooling was then adjusted to give 3 thermally stable phases at thermal pressures between $2000-6000 \;{\rm cm^{-3}}{\rm\,K}$, as required by observations. These stable phases, at temperatures $\sim 10\,$K, $\sim 150\,$K and $\sim 8500\,$K, correspond to the molecular, atomic and warm neutral/ionized phases, respectively. The cooling curve and phase diagram are shown in
\citet{Pittard11}.

The simulation uses a temperature-dependent average particle mass, $\mu$.  In the molecular phase $\mu = 2.36$, reducing to $0.61$ in ionized gas.  The value of $\mu$ is determined from a look-up table of values of $p/\rho$ \citep{Sutherland10}. A temperature independent value of $\gamma$, the ratio of specific heats, is used, and we set $\gamma=5/3$.

Simulations are performed on a $512^{3}$ grid with free outflow boundary conditions.
A number of advected scalars are included to trace the different origins of the gas - stellar wind/SN, GMC clump, and surrounding ambient ISM. 

\subsection{Stellar Evolution}
We assume that the stellar wind feedback into the surrounding GMC clump is dominated by 3 O stars with initial masses of 35\,M$_{\odot}$, 32\,M$_{\odot}$ and 28\,M$_{\odot}$, and Main Sequence (MS) mass-loss rates of 5, 2.5 and 1.5$\times 10^{-7}\,M_{\odot}\,\rm yr^{-1}$ respectively. For each star the MS wind terminal velocity is assumed to be $2000\,\rm km\,s^{-1}$. Three evolutionary phases are considered for each star.  The MS phase of the 35\,M$_{\odot}$ star lasts for 4\,Myr, after which the star becomes a Red Supergiant (RSG) and blows a dense, very slow wind characterized by $\dot{M} = 10^{-4}\,M_{\odot}\,\rm yr^{-1}$ and $v_{\infty} = 50\,\rm km\,s^{-1}$. This phase lasts for 0.1\,Myr after which the star enters the Wolf-Rayet (WR) stage, and a fast, high momentum wind is blown ($\dot{M} = 2 \times 10^{-5}\,M_{\odot}\,\rm yr^{-1}$ and $v_{\infty} = 2000\,\rm km\,s^{-1}$).  The WR phase lasts for a further 0.3\,Myr, after which the star undergoes a supernova explosion.  At this point the star's wind is switched off, and 10$^{51}$ ergs of thermal energy is imparted to the simulation along with 10\,M$_{\odot}$ of ejecta.  The other two stars remain on the MS throughout this entire evolutionary period. The 32\,M$_{\odot}$ star evolves onto the RSG branch 0.1\,Myr after the 35\,M$_{\odot}$ star explodes.  The wind values and lifetimes used are summarized in Table~\ref{evolution} and are intended to be representative of stars of these masses.

\begin{table}
\centering
\caption{Wind properties of the three stars in the cluster as they evolve.}
\begin{tabular}{c@{~}|c@{~}c@{~}c@{~}c@{~}c@{~}}
\hline
Stellar & \multicolumn{5}{|c|}{MS stage} \\
Mass & $\dot{M}$ & v$_{\infty}$ & Duration & Mtm & Energy \\
($\rm\,M_{\odot}$) & ($\rm M_{\odot}\,\rm yr^{-1}$) & (km s$^{-1}$) & (Myr) & (g cm s$^{-1}$) & (ergs) \\
\hline
\hline
35 & 5.0$\times 10^{-7}$ & 2000 & 4.0 & 8.0$\times 10^{41}$ & 8.0$\times 10^{49}$ \\
32 & 2.5$\times 10^{-7}$ & 2000 & 4.5 & 4.5$\times 10^{41}$ & 4.5$\times 10^{49}$ \\
28 & 1.5$\times 10^{-7}$ & 2000 & 5.0 & 3.0$\times 10^{41}$ & 3.0$\times 10^{49}$ \\
\end{tabular}
\begin{tabular}{c@{~}|c@{~}c@{~}c@{~}c@{~}c@{~}}
\hline
Stellar & \multicolumn{5}{|c|}{RSG stage} \\
Mass & $\dot{M}$ & v$_{\infty}$ & Duration & Mtm & Energy \\
($\rm\,M_{\odot}$) & ($\rm M_{\odot}\,\rm yr^{-1}$) & (km s$^{-1}$) & (Myr) & (g cm s$^{-1}$) & (ergs) \\
\hline
\hline
35,32,28 & 1.0$\times 10^{-4}$ & 50 & 0.1 & 1.0$\times 10^{41}$ & 2.5$\times 10^{47}$ \\
\end{tabular}
\begin{tabular}{c@{~}|c@{~}c@{~}c@{~}c@{~}c@{~}}
\hline
Stellar & \multicolumn{5}{|c|}{WR stage} \\
Mass & $\dot{M}$ & v$_{\infty}$ & Duration & Mtm & Energy \\
($\rm\,M_{\odot}$) & ($\rm M_{\odot}\,\rm yr^{-1}$) & (km s$^{-1}$) & (Myr) & (g cm s$^{-1}$) & (ergs) \\
\hline
\hline
35,32,28 & 2.0$\times 10^{-5}$ & 2000 & 0.3 & 2.4$\times 10^{42}$ & 2.4$\times 10^{50}$ \\
\hline
\end{tabular}
\label{evolution}
\end{table}

\subsection{Initial Conditions}
The cluster wind blows into a turbulent and inhomogeneous GMC clump, whose structure is based on the work of \citet{Vasquez-Semadeni08} of turbulent and clumpy molecular clouds (specifically model Ms24J6).  This model has a nominal Mach number of 15, is isothermal, and has no magnetic field.  We scale these results to create a GMC clump of radius 4\,pc and mass 3240\,M$_{\odot}$ (SimA). This gives an average density of $\approx 8 \times 10^{-22}\,{\rm g\,cm^{-3}}$, or a molecular hydrogen number density $n_{\rm H_{2}} \approx 250\,{\rm cm^{-3}}$. The clump initially has a uniform temperature of about 10\,K, and is in rough pressure equilibrium with a surrounding uniform medium of density $3.33\times10^{-25}\,{\rm g\,cm^{-3}}$ ($n_{\rm H} \approx n_{\rm e} \approx 0.2\,{\rm cm^{-3}}$) and temperature 8000\,K.

The hydrodynamic grid covers a cubic region of $\pm 16\,$pc extent centered on the GMC clump. The cluster wind is injected as purely thermal energy within a radius of 0.375\,pc (6 cells). The densest regions cool to 1\,K, which is the imposed temperature floor. 

We also explore another model where the GMC clump radius is 5\,pc and where the density and pressure of the GMC clump and its surroundings are twice as great (SimB). This produces a clump mass of 10,500\,M$_{\odot}$ (due to the inhomogeneity of the clump it is not quite $3.9\times$ as massive as the clump in SimA).  Unless otherwise noted, all results are for SimA.

\subsection{Neglected Processes and Simplifications}
This work is the first step in examining the feedback from a stellar cluster into surrounding molecular material left over from its formation. As such we necessarily make many simplifications and approximations.

Our simulations do not include gravity, thermal conduction, magnetic fields, radiation pressure, photoionization, dust, cosmic rays or radiative losses {\em within} the stellar cluster. We estimate from the parameters noted in Table~\ref{evolution} that the winds have comparable momentum to the radiation fields of the stars (both emit momentum of roughly $10^{43}\,{\rm g\,cm\,s^{-1}}$ over the lifetime of the cluster\footnote{Note that in an optically thick medium where each photon is absorbed and reemitted multiple times, the momentum deposited is limited by the energy rather than by the momentum of the radiation field.}). The key factor in determining the relative importance of the winds and radiation fields to the dynamics of the gas is the strength of the coupling of the radiation field to the gas. If the degree of wind and photon leakage out of the cluster is comparable, then radiation pressure will provide just an order-unity enhancement to the wind pressure \citep{Krumholz09}.
We note that our neglect of direct radiation pressure and gravity offset each other to some degree.
However, the self-gravity of some of the densest structures in our cluster may be important on the timescales that we consider. 

Inclusion of thermal conduction and photoevaporation should speed up the destruction of molecular material. Our simulations reveal that the densest cloud fragments have $\rho \sim 10^{-19}\,{\rm g \, cm^{-3}}$ and radii of about $0.1$\,pc. At a distance of 3\,pc from a stellar cluster emitting $10^{50}$ ionizing photons per second, we estimate that the mass-loss rate from photoevaporation\citep[see, e.g.,][for the relevant equations]{Pittard07} is $\sim 4 \times 10^{20} \,{\rm g \, s^{-1}}$, giving a lifetime of about 1.5\,Myr. This is comparable to the mass-loss rate and lifetime due to hydrodynamic ablation. So we should expect slightly quicker destruction of molecular material than occurs in our simulations through ablation alone. Having said this, photoevaporation may be suppressed in regions where the ram or thermal pressure of the surrounding medium is greater than the pressure of the evaporating flow \citep{Dyson94}. Given these considerations, we believe that our simulations should be reasonably representative even with our neglect of photoevaporation. Neglecting thermal conduction may also make little difference to our results given that it will be confined to regions where there is a temperature gradient parallel to the magnetic field.

We also note that dust can dramatically affect the cooling within hot bubbles if it can be continuously replenished, perhaps by the evaporation/destruction of dense clumps \citep{Everett10} which will also mass-load the bubble \citep[see, e.g.,][]{Pittard01a,Pittard01b}. The presence of dust will also affect the photoionization rate throughout the cluster. Clearly the effects of dust warrant study in future work.

Radiative losses {\em within} the cluster can significantly change the energy flux into the surrounding environment \citep{Silich04}. However, this is not a significant effect for the cluster parameters we have considered, and is likely important only for the most massive clusters and super star clusters.

A final consideration concerns the initial conditions. In this first work we have used the results of simulations of a turbulent ISM, truncated at a radius of 4 or 5\,pc around the stellar cluster.
In future we will attempt to construct models which are more self-consistent, by investigating feedback into structures formed from colliding flows, clouds and filaments.

\section{Results}
\label{sec:results}

In this section we present our results. We examine the initial blowout (Sec.~\ref{sec:blowout}), then the evolution up to $t=4\,$Myr (Sec.~\ref{sec:ms}). In Sec.~\ref{sec:rsgwrevol} we focus on the feedback during the evolution of the stars through their various evolved stages (RSG and WR) and their subsequent explosions as SNe. We examine the mass and energy fluxes into the wider surroundings in Sec.~\ref{sec:fluxes} and the evolution of the column density in Sec.~\ref{sec:columndensity}. A comparison between SimA and SimB, plus the evolution of the molecular mass is made in Sec.~\ref{sec:molecularmass}.

\subsection{Initial Blowout}
\label{sec:blowout}
\begin{figure*}
\centering
\includegraphics[width=0.31\textwidth, height=0.31\textwidth]{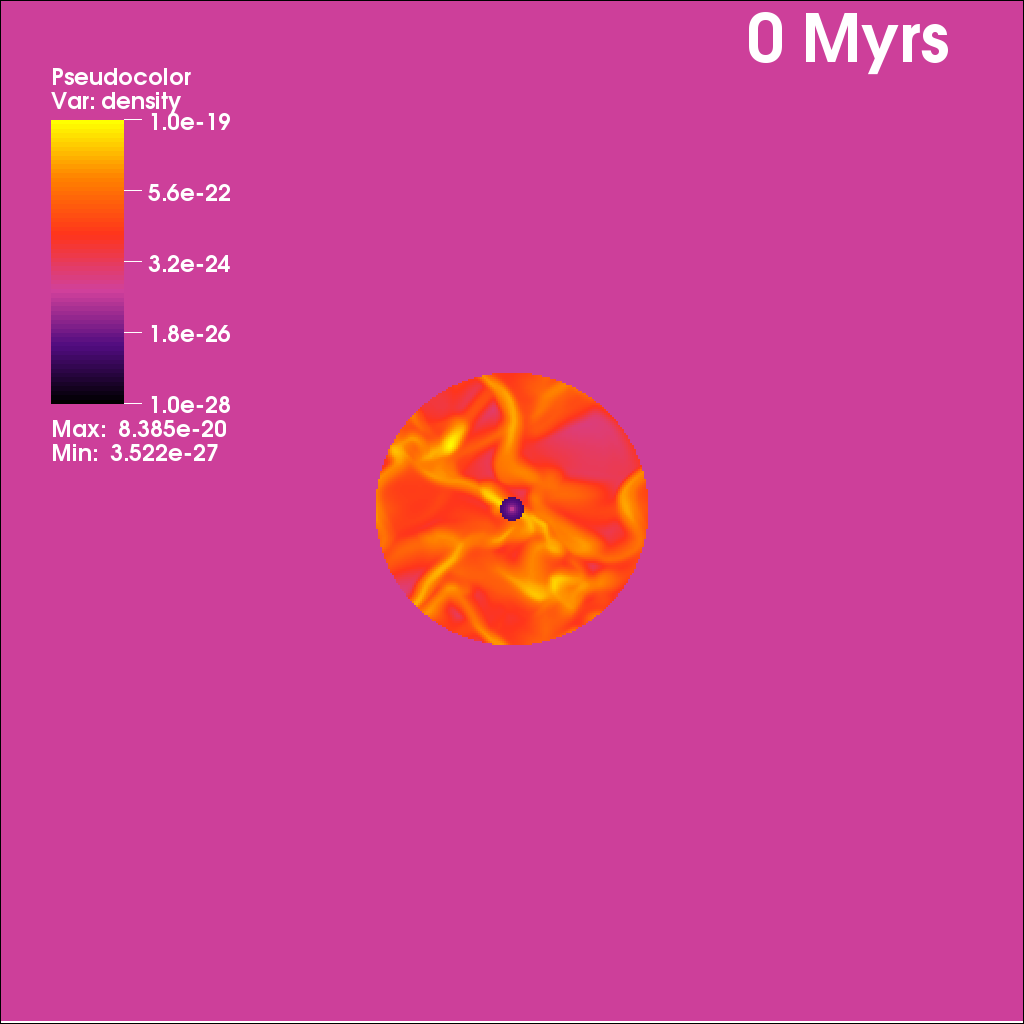}
\includegraphics[width=0.31\textwidth, height=0.31\textwidth]{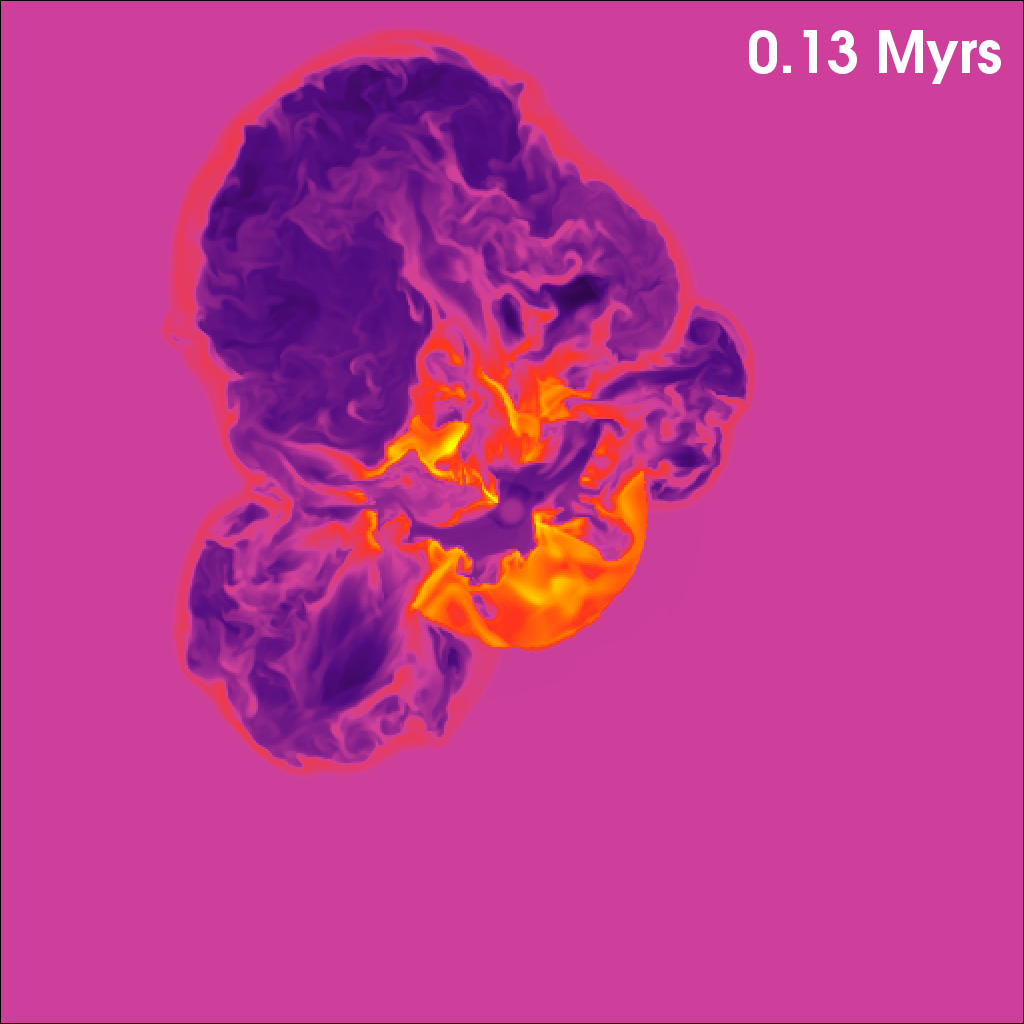}
\includegraphics[width=0.31\textwidth, height=0.31\textwidth]{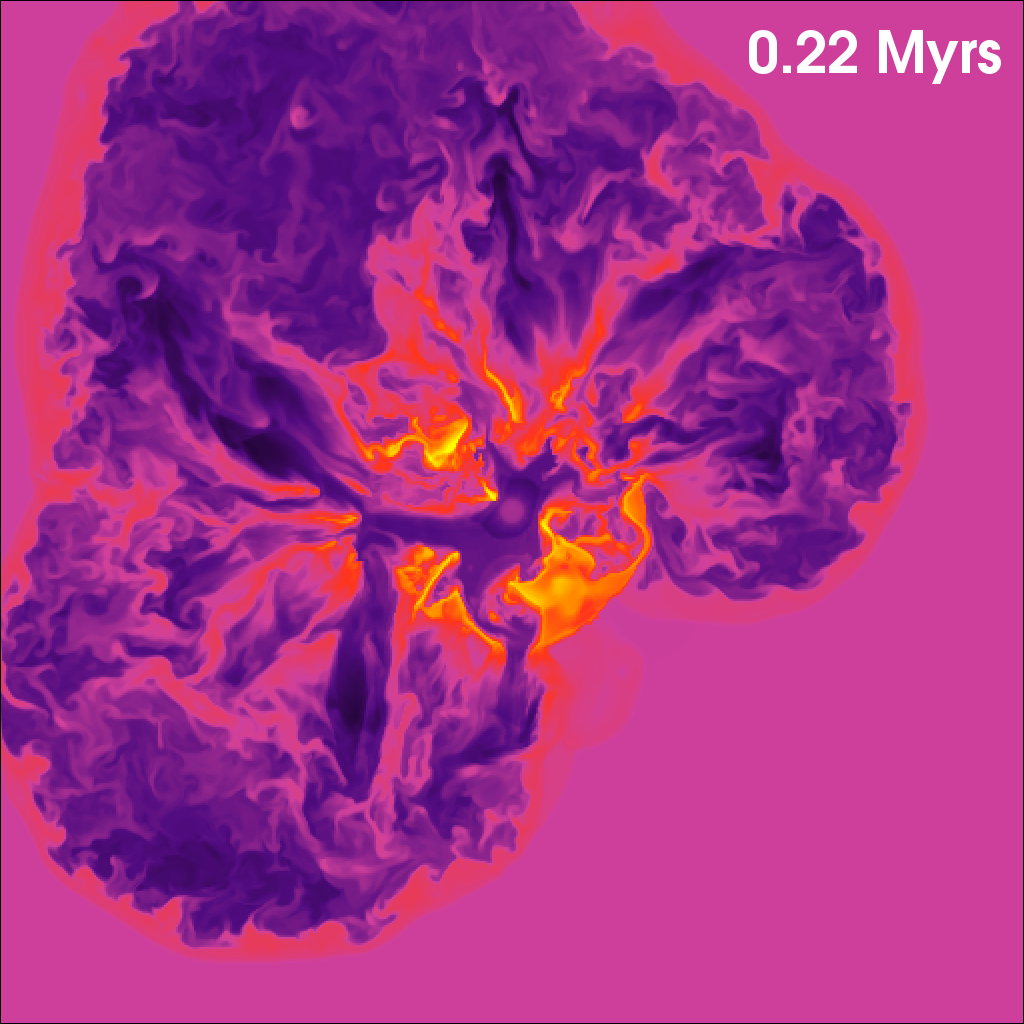}
\caption{Density slices through the 3D simulation (SimA) in the xy-plane during the initial interaction of the cluster wind with the GMC clump.  The density scale is shown in the top left panel. \label{initialblowout}}
\end{figure*}

\begin{figure*}
\centering
\includegraphics[width=0.31\textwidth, height=0.31\textwidth]{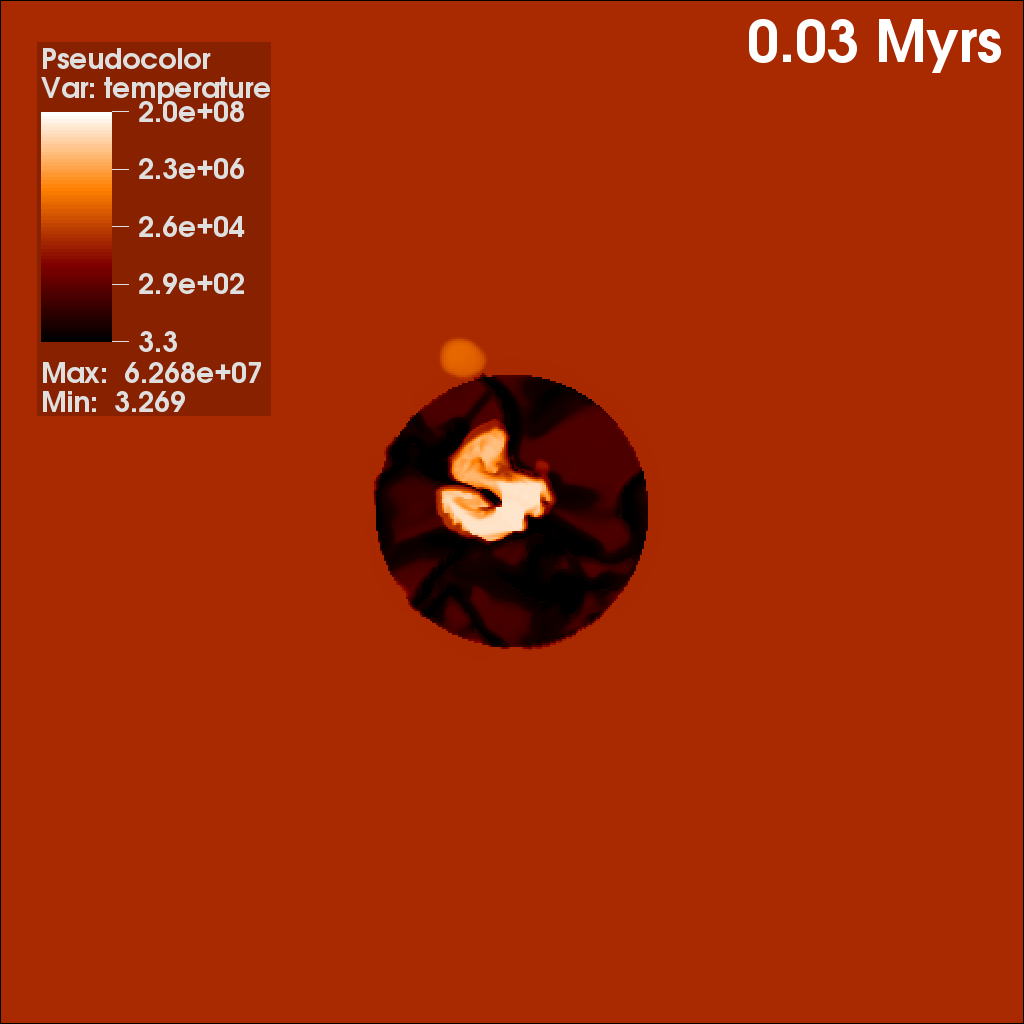}
\includegraphics[width=0.31\textwidth, height=0.31\textwidth]{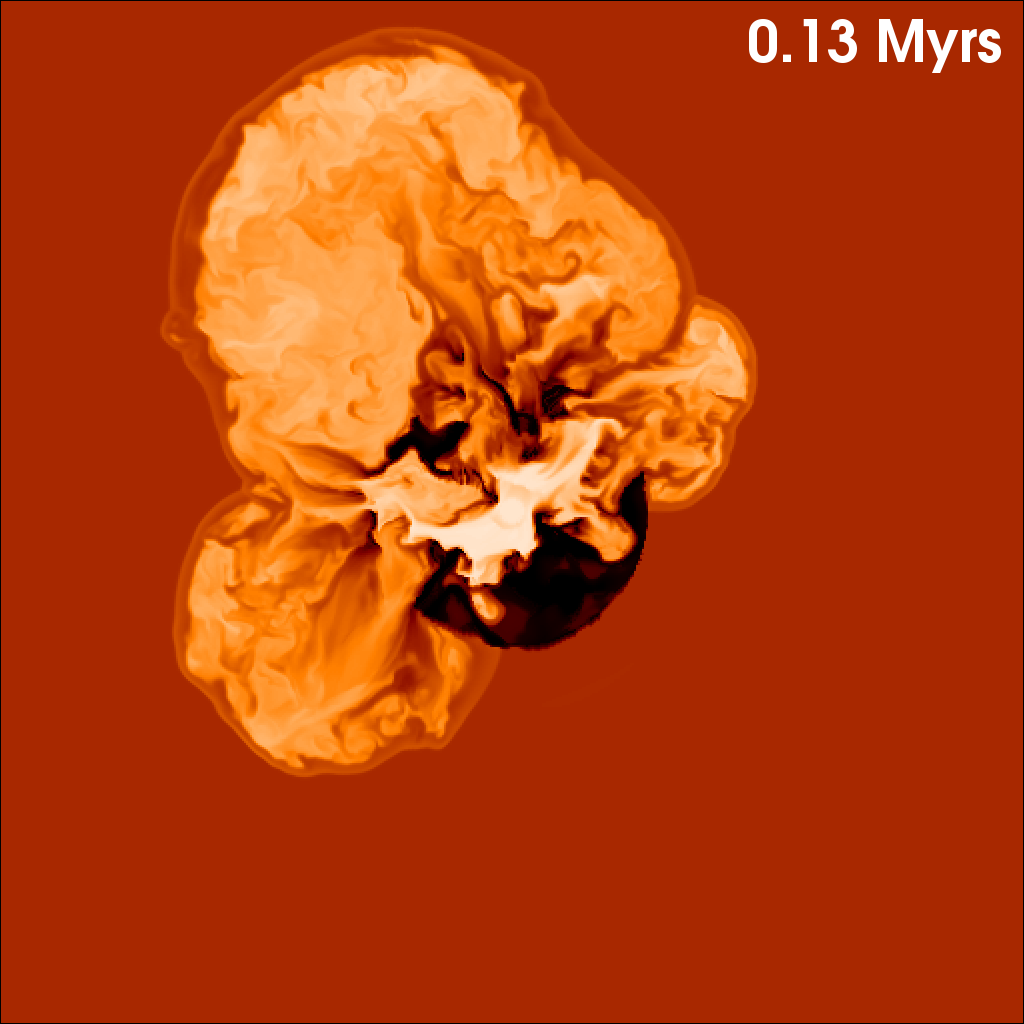}
\includegraphics[width=0.31\textwidth, height=0.31\textwidth]{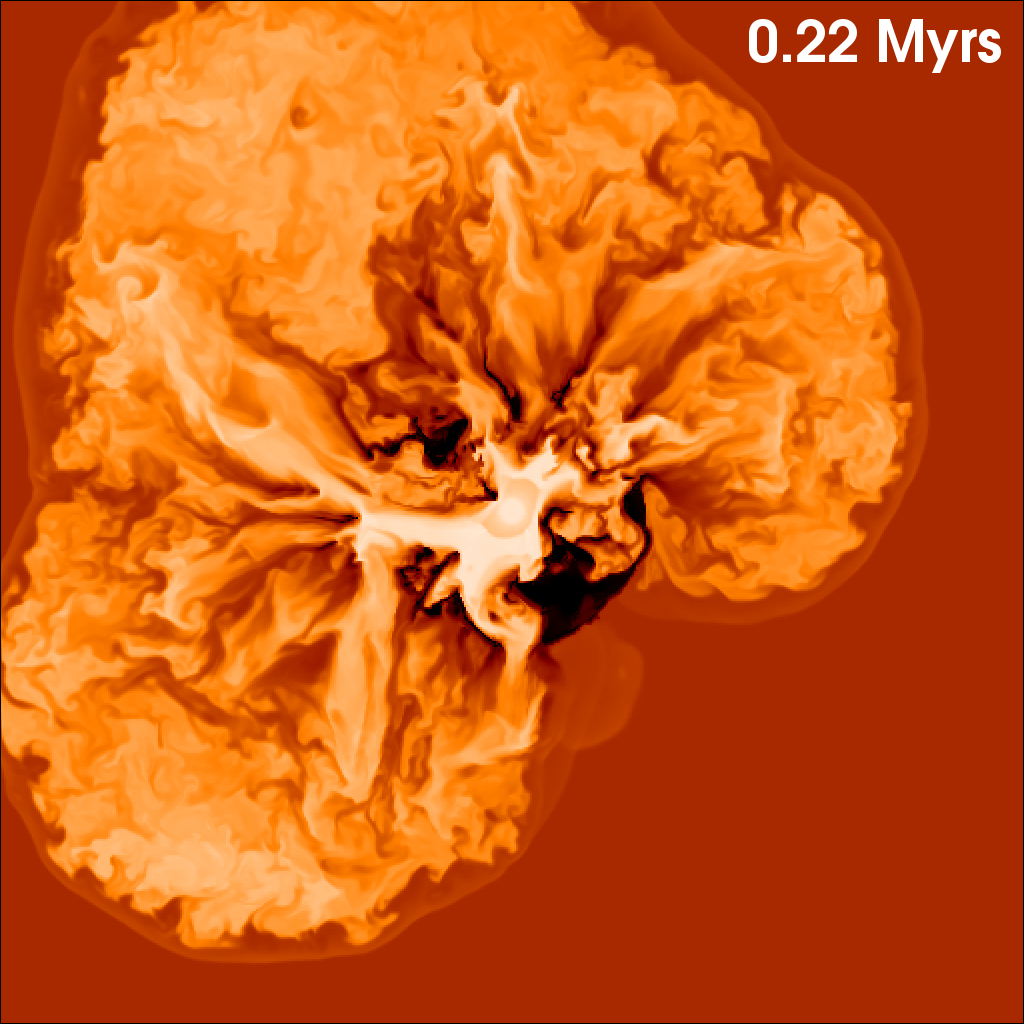}
\caption{Temperature slices through the 3D simulation (SimA) in the xy-plane during the initial interaction of the cluster wind with the GMC clump.  The temperature scale is shown in the left panel. \label{blowouttemp}}
\end{figure*}

The cluster wind creates a high pressure and high temperature bubble within the GMC clump which expands most rapidly into regions of lower density.  The initial blowout of this bubble into the lower density medium surrounding the GMC clump occurs at $t \sim 0.03$\,Myr, and is very aspherical due to inhomogeneity of the GMC clump (Fig.~\ref{initialblowout}). This blowout occurs much faster than for a GMC clump with an equivalent uniform density - in such a case a 1D spherical bubble expands to a radius of 4\,pc in 0.29\,Myr, which is $\sim$\,10 times slower than for the inhomogeneous GMC clump used in our models.  Isolated blowouts at a number of distinct positions around the surface of the GMC clump rapidly grow and merge so that the entire clump is quickly surrounded by hot, expanding wind material. As this wind material streams out through low density channels, clump material is ablated into the flow and the clump gradually loses mass.  Fig.~\ref{blowouttemp} shows snapshots of the temperature evolution during this initial phase at $t = 0.03$, 0.13 and 0.22\,Myr.  Here the hot wind is clearly visible streaming through the cold GMC.  It is noticeable that dense parts of the clump can shield and protect less dense material in their ``shadow'', though the ability of the hot, high pressure gas to flow around denser objects mitigates this effect to some extent.  This behaviour is apparent in Fig.~\ref{initialblowout} where material towards the bottom right of the clump is protected against ablation from the cluster wind by intervening dense material.  

A swept-up shell around the outer edge of the bubble is visible in both Figs.~\ref{initialblowout} and~\ref{blowouttemp}. In this simulation the shell is reasonably thick and confines the hotter gas in the bubble interior, but for bubbles expanding into a higher density medium the shell is thinner and less stable. In such situations the shell fragments, and mini blowouts occur. These expand for a certain distance, before stalling, and merging back into the ``global'' shell \citep[see also][]{Meaburn88}.

At this early stage by far the strongest shock in the simulation is the reverse shock near the central stellar cluster. The reverse shock is not spherical - instead its position is largely defined by the presence and location of the dense molecular material nearest to the stellar cluster. The shock heated cluster wind experiences a large range of densities, temperatures and velocities as it flows out of the GMC clump. At times it interacts subsonically with its environment, and at other times this interaction is supersonic. Relatively weaker shocks occur within the shock heated wind as it flows into and past the densest molecular gas. The outflow is generally quite turbulent in nature. This can affect the rate at which material is stripped from clouds \citep{Pittard09}. The rate at which clouds lose mass through hydrodynamic ablation has been investigated in detail by \citet{Pittard10}.

\subsection{Evolution during the star's MS stage}
\label{sec:ms}
\begin{figure*}
\centering
\includegraphics[width=0.31\textwidth, height=0.31\textwidth]{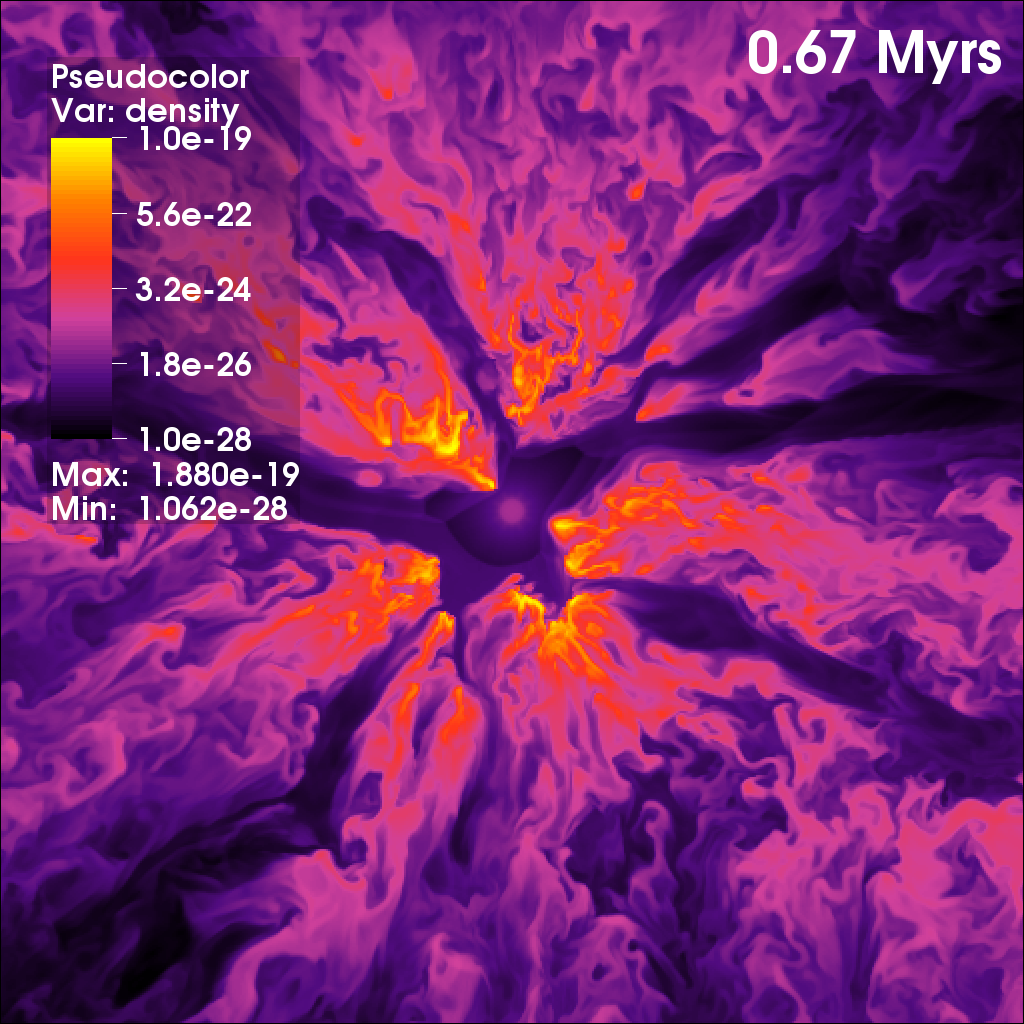}
\includegraphics[width=0.31\textwidth, height=0.31\textwidth]{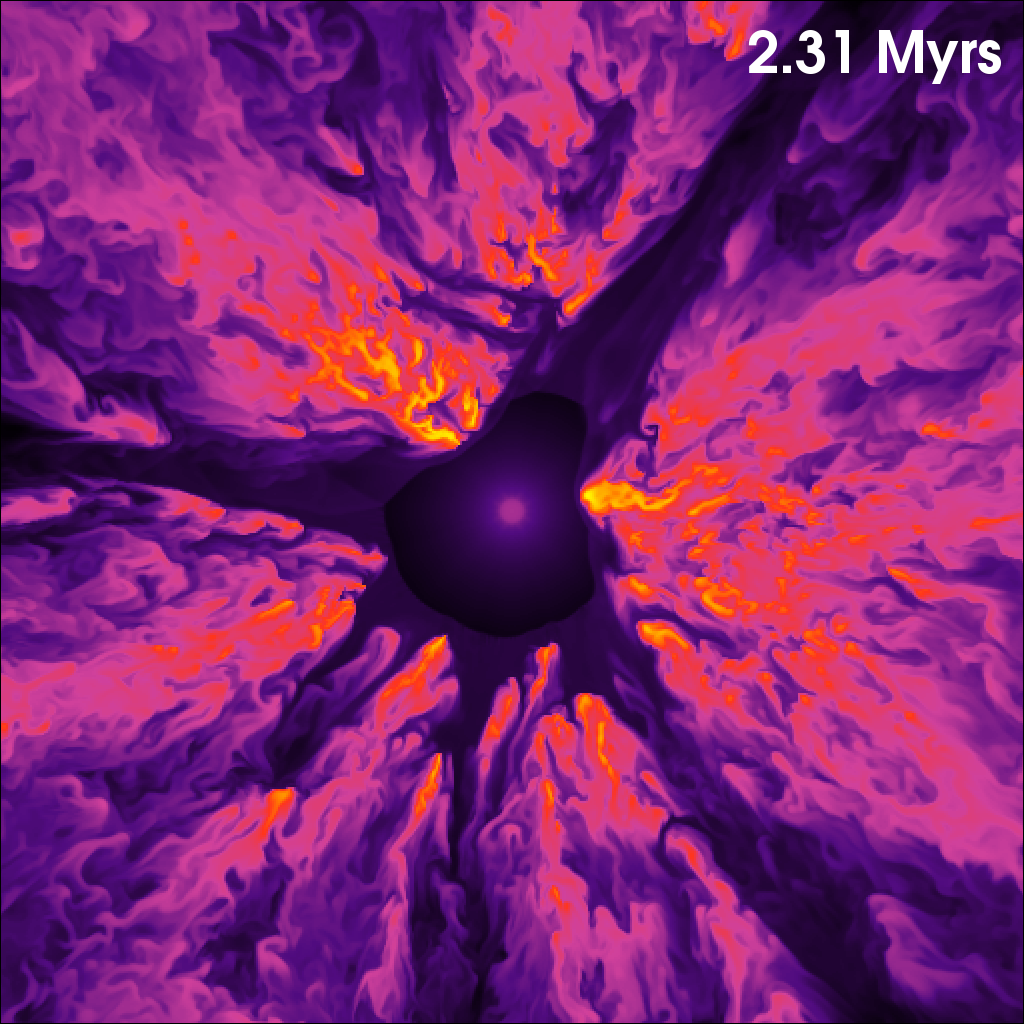}
\includegraphics[width=0.31\textwidth, height=0.31\textwidth]{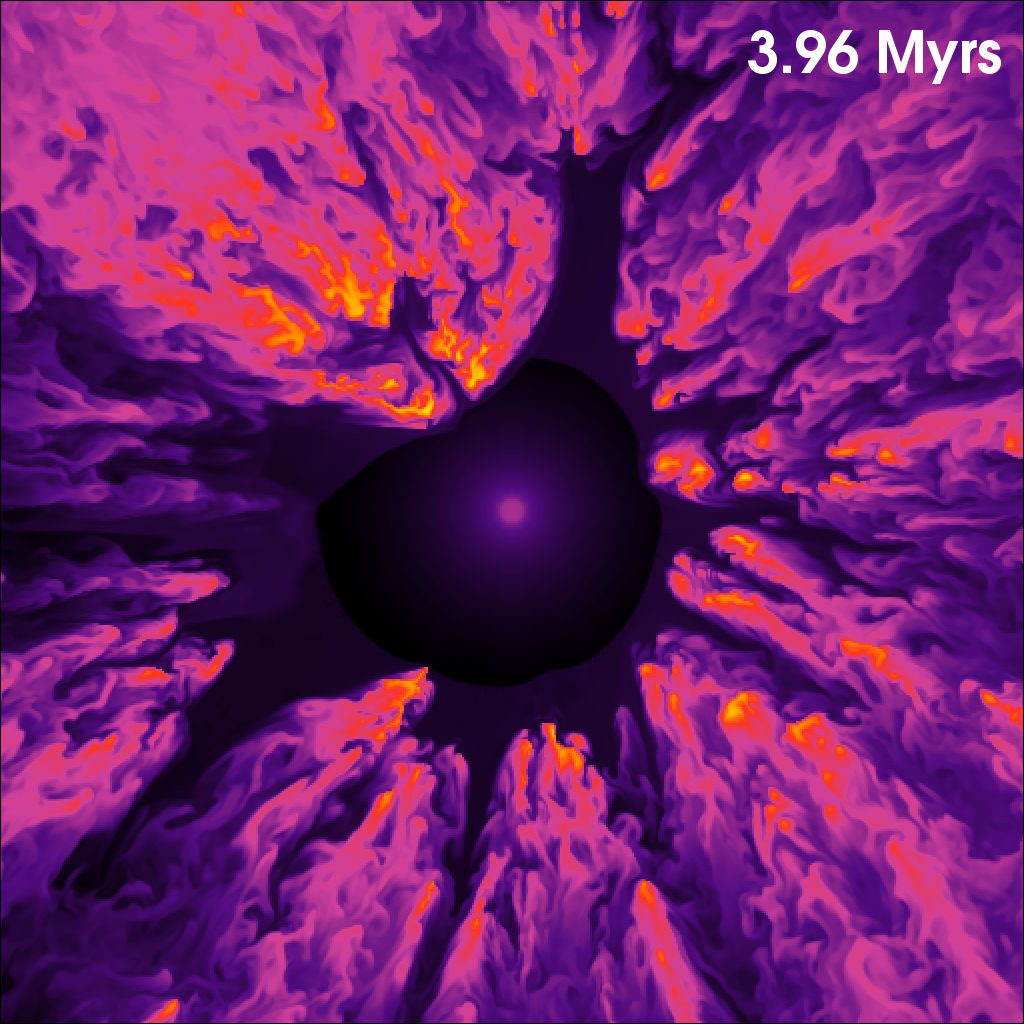}
\caption{Density slices during the MS phase of the simulation (SimA) in the xy-plane.  The last panel shows the density of gas in the cluster environment shortly before the most massive star transitions to a RSG.  The channels carved by the cluster wind in the GMC clump structure slowly evolve over this period.  The density scale is shown in the left panel. \label{MS1}}
\end{figure*}

Fig.~\ref{MS1} shows the early evolution of the cluster environment after the initial blow out has occured but while all three stars remain on the MS.  The low density channels through the GMC clump have left their imprint on the outflowing cluster wind as similarly low density channels.  These channels contain hot ($\sim\,\rm10^7\rm\,K$), fast flowing gas which is relatively unimpeded by dense gas along its route, with a typical velocity of around 1000 $\rm\,km\,s^{-1}$.  The flows are mass-loaded as denser material along their edges is mixed in.  The orientation of the channels alters slightly as the simulation progresses due to the small velocity dispersion ($\approx$ few km\,s$^{-1}$) of the dense molecular material in the GMC clump causing its structure to change with time. By about $t = 3$\,Myr the position of the channels seems to have settled and they are reasonably stable.

\begin{figure*}
\centering
\includegraphics[width=0.31\textwidth, height=0.31\textwidth]{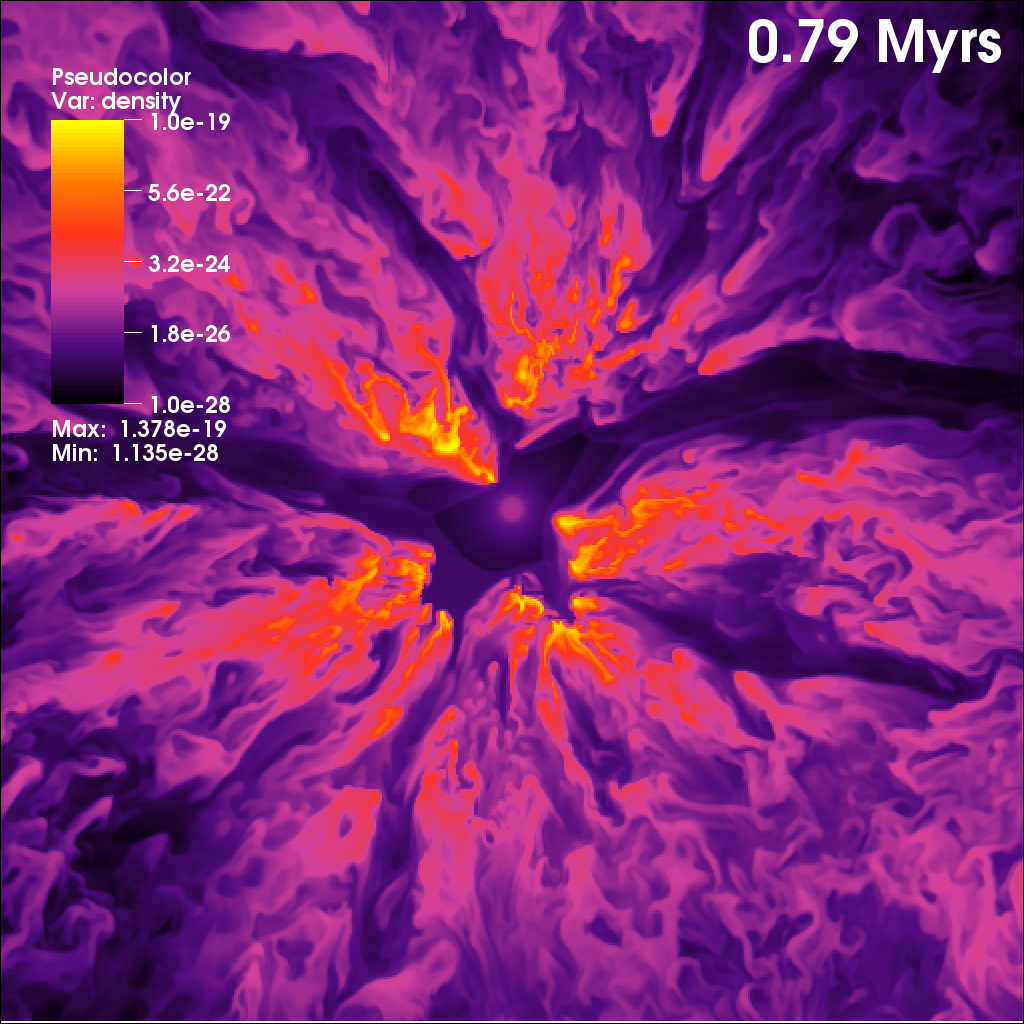}
\includegraphics[width=0.31\textwidth, height=0.31\textwidth]{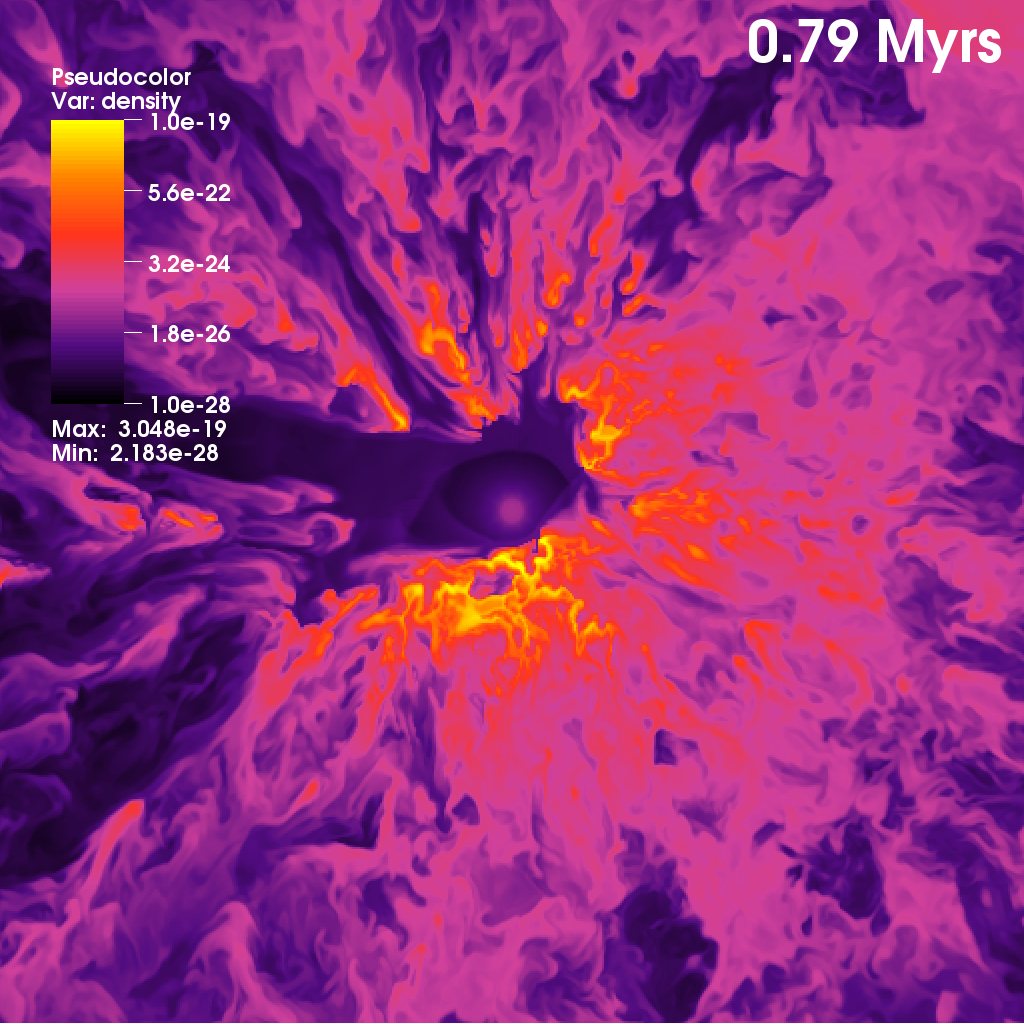}
\includegraphics[width=0.31\textwidth, height=0.31\textwidth]{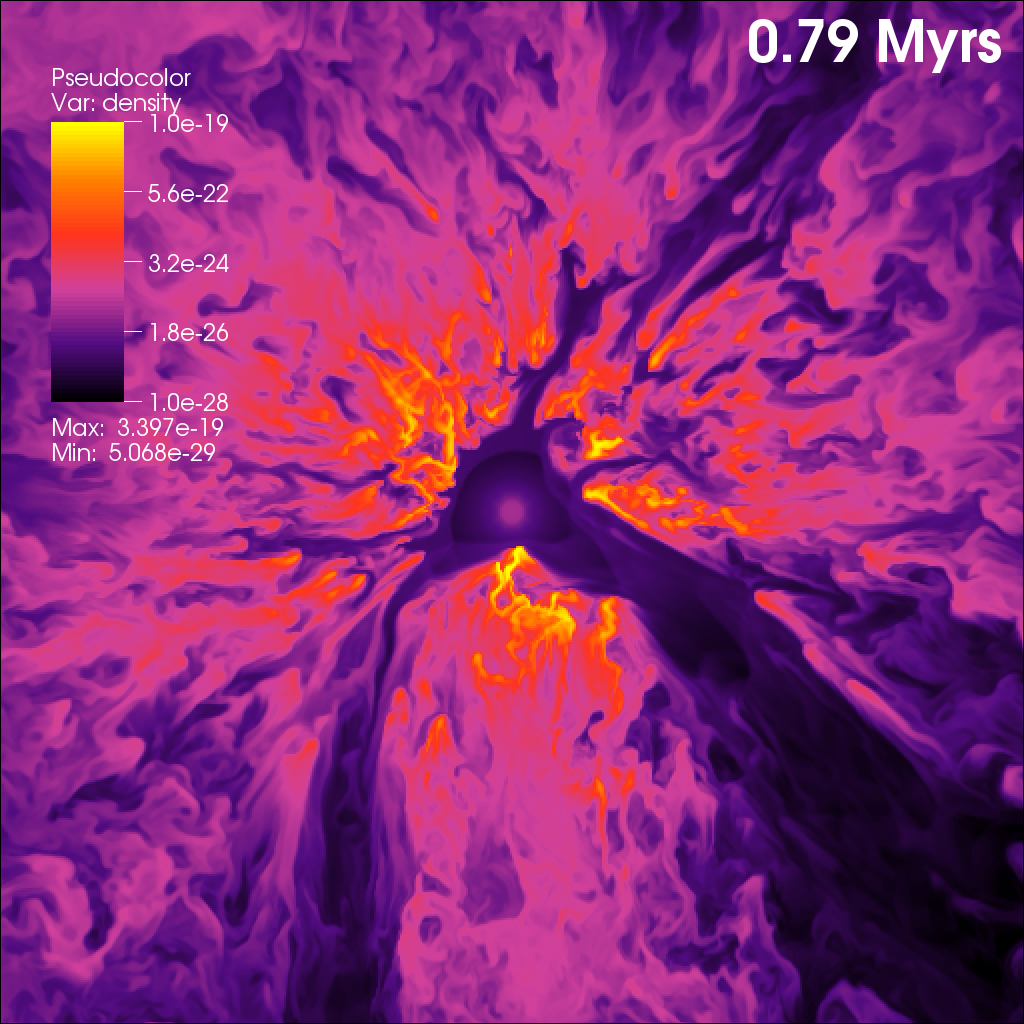}
\caption{Density slices in three planes from SimA at $t = 0.79$\,Myr. [Left]: xy [Middle]: xz and [Right]: yz. \label{compearly}}
\end{figure*}

\begin{figure*}
\centering
\includegraphics[width=0.31\textwidth, height=0.31\textwidth]{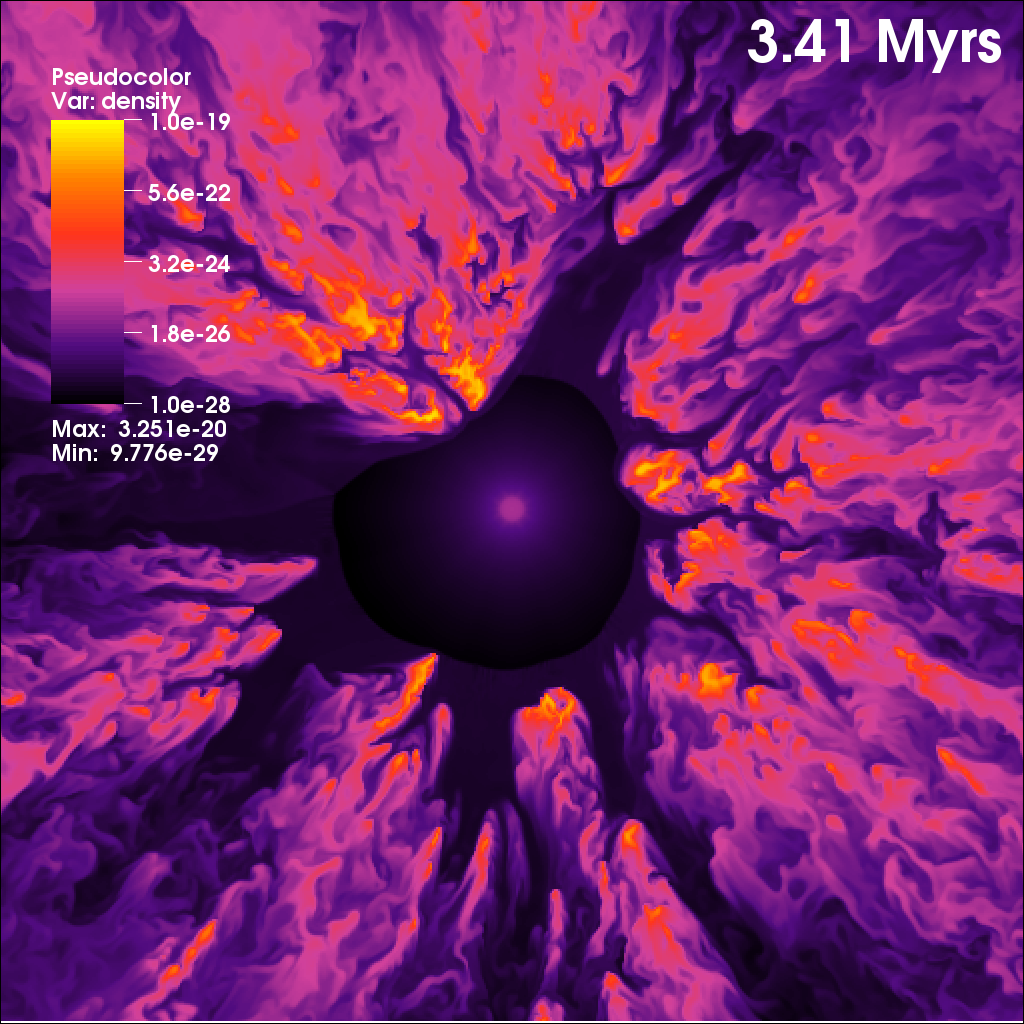}
\includegraphics[width=0.31\textwidth, height=0.31\textwidth]{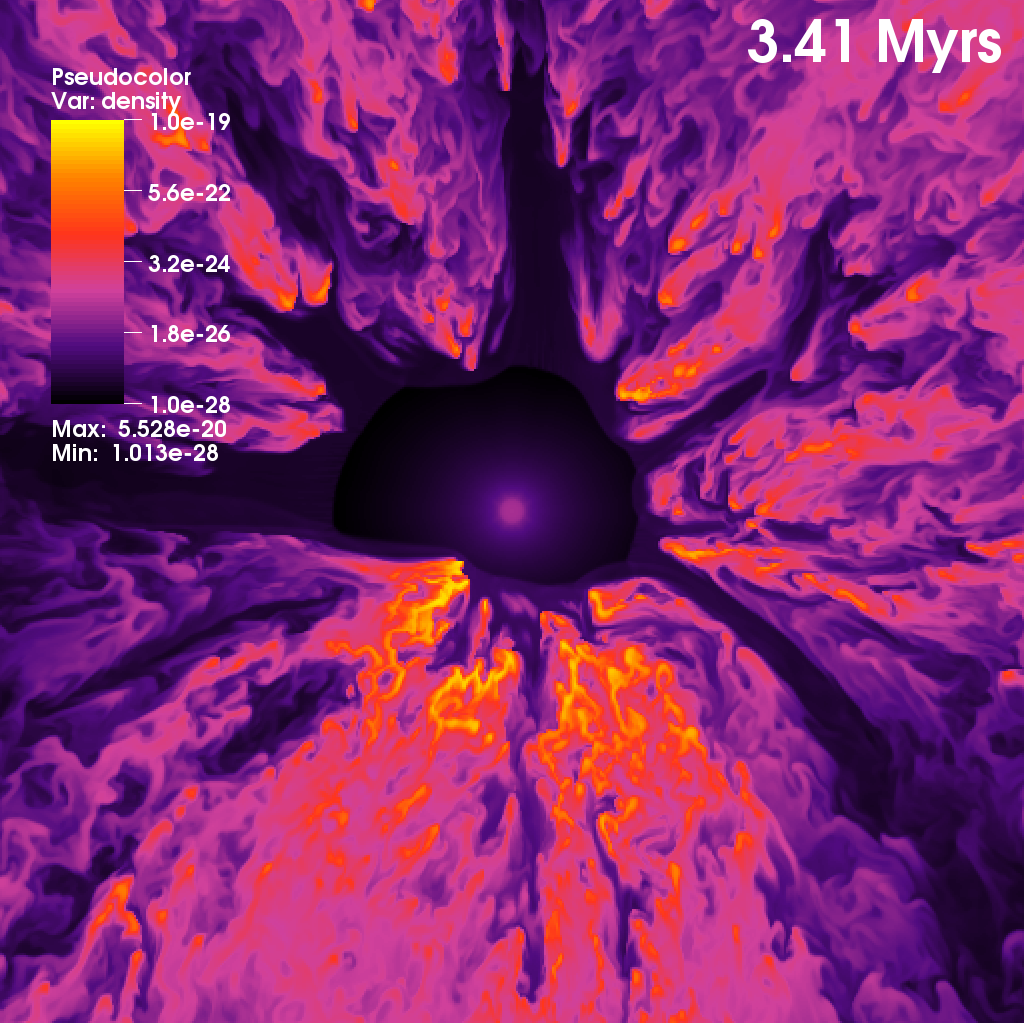}
\includegraphics[width=0.31\textwidth, height=0.31\textwidth]{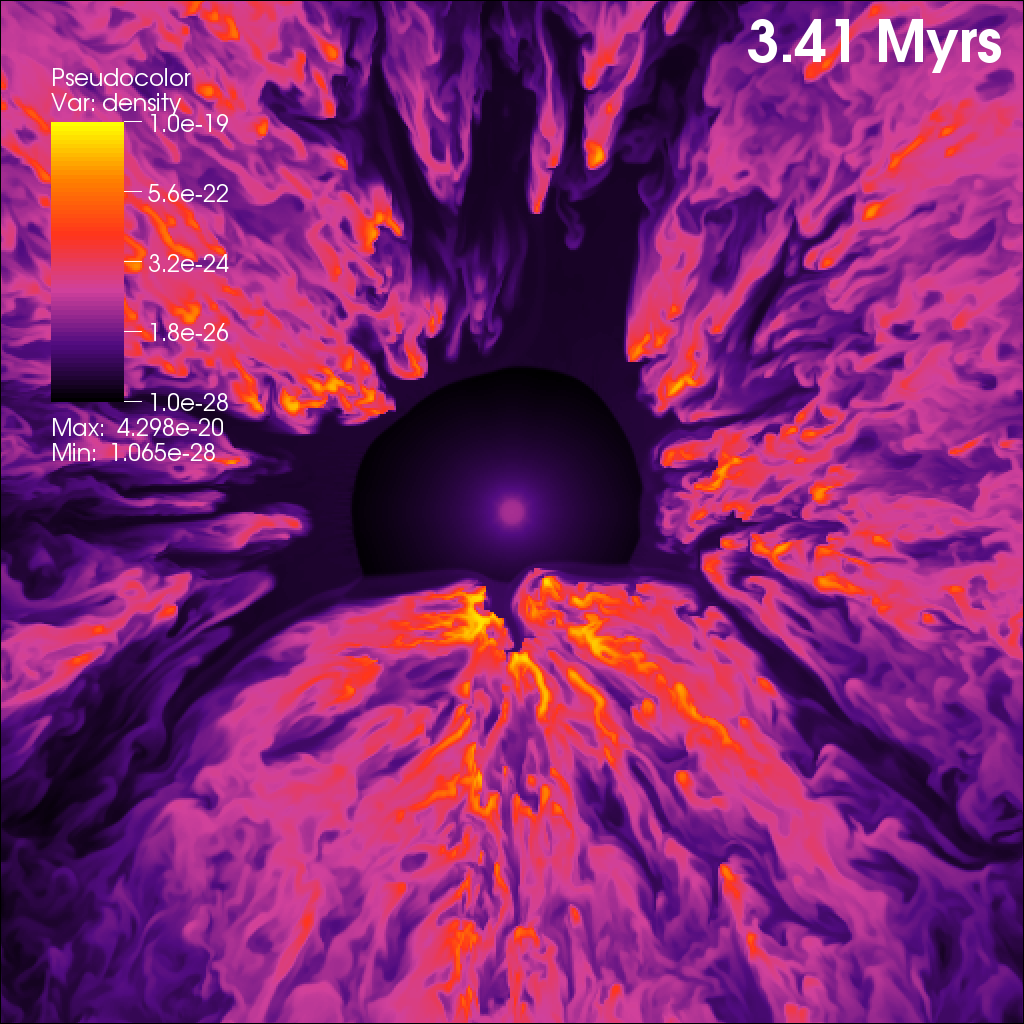}
\caption{Density slices in three planes from SimA at $t = 3.41$\,Myr. [Left]: xy [Middle]: xz and [Right]: yz. \label{complate}}
\end{figure*}

Fig.~\ref{compearly} shows slices of the simulation at $t = 0.79$\,Myr in three different planes, while Fig.~\ref{complate} shows the situation at $t = 3.41$\,Myr.  The channels carved by the cluster wind are also evident in the panels in these figures, and the changes in orientation of the channels is apparent between the two times.  The left panels in Figs.~\ref{compearly} and~\ref{complate} correspond to the xy-plane, which is the default used for the other 2D slices within this paper.  However, the center and right panels in Figs.~\ref{compearly} and~\ref{complate} show that dense clump material remains closer, for longer, to the central star cluster in the xz and yz-planes. 

Together, Figs.~\ref{MS1}-\ref{complate} reveal that the reverse shock expands and becomes gradually more spherical with time as the cluster wind drives out more material and high density material continues to ablate away. By $t = 4$\,Myr the radius of the reverse shock has increased to $\sim 5$\,pc, though its radius is $\approx 3$\,pc at the position of the closest dense cloud to the centre of the cluster.

\begin{figure}
\centering
\includegraphics[width=0.45\textwidth, height=0.20\textwidth]{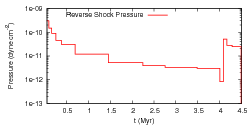}
\caption{Pressure at the reverse shock as a function of time. \label{RSpressure}}
\end{figure}

Fig.~\ref{RSpressure} shows the pressure at the reverse shock as a function of time.
The pressure steadily declines throughout the main sequence phase of the most massive star, reaching a value of $3 \times 10^{-12}\,{\rm dyn\,cm^{-2}}$ at $t = 4$\,Myr. It is difficult to know a priori what value to expect for the reverse shock pressure. At $t=1$\,Myr, Eq.~22 of \citet{Weaver77} gives $P_{bub} = 1.6 \times 10^{-10}\,{\rm dyn\,cm^{-2}}$ assuming an ambient density of $9\times10^{-22}\,{\rm g\,cm^{-3}}$ (roughly the average density of our GMC clump), and $P_{bub} = 1.4 \times 10^{-12}\,{\rm dyn\,cm^{-2}}$ assuming an ambient density of $3.33\times10^{-25}\,{\rm g\,cm^{-3}}$ (the density of the medium outside of our GMC clump). Compared to our measured pressure of $\approx 10^{-11}\,{\rm dyn\,cm^{-2}}$ at this time, we see that the former estimate is too high, while the latter is too low.
Fig.~5 of \citet{Harper-Clark09} reveals that $P_{bub}$ is about a factor of $4-5$ lower, when the covering fraction $C_{f} \sim 0.3-0.6$, than the \cite{Weaver77} estimate. This implies that our finite-sized and porous clump has an effective covering fraction $C_{f} < 0.3$.

The way that $P_{bub}$ is set is fundamental to the evolution of the flow and its affect on the GMC clump. \citet{Harper-Clark09} claim that the dynamics of a leaky bubble is set by $P_{\rm HII}$, the pressure of the ionized gas component. Their argument is that when $P_{RS}$ drops to $\approx P_{\rm HII}$, the escape of hot gas through gaps in the bubble shell slows as the hot gas is impeded by the cooler gas. However, it seems more likely that it is simply determined by the covering fraction of the shell $C_{f}$ and the ram pressure of the wind at the shell. If $C_{f}$ is reasonably high, then individual bow shocks around the shell fragments/dense clumps merge to create a single reverse shock. The reverse shock will have an increasingly large stand-off distance (and thus smaller distance from the cluster) as $C_{f}$ approaches unity. On the other hand, if $C_{f}$ is reasonably low, then the bow shocks around individual clumps maintain their identity for longer, only merging downstream to create a global reverse shock at larger radii from the cluster. Such behaviour can be identified in the simulations presented in \citet{Pittard05} and \citet{Aluzas12}.

\subsection{Later evolutionary stages}
\label{sec:rsgwrevol}

\subsubsection{Response due to the evolution of the 35$\,M_{\odot}$ star}
\begin{figure*}
\centering
\includegraphics[width=0.31\textwidth, height=0.31\textwidth]{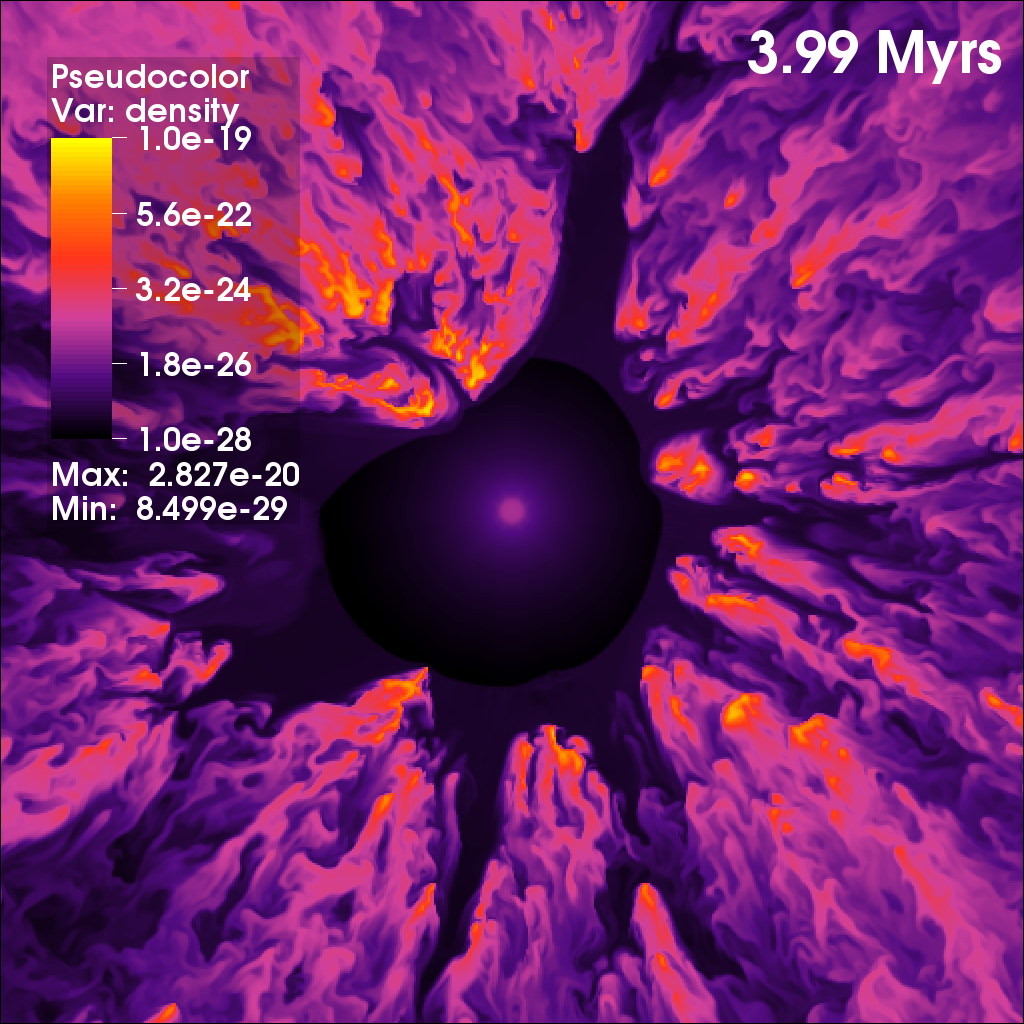}
\includegraphics[width=0.31\textwidth, height=0.31\textwidth]{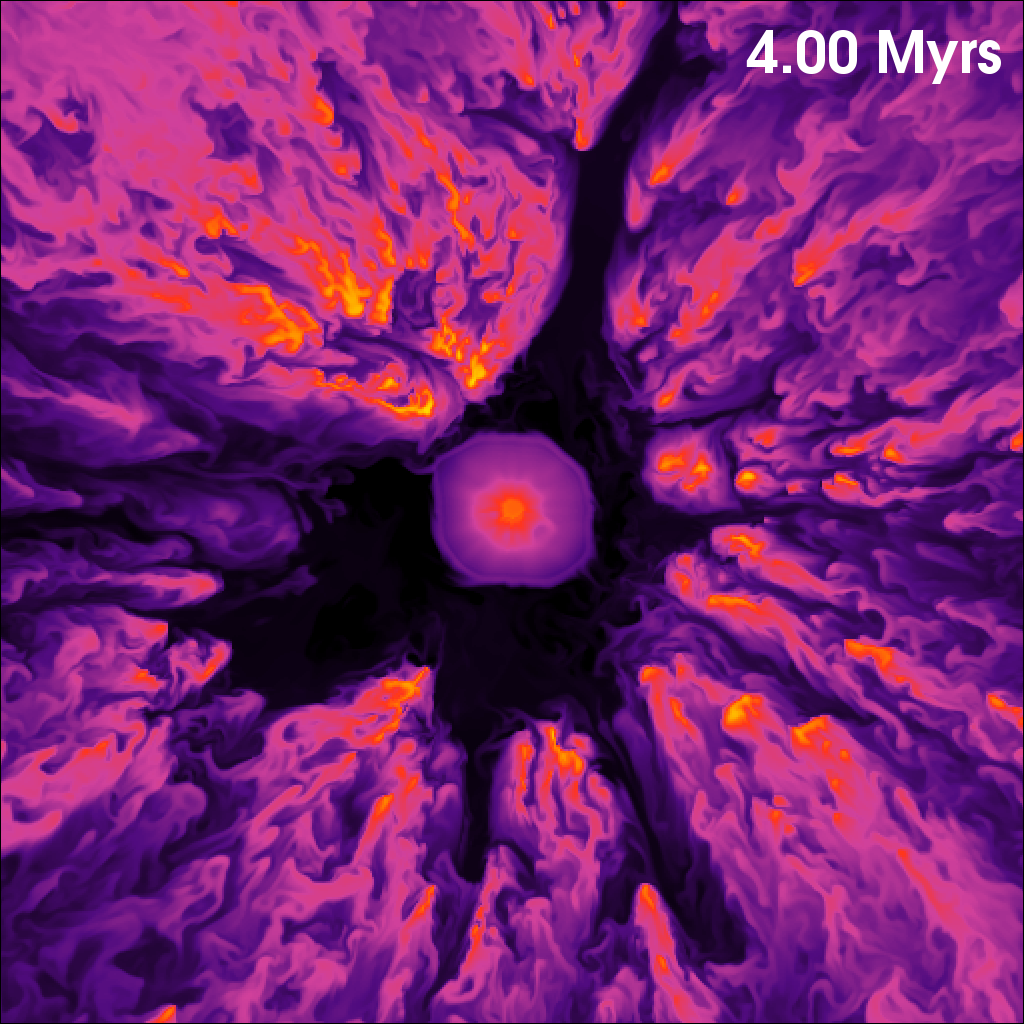}
\includegraphics[width=0.31\textwidth, height=0.31\textwidth]{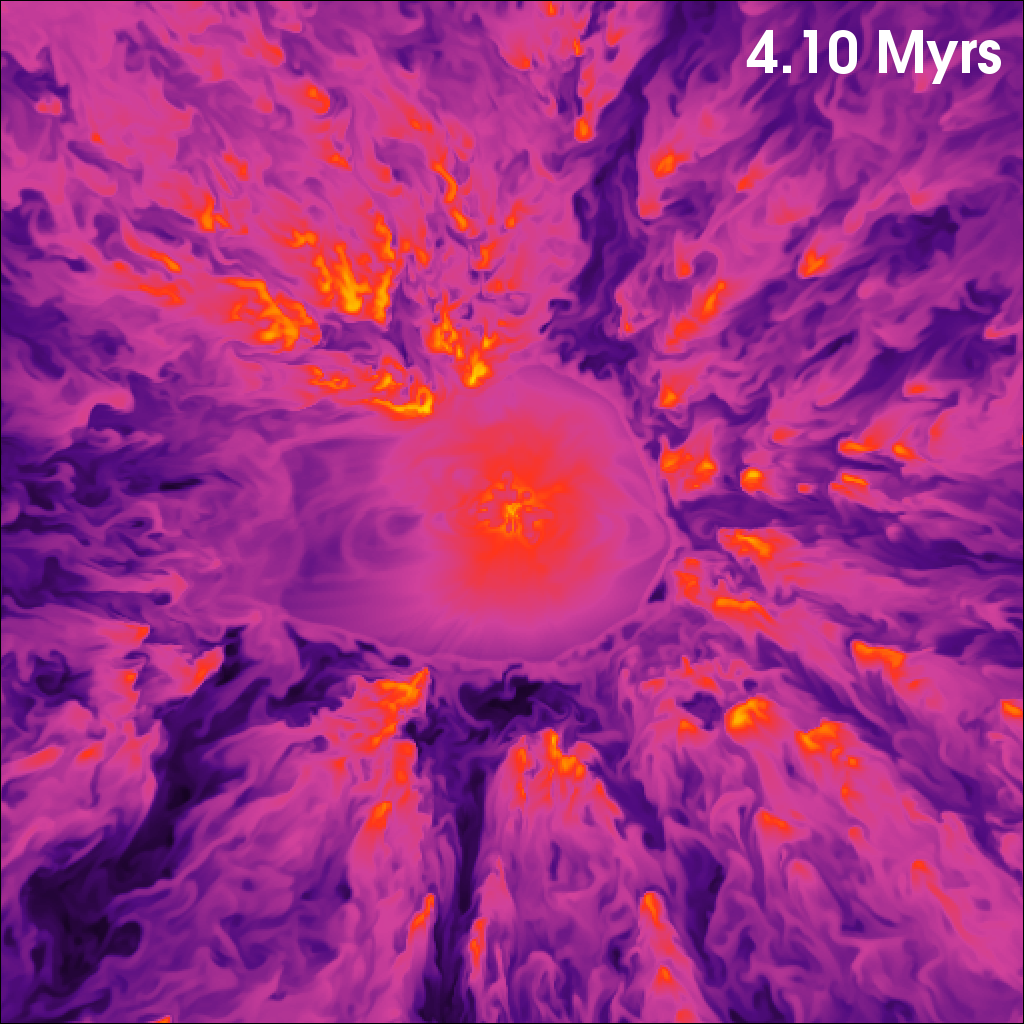}
\includegraphics[width=0.31\textwidth, height=0.31\textwidth]{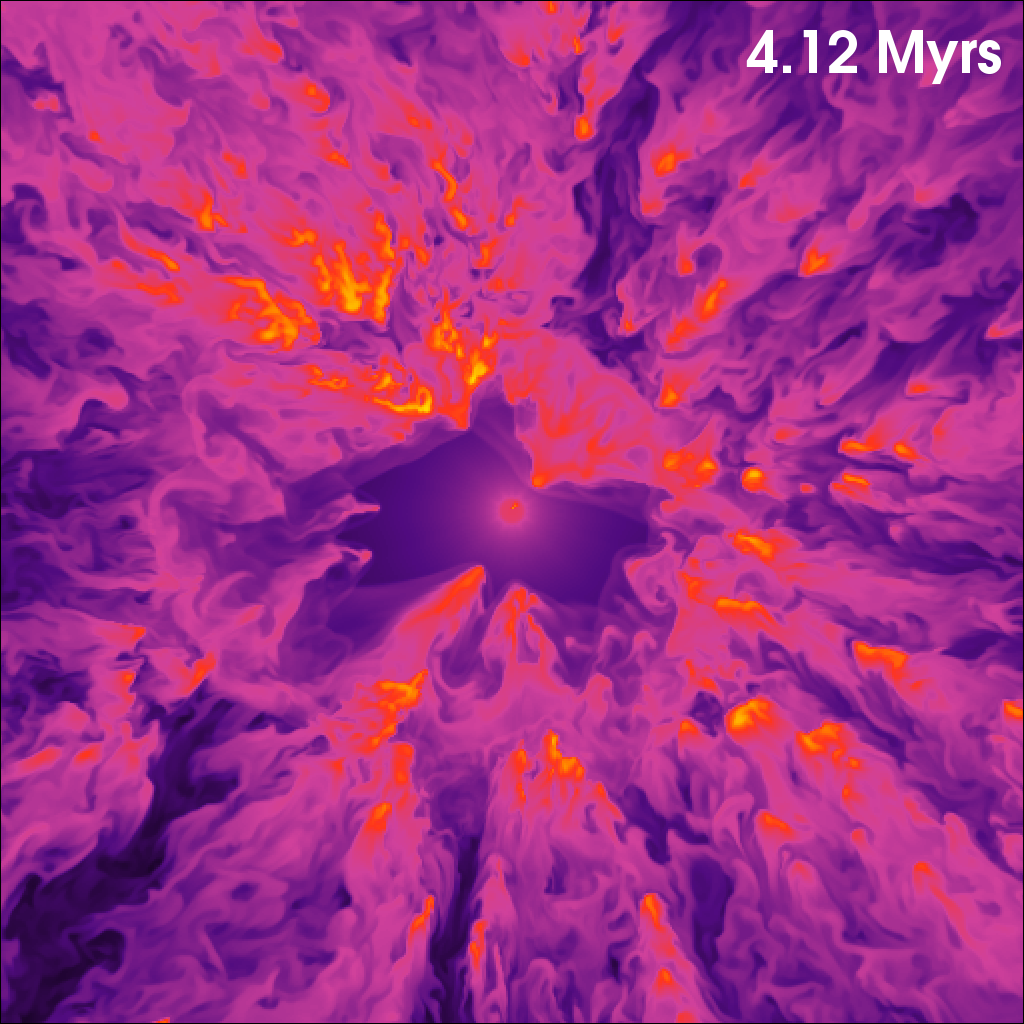}
\includegraphics[width=0.31\textwidth, height=0.31\textwidth]{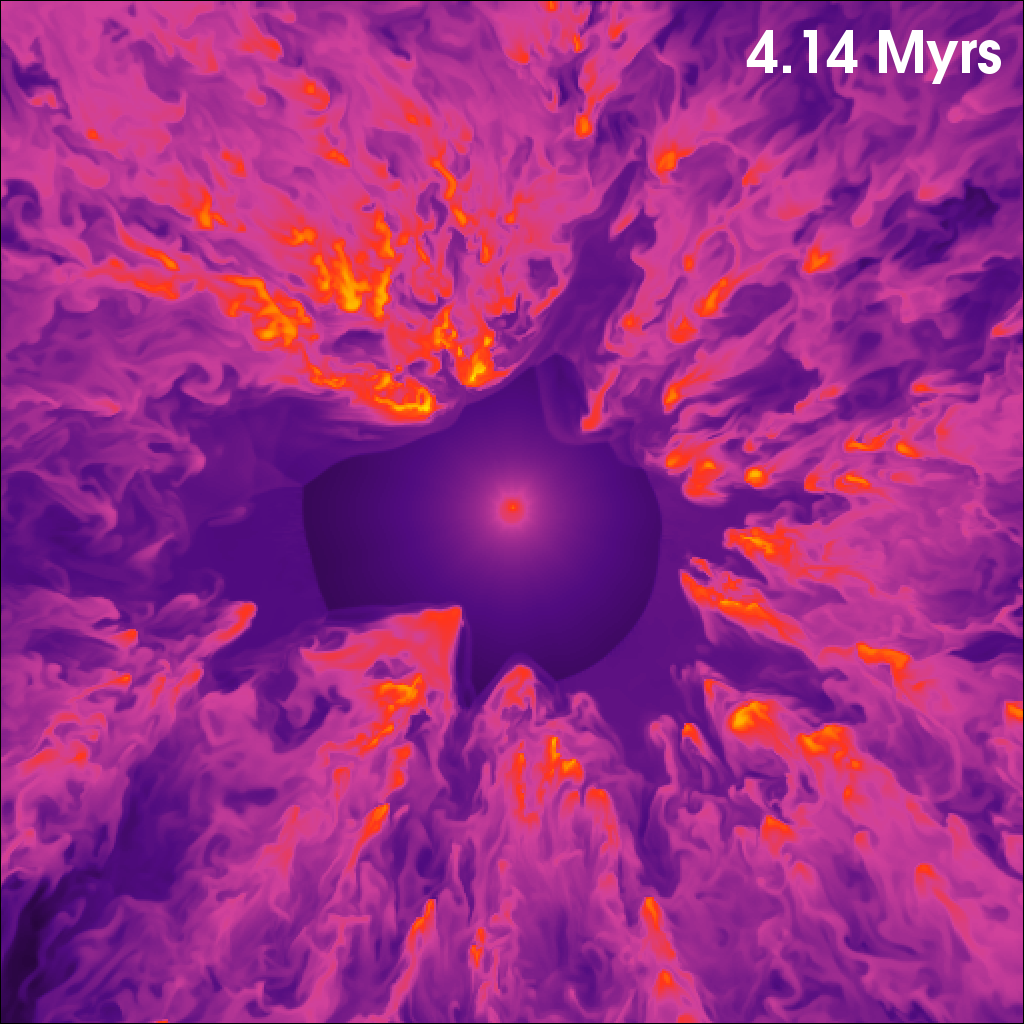}
\includegraphics[width=0.31\textwidth, height=0.31\textwidth]{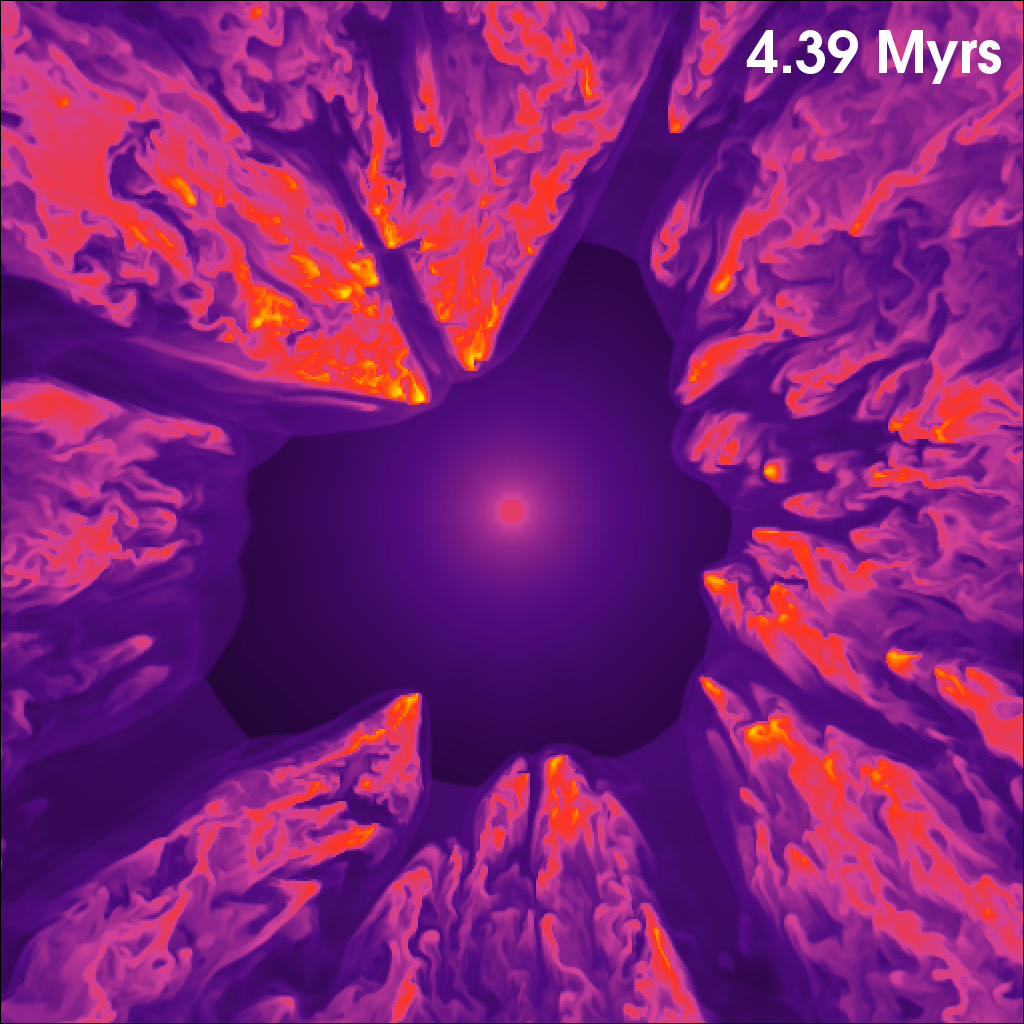}
\includegraphics[width=0.31\textwidth, height=0.31\textwidth]{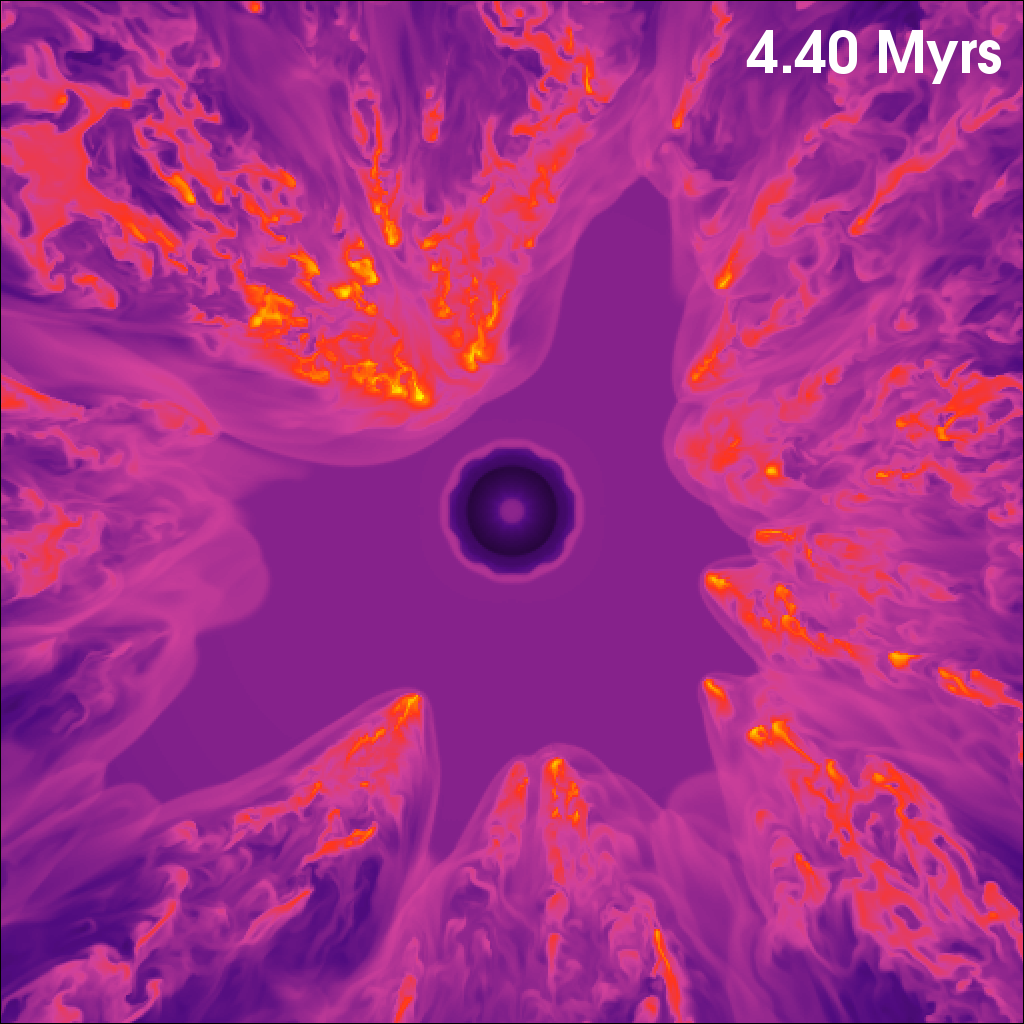}
\includegraphics[width=0.31\textwidth, height=0.31\textwidth]{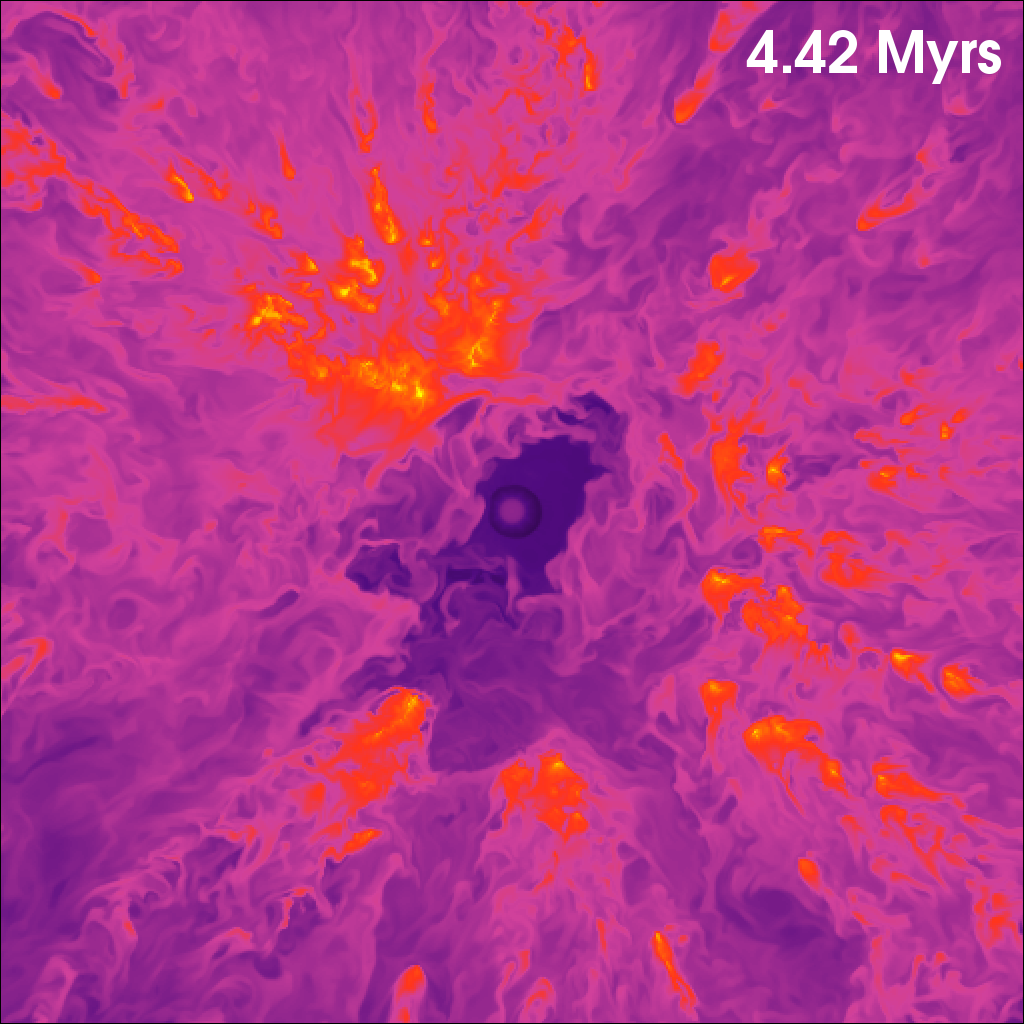}
\includegraphics[width=0.31\textwidth, height=0.31\textwidth]{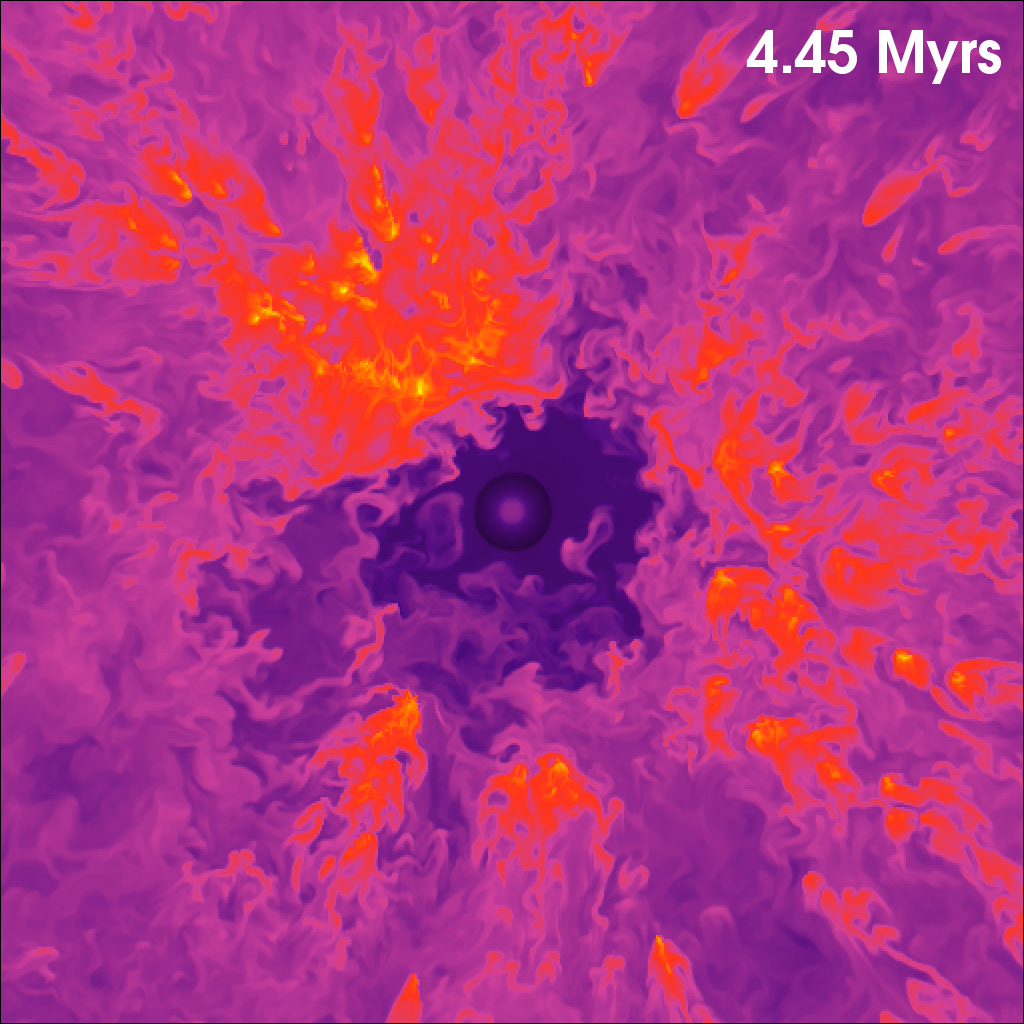}
\caption{Density slices from SimA in the xy-plane during the RSG, WR and SN stages of the highest mass star.  This star transitions from the MS to the RSG stage at $t = 4.0$\,Myr (top middle panel), from the RSG to the WR stage at $t = 4.1$\,Myr (top right panel), and explodes at $t = 4.4$\,Myr (bottom left panel).  The other two stars remain on the MS during this time.  The density scale is shown in the top left panel. \label{evo1}}
\end{figure*}

\begin{figure*}
\centering
\includegraphics[width=0.31\textwidth, height=0.31\textwidth]{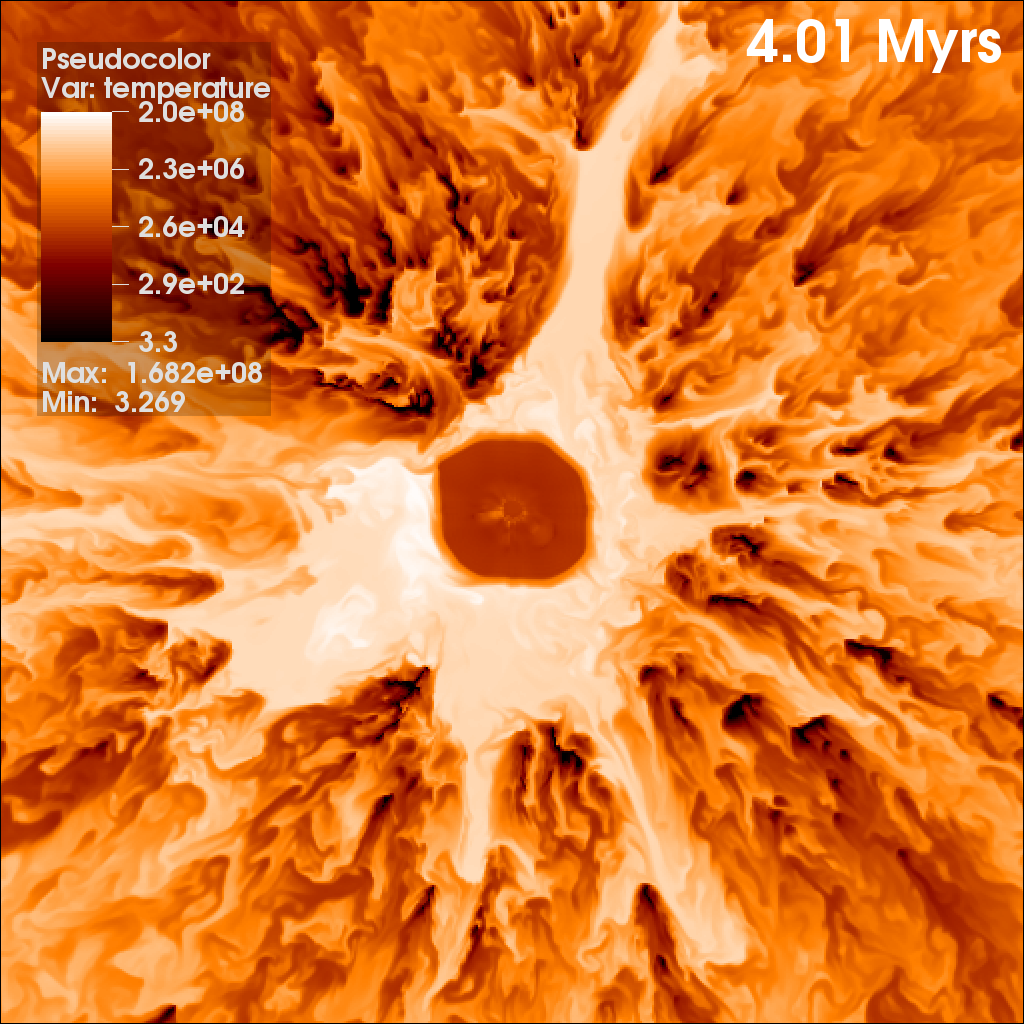}
\includegraphics[width=0.31\textwidth, height=0.31\textwidth]{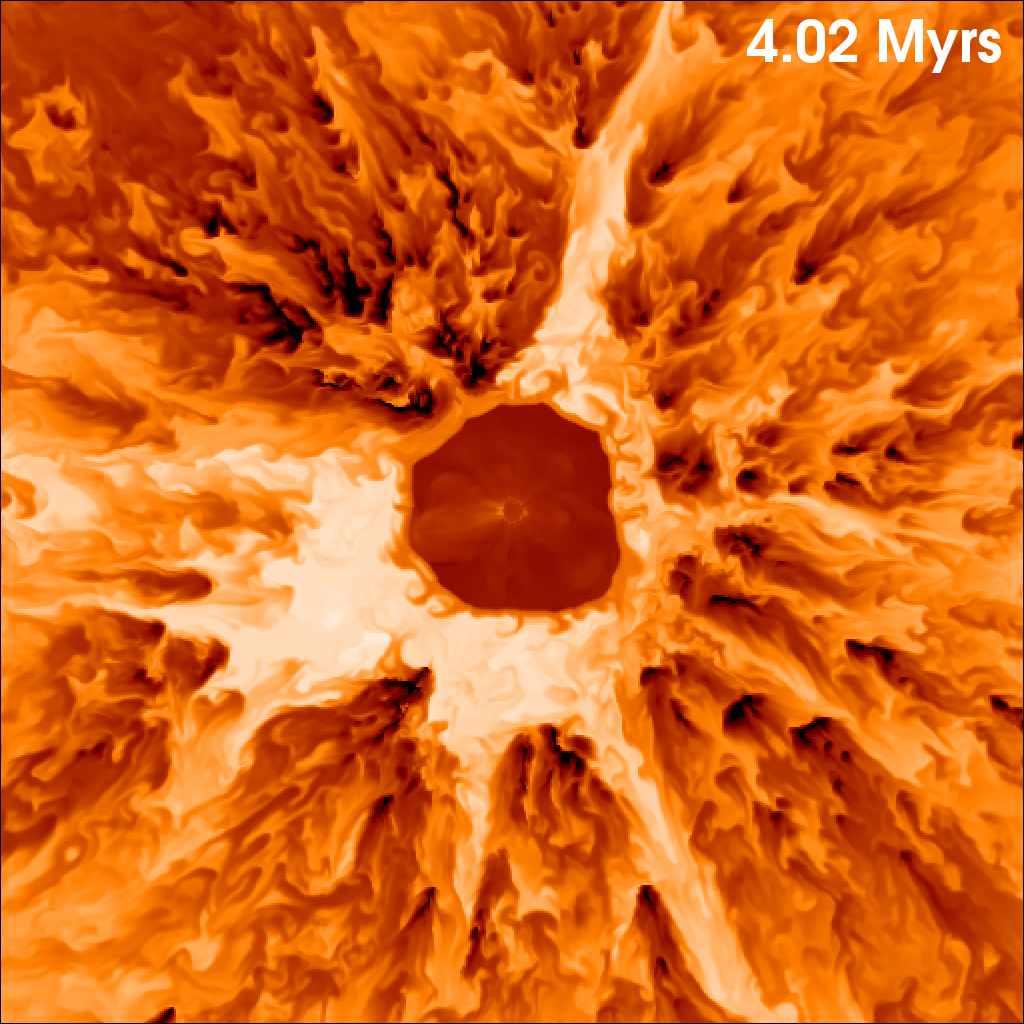}
\includegraphics[width=0.31\textwidth, height=0.31\textwidth]{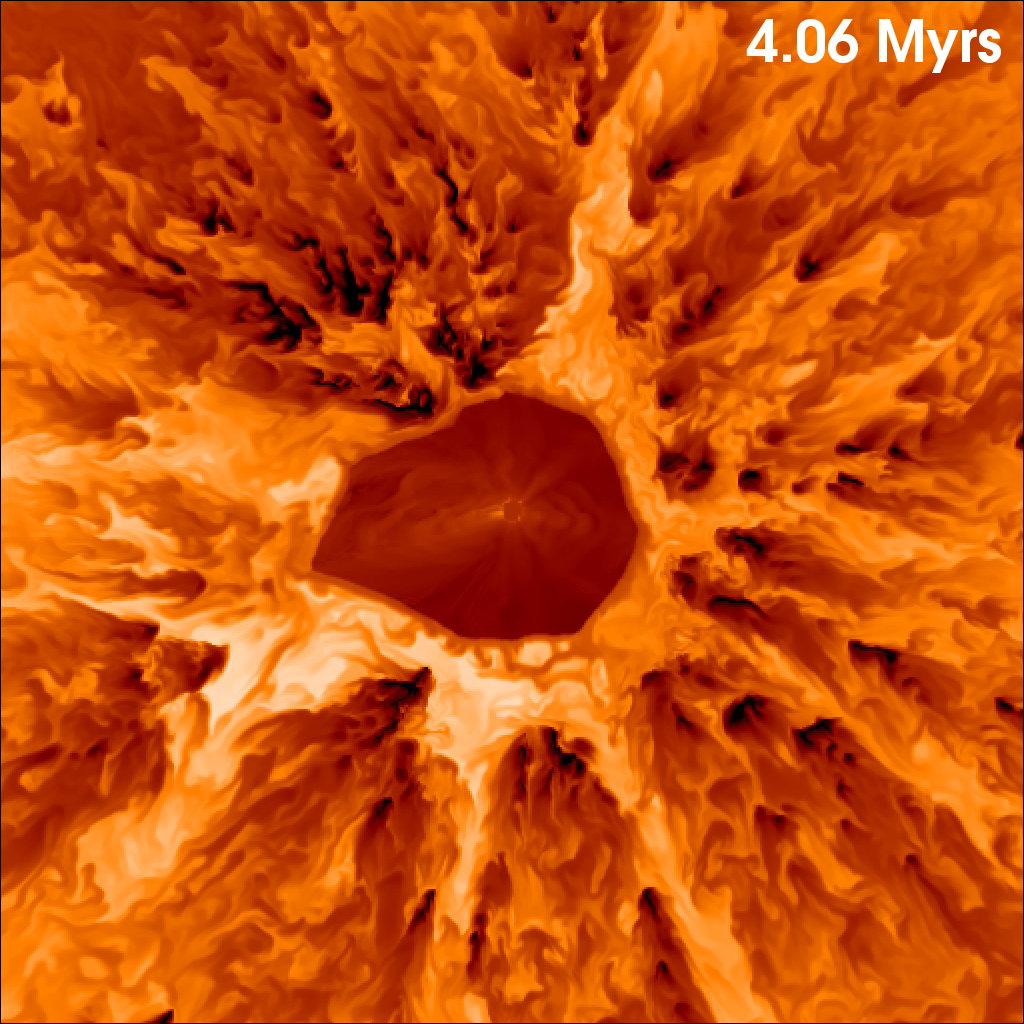}
\includegraphics[width=0.31\textwidth, height=0.31\textwidth]{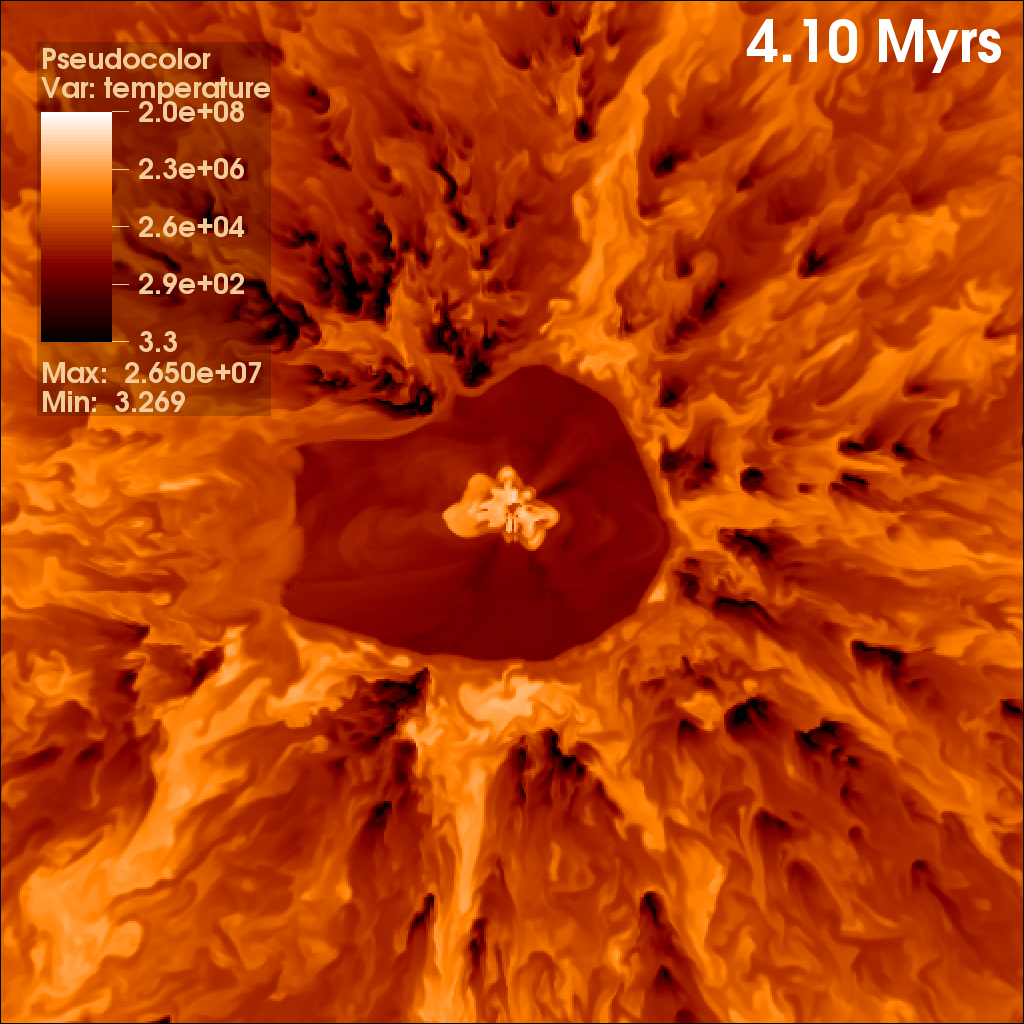}
\includegraphics[width=0.31\textwidth, height=0.31\textwidth]{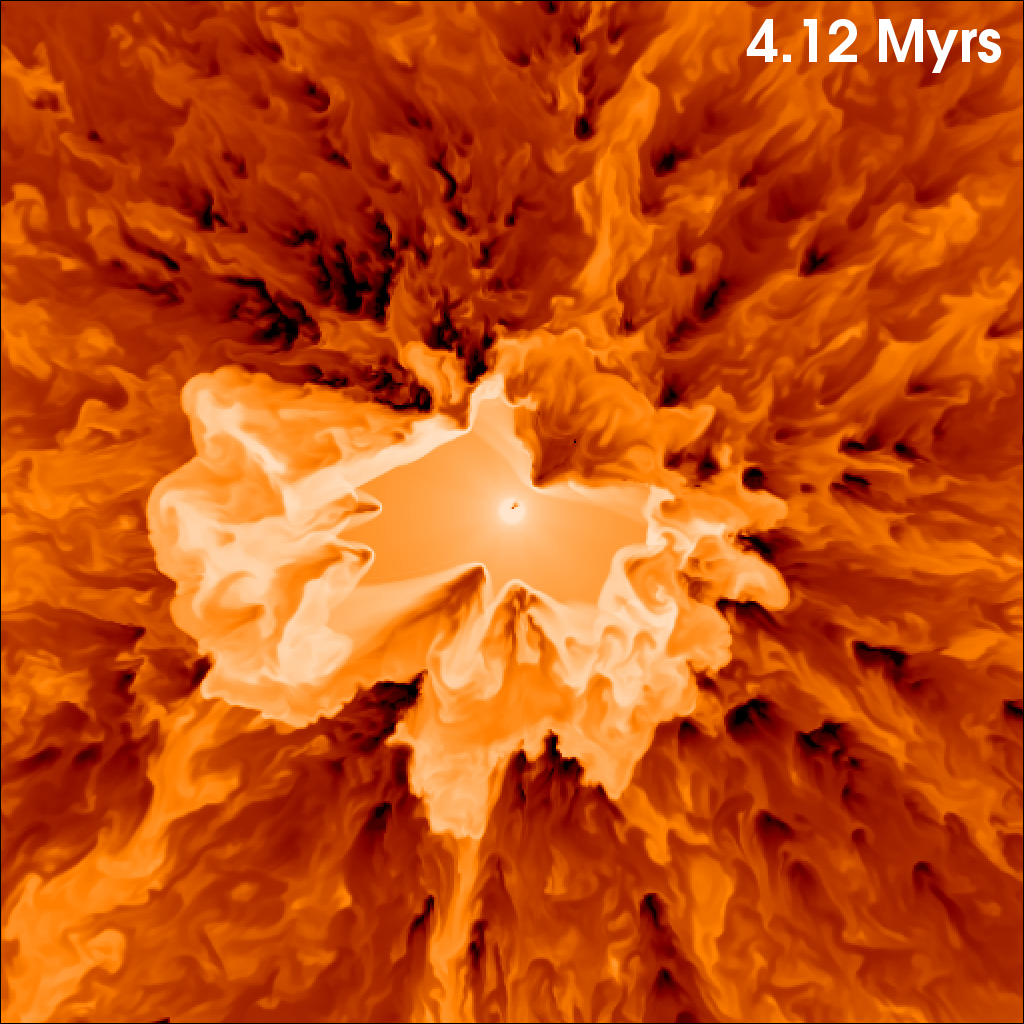}
\includegraphics[width=0.31\textwidth, height=0.31\textwidth]{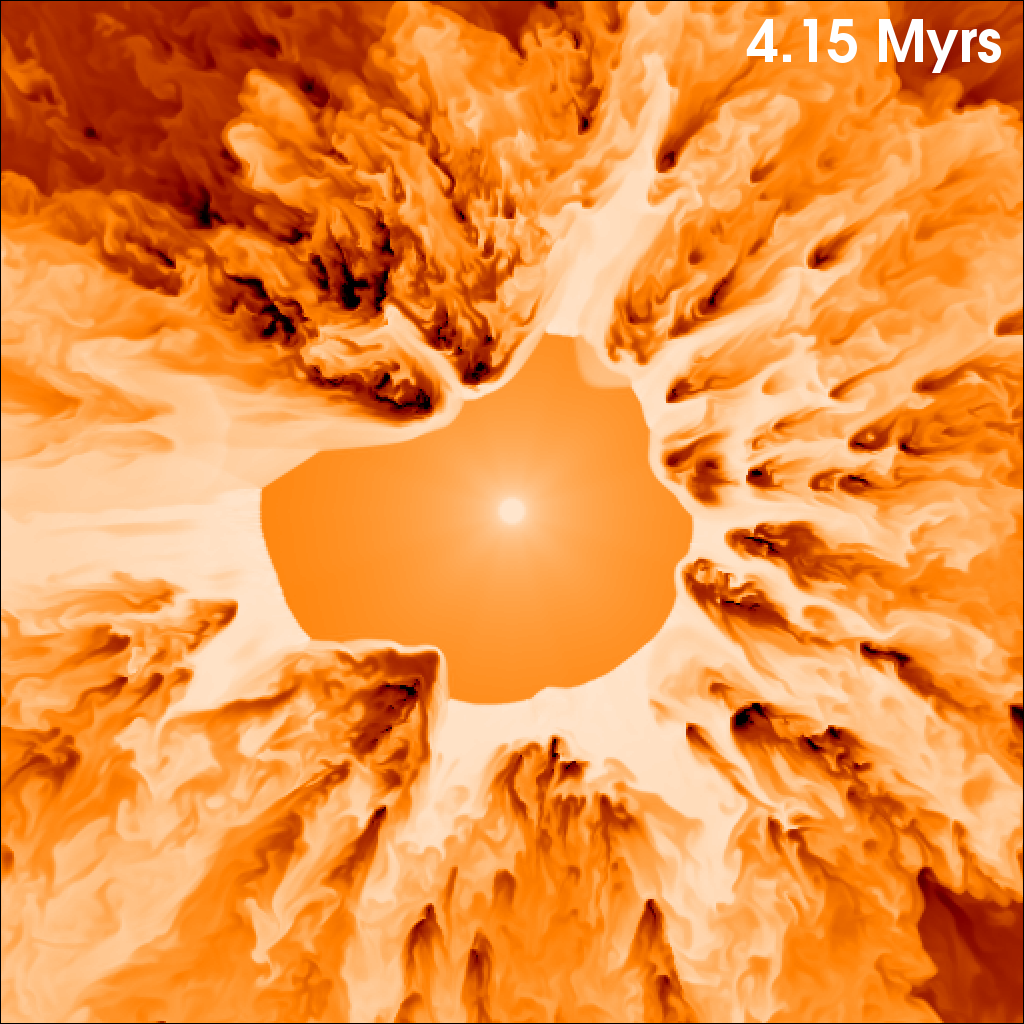}
\includegraphics[width=0.31\textwidth, height=0.31\textwidth]{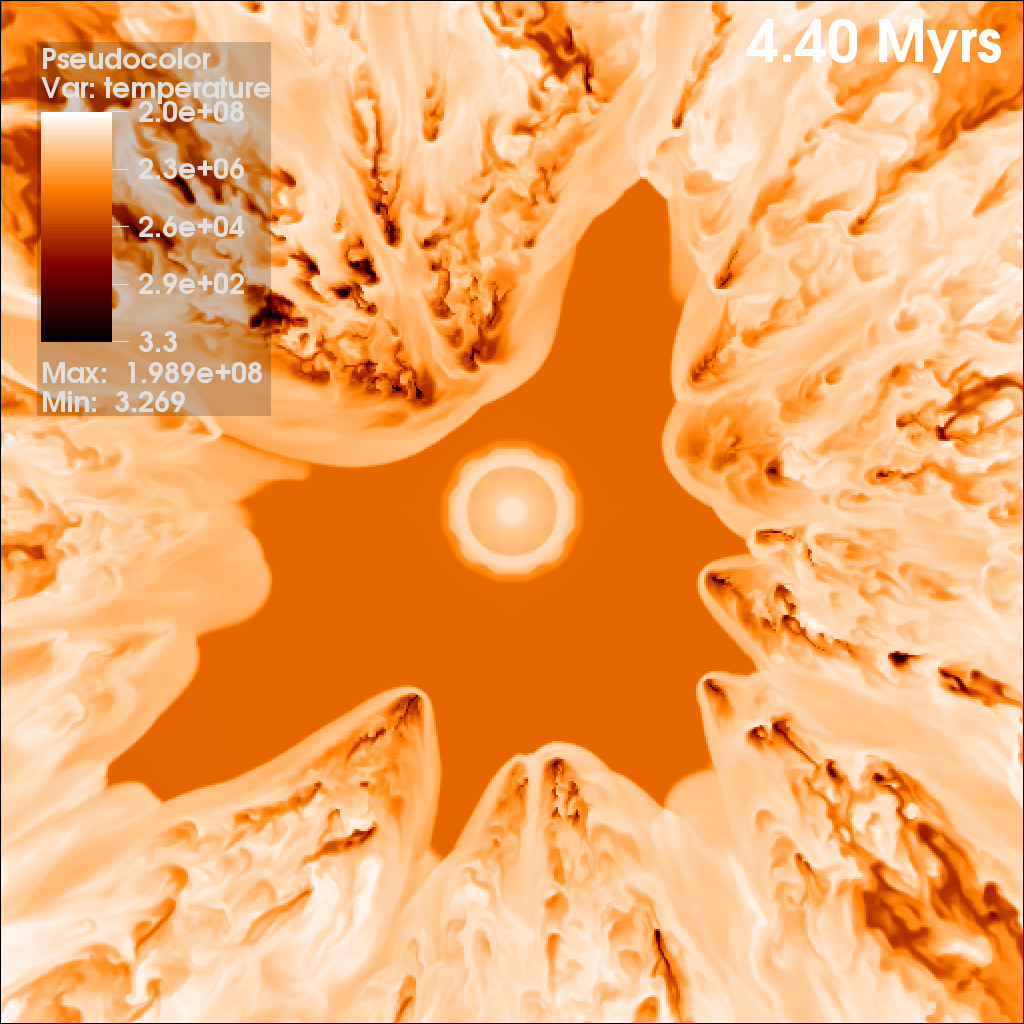}
\includegraphics[width=0.31\textwidth, height=0.31\textwidth]{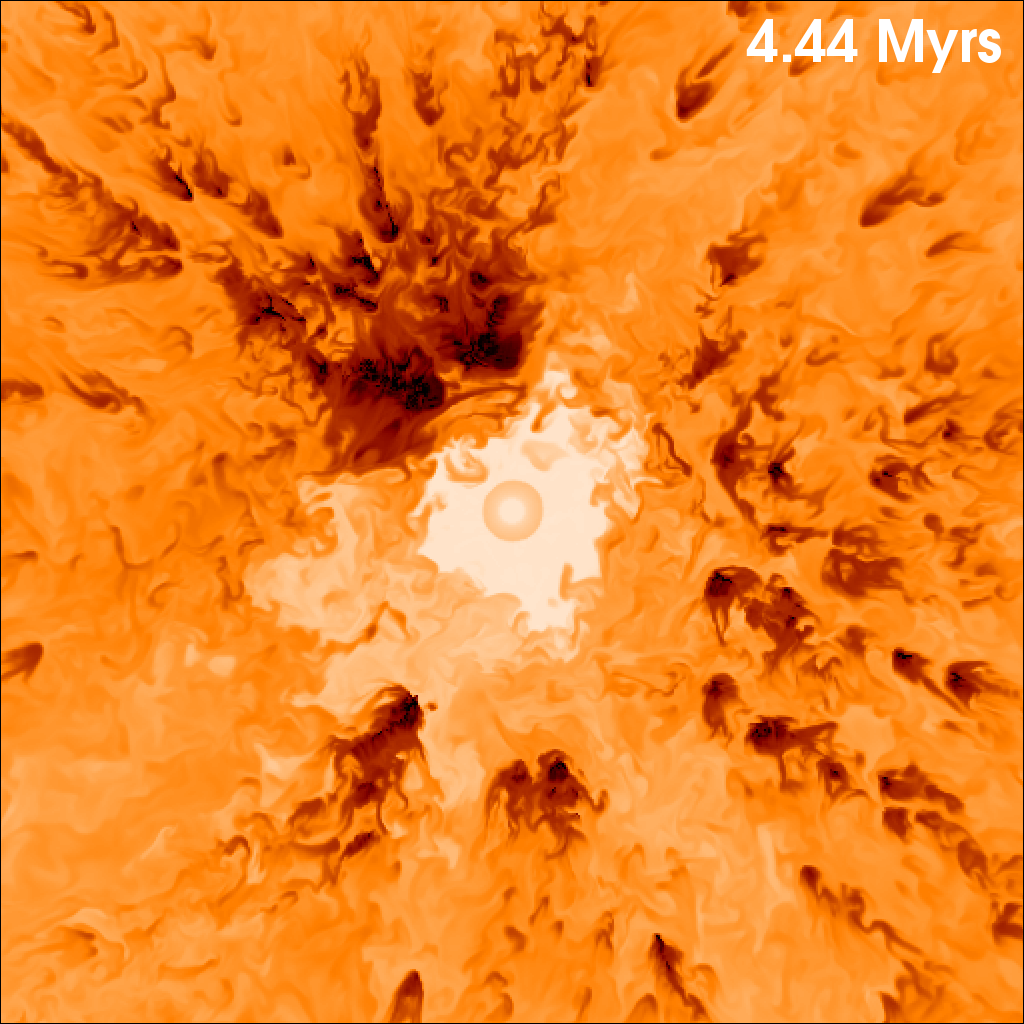}
\includegraphics[width=0.31\textwidth, height=0.31\textwidth]{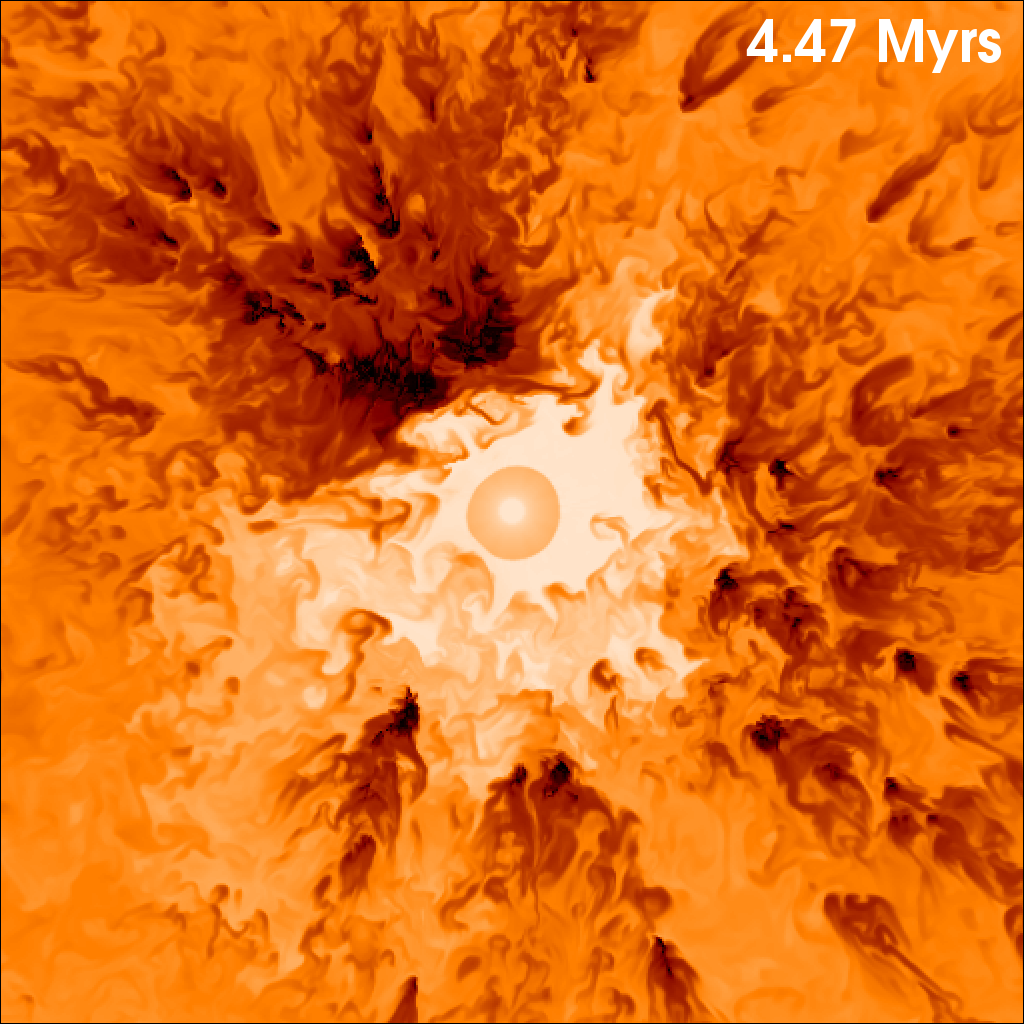}
\caption{Temperature slices from SimA in the xy-plane during the RSG, WR and SN stages for the highest mass star.  The first panel corresponds to just after this star has become a RSG.  The star transitions to a WR star at $t = 4.1$\,Myr (left panel middle row) and explodes as a SN at $t = 4.4$\,Myr (left panel bottom row).  The other stars remain on the MS during this time.  The temperature scale is shown in the left panels. \label{evotemp}}
\end{figure*}

The most massive star evolves to a RSG after 4\,Myr.  At this point its wind speed decreases to $v_{\infty} = 50\,\rm km{\,\rm s^{-1}}$ and its mass loss rate increases to $\rm{\dot{M}} = 10^{-4} {\rm\, M_{\odot}\,yr^{-1}}$ (see Table~\ref{evolution}).  This change results in a slower and denser cluster wind.  The total kinetic power of the cluster wind reduces by about a half, from $1.14\times10^{36} \,\rm ergs{\,\rm s^{-1}}$ to $5.87\times10^{35}\,\rm ergs{\,\rm s^{-1}}$, while the cluster wind becomes dominated by RSG material. This transition is shown in the top rows of Fig.~\ref{evo1} and Fig.~\ref{evotemp}.  The reverse shock moves inward to reestablish pressure equilibrium with the weaker cluster wind.  This depressurises the previously shocked gas and leads to a rapid fall in temperature of the hottest gas in the simulaton. 

The RSG-enhanced cluster wind is much denser than the wind blown when all three stars were on the MS. As it interacts with the surrounding gas it is compressed into a thin shell which is Rayleigh-Taylor (RT) unstable.  RT fingers are visible in the top right panel in Fig.~\ref{evo1} and Fig.~\ref{evotemp}.  These are short lived, lasting approximately 0.04\,Myr.  The most massive star remains in the RSG phase for 0.1\,Myr, at which point the RSG-enhanced cluster wind has expanded to a typical radius of $\sim 5$\,pc. 

The most massive star then evolves into a Wolf Rayet star, with a mass-loss rate of $2\times 10^{-5} {\,\rm M_{\odot}\,yr^{-1}}$ and a wind speed of $2000 \rm km{\,\rm s^{-1}}$.  This change results in a much faster and more powerful cluster wind.  The total kinetic power of the cluster wind increases by nearly two orders of magnitude to $2.59\times10^{37}\rm\,ergs\,s^{-1}$.  This transition occurs at $t = 4.1$\,Myr and can be seen in the middle row of Fig.~\ref{evo1} and Fig.~\ref{evotemp}.  The more powerful cluster wind forcefully pushes back the dense RSG material to beyond the position of the reverse shock during the previous MS phase.  The typical radius of the reverse shock increases from about 5\,pc at $t = 4.14$\,Myr to $\approx 8$\,pc at $t = 4.4$\,Myr (see middle row of Fig.~\ref{evo1}).  The shocked cluster wind is $\approx 10^3$ times hotter than was the case when the cluster wind was ``RSG-enhanced''.  Hot gas pervades almost completely the computational volume by $t = 4.15$\,Myr (see Fig.~\ref{evotemp}). 

\subsubsection{Impact of the first supernova explosion}
 After 4.4\,Myrs of evolution the most massive star explodes as a supernova (see the bottom rows of Figs.~\ref{evo1} and~\ref{evotemp}), adding $10\,{\rm M_{\odot}}$ of ejecta and 10$^{51}$\,ergs of energy into the surroundings.  The SN ejecta sweeps up the WR-dominated cluster wind into a thick shell, which propagates at high speed through the lower density regions surrounding the explosion, heating the gas to very high temperatures. The forward shock has already propagated off the grid by the time of the snapshot in the bottom left panel of Fig.~\ref{evo1}.  The highly aspherical reverse shock seen in this panel is caused by the forward shock reflecting off dense clumps of gas in the simulation.  The reverse shock then moves inwards towards the central cluster and nearly overwhelms the MS-dominated cluster wind from the remaining two stars (bottom middle panel of Fig.~\ref{evo1}), which is seen sweeping up the cold SN ejecta in the bottom left panel.  The reverse shock of the cluster wind is forced back to a radius of $\approx 0.9$\,pc by $t = 4.42$\,Myr, but slowly begins to  expand to $\approx 1.5$\,pc by $t = 4.47$\,Myr as the pressure within the supernova remnant diminishes.  One of the main effects of the SN is to fill the low-density wind-carved channels through the GMC clump with denser material which the cluster wind must again clear.  The very densest fragments left over from the original GMC clump are not only resistant to ablation from the cluster wind, but also are affected very little by the propagation of the SN shockwave. 

\begin{figure}
\centering
\includegraphics[width=0.49\textwidth, height=0.25\textwidth]{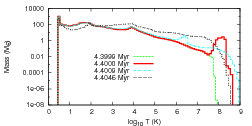}
\includegraphics[width=0.49\textwidth, height=0.25\textwidth]{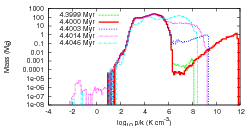}
\includegraphics[width=0.49\textwidth, height=0.25\textwidth]{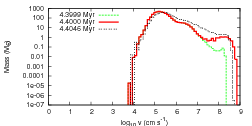}
\caption{Histogram of the gas temperature (top), pressure (middle), and velocity (bottom) during the first supernova explosion.  In each panel the solid red line corresponds to a time just after the explosion occurs.\label{hist}}
\end{figure}

Fig.~\ref{hist}a) shows the temperature of the gas in the simulation during the 5000 years after the most massive star goes supernova.  The green dashed line shows the state just before the supernova explosion, and the thick, red, solid line corresponds to just afterwards.  The maximum temperature on the grid jumps by almost an order of magnitude in the first 100 years (from $\sim10^{8}$ to $\sim4\times10^{8}$\,K) and the amount of material between $10^7 - 10^8$\,K increases from about $1\,\rm\,M_{\odot}$ to $10\,\rm\,M_{\odot}$.  The maximum temperature continues to increase up to 10$^9$\,K as the shockwave propagates through the cloud.  Gas with $T\, \lesssim \rm100\,K$ responds more slowly to the SN explosion, but it is clear that some of it is heated to $T\,\approx 10^4\rm\,K$ by the passage of the shock wave.  A significant proportion of the coldest gas with $T\,\lesssim\,\rm10\,K$ is hardly affected, however.  This very cold gas is situated in the very densest regions of the remaining GMC clump.  As the shockwave passes through the simulation, the less dense material surrounding these regions is heated and ionized, but the densest parts are relatively untouched. This is because the transmitted shock speed through the densest clumps is as low as a few 10's of ${\rm km\,s^{-1}}$, and the gas cools back to its original pre-shock temperature on timescales as short as a few 10's of yrs. More typical cooling times are $10^{3}$\,yrs or so, but it is clear that the transmitted shocks into the densest clumps are highly radiative (the crossing time of the transmitted shocks through these fragments is $\sim 10^{4}$\,yrs). The strong cooling of this gas is responsible for the rapid rise in the $H_{2}$ mass following its initial drop after each SN explosion (see Sec.~\ref{sec:molecularmass}). The SN ejecta first begins to leave the grid approximately 4000 years after the explosion. 

Fig.~\ref{hist}b) shows the pressure ($p$/k) during this time.  After the SN explosion the pressure increases by four orders of magnitude from $10^8 - 10^{12}\rm\,K\,cm^{-3}$.  The maximum pressure then slowly decreases as the remnant expands adiabatically.  By $t = 4.4046$\,Myr (black double-dashed line) the maximum pressure on the grid is back to $\approx 10^8\rm\,K\,cm^{-3}$, but now there is $\approx 860\rm\,M_{\odot}$ of material at 10$^7\leq \rm{p/k} \leq 10^8\rm\,K\,cm^{-3}$, compared to the 0.02$\rm\,M_{\odot}$ of gas in this pressure range prior to the SN. $50,000$\,yrs after the SN explosion the net effect is a shift in gas from lower ($10^3 \lesssim \rm p/k \lesssim 10^6 \rm\,K\,cm^{-3}$) to higher  ($10^6 \lesssim \rm p/k \lesssim 10^8 \rm\,K\,cm^{-3}$) pressure.

The maximum gas velocity also increases rapidly in the first 100 years after the supernova explosion, as shown in Fig.~\ref{hist}c).  The maximum gas velocity increases from $2000\,\rm km\,s^{-1}$ to $10,000\,\rm km\,s^{-1}$ as the hot, high pressure ejecta starts its expansion.  Over the next 5000\,yrs the maximum velocity drops to $\approx 5000\rm\,km\,s^{-1}$ and there is significantly more mass with $v \geq 100\rm\,km\,s^{-1}$ than was the case pre-SN.  The majority of the gas continues to have a velocity of $\sim 1-10\rm\,km\,s^{-1}$, however, which highlights the weak coupling of the SN explosion to the densest parts of the surrounding gas.

\subsubsection{Response due to the evolution of the 32$\,M_{\odot}$ and 28$\,M_{\odot}$ stars}
$0.1$\,Myr after the explosion of the most massive star, the 32$\rm\,M_{\odot}$ star evolves off the MS and onto the RSG branch (see Table~\ref{evolution}), decreasing the kinetic power of the cluster wind still further to $2.7\times 10^{35}\rm\,ergs\,s^{-1}$.  This star follows the same post-MS evolution as the most massive star, and blows a dense slow moving RSG wind followed by a high momentum WR wind before finally exploding.  The SN explosion again imparts $10\rm\,M_{\odot}$ of material and 10$^{51}\rm\,ergs\,s^{-1}$ of energy into the surroundings.  After a further 0.1\,Myr the lowest mass (and only remaining) star considered in our model evolves onto the RSG branch, reducing the kinetic power of the cluster wind to $7.9 \times 10^{34}\rm\,ergs\,s^{-1}$.  This star explodes as a supernova at $t = 5.4$\,Myr.  At this moment there are no wind-blowing stars remaining in our model, and the ambient and enriched gas gradually disperses and flows off the grid boundaries.

\subsection{Mass and energy fluxes into the wider environment}
\label{sec:fluxes}
\begin{figure}
\centering
\includegraphics[width=0.49\textwidth, height=0.25\textwidth]{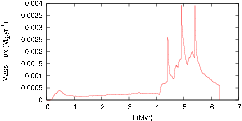}
\caption{Total mass flux off the grid as a function of time. \label{massflux}}
\end{figure}

Fig.~\ref{massflux} shows the total mass flux off the grid as a function of time. The mass flux is zero until the first blowouts reach the edge of the grid and then steadily increases up until $t\approx 0.4$\,Myr as various shells of swept up material reach the outer boundary of the simulation.  The material behind the shells is less dense, and therefore the mass flux declines as the shells leave the grid. The mass flux stabilizes at $t \approx 1.15$\,Myr and then begins a slow, almost linear, increase from $1.6\times10^{-4} \rm\,M_{\odot}\,yr^{-1}$, reaching nearly $3\times10^{-4} \rm\,M_{\odot}\,yr^{-1}$ at $t = 4.0$\,Myr. In comparison,  the mass-loss rates of the three stars during the MS is $5\times10^{-7}, 2.5\times10^{-7}$ and $1.5\times10^{-7}\rm\,M_{\odot}\,yr^{-1}$ for the 35\,M$_{\odot}$, 32\,M$_{\odot}$ and 28\,M$_{\odot}$ stars, respectively, giving a cluster mass-loss rate of $9\times10^{-7}\rm\,M_{\odot}\,yr^{-1}$. This indicates that the cluster wind is ``mass-loaded''\footnote{Although we use the term ``mass-loaded'' here, the molecular material ablated by the cluster wind is not fully mixed into the flow by the time that it leaves the grid. Since distinct phases are still identifiable, ``mass entrainment'' may be a more appropriate term.} by factors of $200-300$ as it streams through the molecular material in the GMC clump, and that
less than 1 per cent of the material leaving the grid during this period originated in the stellar winds.

There is a slight decrease in the mass flux off the grid between $t = 4.0$ and 4.1\,Myr, during the RSG stage of the highest mass star in which the flow de-pressurizes. This is followed by an increase in the mass flux to $\approx 7\times10^{-4} \rm\,M_{\odot}\,yr^{-1}$ during the subsequent WR stage. This represents a ``mass-loading'' factor $35\times$ the mass-loss rate of the cluster wind during this period ($2.04\times10^{-5}\rm\,M_{\odot}\,yr^{-1}$). Therefore, only 3 per cent of the material leaving the grid is from the stellar winds. The mass flux jumps to $2.5\times10^{-3} \rm\,M_{\odot}\,yr^{-1}$ following the first SN explosion, and peaks at $3.9\times10^{-3} \rm\,M_{\odot}\,yr^{-1}$ following each of the second and third explosions. In between the SN explosions, the mass flux peaks at $2.0\times10^{-3} \rm\,M_{\odot}\,yr^{-1}$ at $t=5.31$\,Myr. Clearly the WR winds and SN explosions act to speed up the rate at which molecular material is cleared from the cluster surroundings.

\begin{figure}
\centering
\includegraphics[width=0.49\textwidth, height=0.25\textwidth]{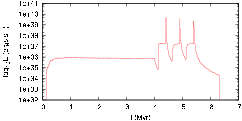}
\includegraphics[width=0.49\textwidth, height=0.25\textwidth]{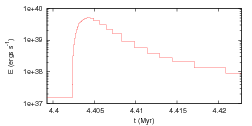}
\caption{Total energy flux off the grid as a function of time. Top: Entire simulation. Bottom: Focussed on the first SN explosion.  The steps are caused by the cadence of the timesteps analysed.\label{energyflux}}
\end{figure}

The top panel of Fig.~\ref{energyflux} shows the total energy flux off the grid as a function of time. The energy flux increases from zero at the time when the shell first encounters the grid boundary, to $\approx 8.5 \times 10^{35}\,\rm{ergs\,s^{-1}}$ at $t=1$\,Myr, with a gradual and linear decline to $7.5 \times 10^{35}\,\rm{ergs\,s^{-1}}$ at $t=4$\,Myr. The linear decline correlates with the gradual increase in the ablation rate during this period. These values for the energy flux compare to the total kinetic power of the cluster wind of $1.14 \times10^{36}\,\rm{ergs\,s^{-1}}$. It is clear, therefore, that the shocked cluster wind is largely adiabatic, but that nevertheless about one quarter to one third of the injected wind power is lost to radiative processes. It is interesting to note that \citet{Bruhweiler10} invoke substantial radiative losses due to the turbulent interaction of stellar winds with inhomogeneous surroundings in order to explain their observations of the Rosette Nebula. 

The combined energy input by the 3 O-stars during the first 4\,Myr is $1.44 \times 10^{50}\,\rm{ergs}$.
When the most massive star evolves to the RSG phase there is a decrease in the kinetic power of the cluster wind to $5.87\times10^{35}\rm\,ergs\,s^{-1}$, and this is reflected in a corresponding decrease in the energy flux off the grid, which, however, drops below this value. This ``overshoot'' is likely due to some combination of enhanced cooling in the denser cluster wind and possible overstable behaviour of the flow. The energy flux off the grid increases by two orders of magnitude to $1.8 \times10^{37}\rm\,ergs\,s^{-1}$ when the most massive star has evolved to a WR star. Comparing to the kinetic power of the cluster wind at this time ($2.59\times10^{37}\rm\,ergs\,s^{-1}$), we find that radiative losses are again about 30 per cent.  The combined energy input by the WR star and the 2 remaining O-stars during the period $4.1-4.4$\,Myr is $2.46 \times 10^{50}\,\rm{ergs}$. Altogether, the most massive star injects $\approx 3.9 \times 10^{50}\,\rm{ergs}$ of energy via its wind during its lifetime, and about $70$ per cent of this escapes to large distances.

The bottom panel of Fig.~\ref{energyflux} shows the total energy flux off the grid around the time of the first SN explosion which occurs at $t=4.4$\,Myr. The energy flux rises steeply as the blast shock propagates off the grid, and peaks at $\approx 5\times10^{39}\rm\,ergs\,s^{-1}$ about 5000\,yr after the explosion. Thereafter the energy flux steadily decreases. The integrated energy off the grid during this time reveals that greater than 99 per cent of the SN energy flows off the grid, and less than 0.5 per cent is radiated. Clearly the SN energy propagates through the environment relatively unimpeded by the dense clumps.

The subsequent energy flux from the grid is dominated by the WR stages from the two lower mass stars and their subsequent explosions, and drops below $10^{35}\rm\,ergs\,s^{-1}$ as relic hot gas expands and dissipates.

\subsection{Evolution of column densities}
\label{sec:columndensity}

\begin{figure}
\centering
\includegraphics[width=0.45\textwidth, height=0.20\textwidth]{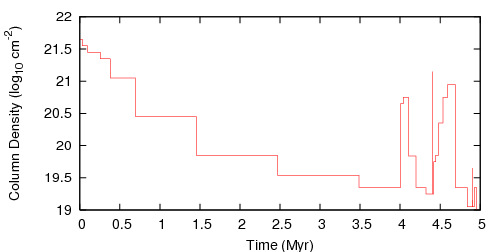}
\caption{The evolution of the average column density from the centre of the cluster. \label{columndensity}}
\end{figure}

Fig.~\ref{columndensity} shows the time evolution of the average column density, $\bar{N}_{\rm H}$, from the centre of the cluster. This is calculated over $10^{4}$ individual sight lines spaced equally in solid angle and traced out to the edge of the grid. The column density is greatest at $t=0$, but diminishes with time as the stellar winds push the clump material away from the cluster (some of this decrease is also due to material leaving the grid). A factor of 100 reduction from an initial value of $\approx 4 \times 10^{21}\,{\rm cm^{-2}}$ occurs by $4$\,Myr. Then, as the most massive star enters its RSG stage, $\bar{N}_{\rm H}$ increases by over 1 dex to nearly $10^{21}\,{\rm cm^{-2}}$. This increase is short-lived however, because the density of the cluster wind drops significantly once the most massive star enters its WR stage. Immediately prior to the first SN explosion, $\bar{N}_{\rm H} \approx 2 \times 10^{19}\,{\rm cm^{-2}}$. The $10\,{\rm M_{\odot}}$ of ejecta from the explosion momentarily increases the average column density to more than $10^{21}\,{\rm cm^{-2}}$, but this rise is extremely short-lived and less than a few hundred years in duration. $\bar{N}_{\rm H}$ then gradually increases as the cluster wind again begins to fill the nearby environment with mass, especially once the second most massive star enters its RSG stage. Similar behaviour in the evolution of $\bar{N}_{\rm H}$ then occurs as the remaining massive stars evolve in turn through their various wind and SN stages.

\begin{figure*}
\centering
\includegraphics[width=0.32\textwidth]{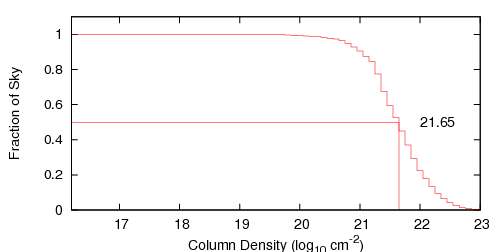} 
\includegraphics[width=0.32\textwidth]{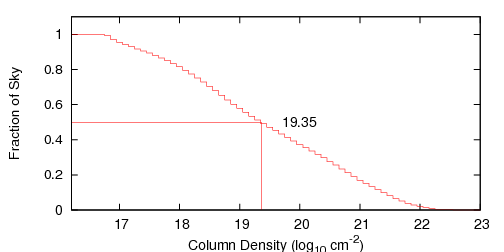} 
\includegraphics[width=0.32\textwidth]{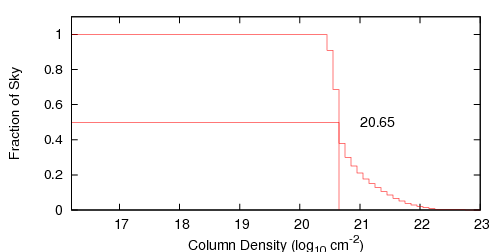} 
\includegraphics[width=0.32\textwidth]{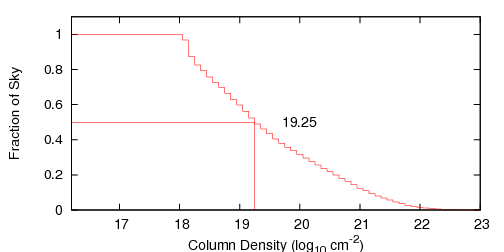} 
\includegraphics[width=0.32\textwidth]{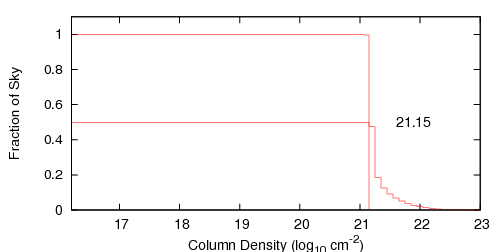} 
\includegraphics[width=0.32\textwidth]{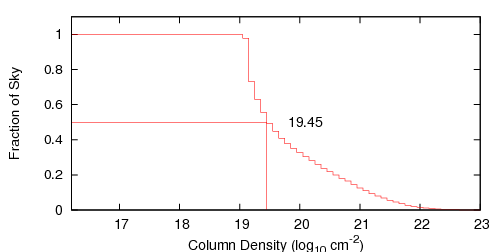} 
\caption{The evolution of the average column density from the centre of the cluster. From left to right and top to bottom the plots are at times $t=0$, $t=3.99$, $t=4.02$, $t=4.39$, $t=4.40000$ and $t=4.40110$\,Myr. The numerical value in each panel notes the value of log$_{10}\,{\rm N_{H}}$ at which 50 per cent of the sky has a smaller or larger column. \label{columndensitydistributions}}
\end{figure*}

Fig.~\ref{columndensitydistributions} shows histograms of the column density distribution at specific times in the simulation. At $t=0$\,yr, columns up to $\approx 10^{23}\,{\rm cm^{-2}}$ exist for sightlines through the densest parts of the initial GMC clump, but a very small fraction of sightlines from the cluster experience column densities as low as $10^{20}\,{\rm cm^{-2}}$. At $t=3.99$\,Myr, the distribution of column densities is much broader, with those passing through dense and relatively nearby regions having column densities up to $10^{22}\,{\rm cm^{-2}}$ and those passing through the lowest density material having $N_{\rm H} \approx 10^{17}\,{\rm cm^{-2}}$. It is interesting to see how the distribution has changed shape by $t=4.02$\,Myr, once the most massive star has entered its RSG stage. The RSG-enhanced cluster wind ``fills in'' all of the low column density sightlines so that the minimum value is now $N_{\rm H} \approx 3 \times 10^{20}\,{\rm cm^{-2}}$.
The impact of the WR-enhanced cluster wind is seen at $t=4.39$\,Myr. The higher speed and reduced density of the cluster wind now reduces the low column density sightlines to $\approx 10^{18}\,{\rm cm^{-2}}$. The column density distribution at a time just after the first SN explosion is shown at $t=4.40000$\,Myr. The relatively dense ejecta causes the minimum value of $N_{\rm H}$ to rise to $\approx 10^{21}\,{\rm cm^{-2}}$, yet this drops to just $10^{19}\,{\rm cm^{-2}}$ 1100\,yrs later.

\begin{figure*}
\centering
\includegraphics[width=0.32\textwidth]{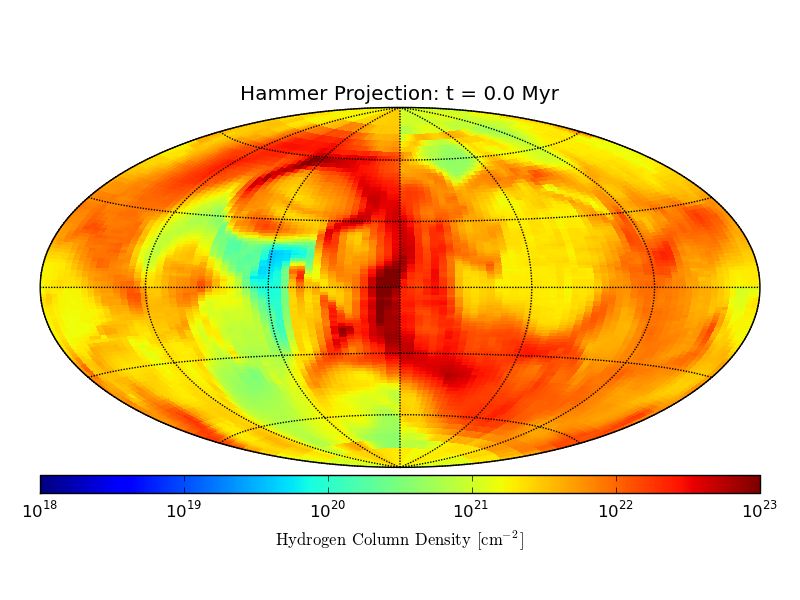} 
\includegraphics[width=0.32\textwidth]{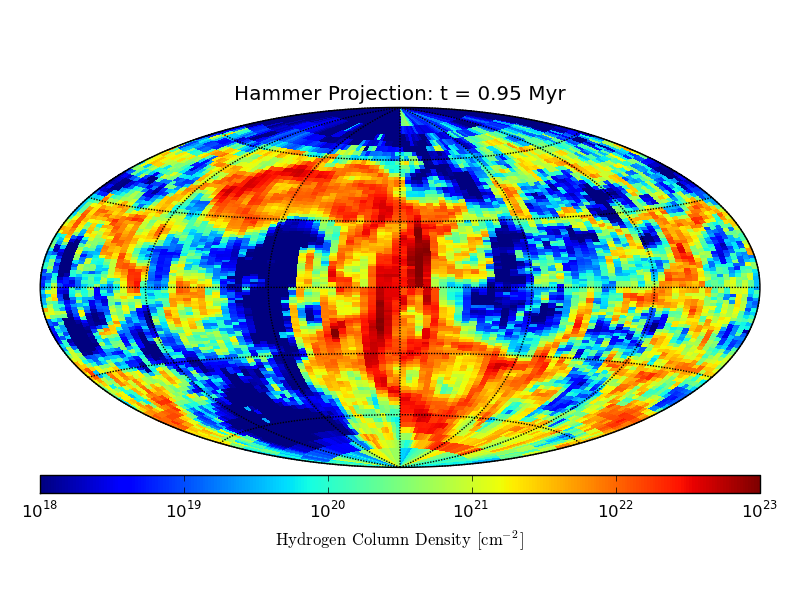} 
\includegraphics[width=0.32\textwidth]{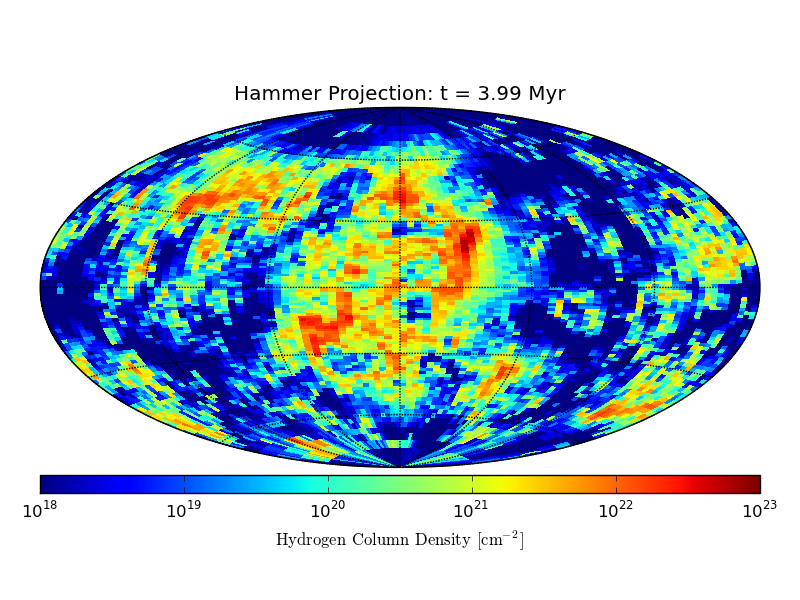} 
\includegraphics[width=0.32\textwidth]{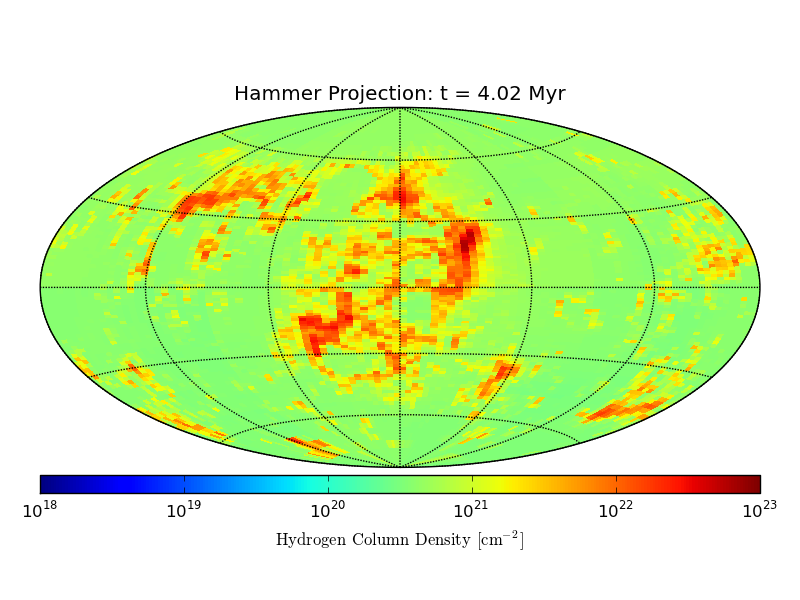} 
\includegraphics[width=0.33\textwidth]{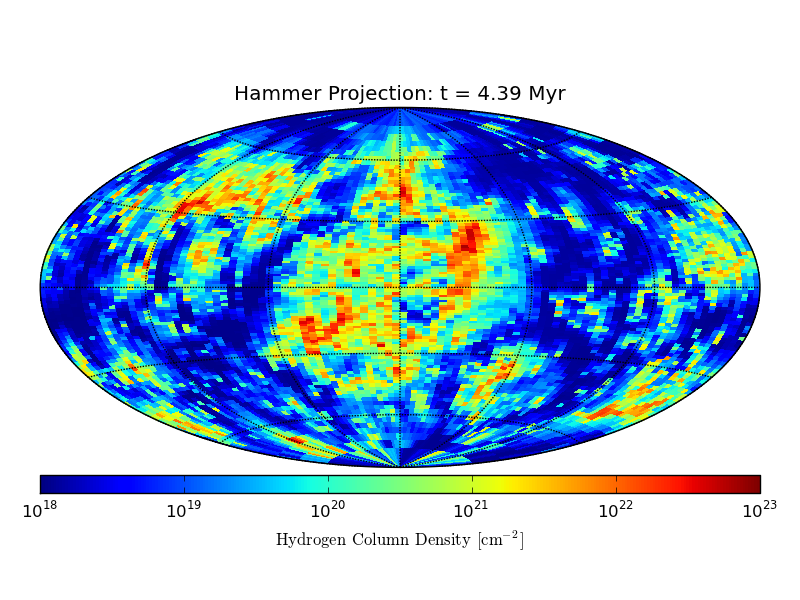} 
\includegraphics[width=0.33\textwidth]{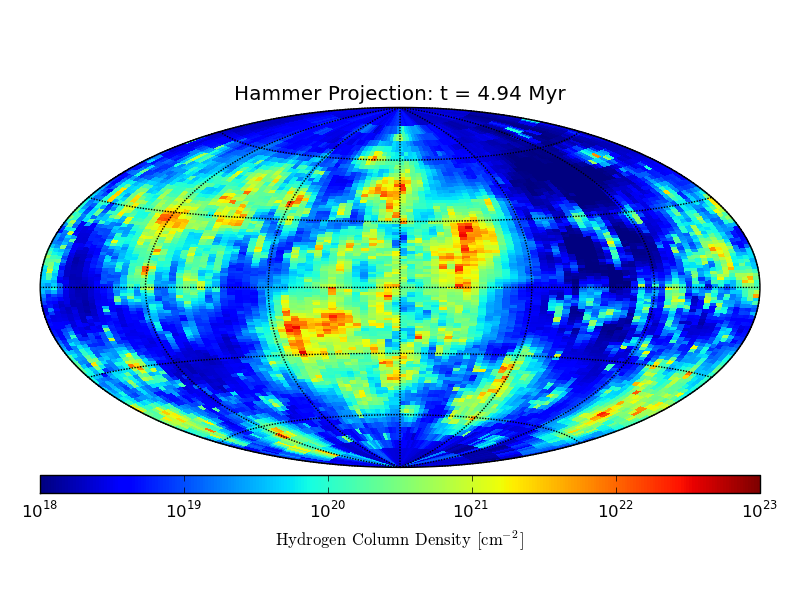} 
\caption{Hammer projections of the column density at specific times.\label{hammerplotNH}}
\end{figure*}

Fig.~\ref{hammerplotNH} shows Hammer projections of the column density at specific times.
The directions with the highest (lowest) initial column density maintain their positions throughout the simulation. It is clear that the regions of highest column density become increasingly ``porous'' and ``shredded'' by the action of the cluster wind and supernovae. The ``filling in'' of the column density during the time of the RSG-enhanced cluster wind ($t = 4.02$\,Myr in Fig.~\ref{hammerplotNH}) is readily apparent. 

\begin{figure*}
\centering
\includegraphics[width=0.32\textwidth]{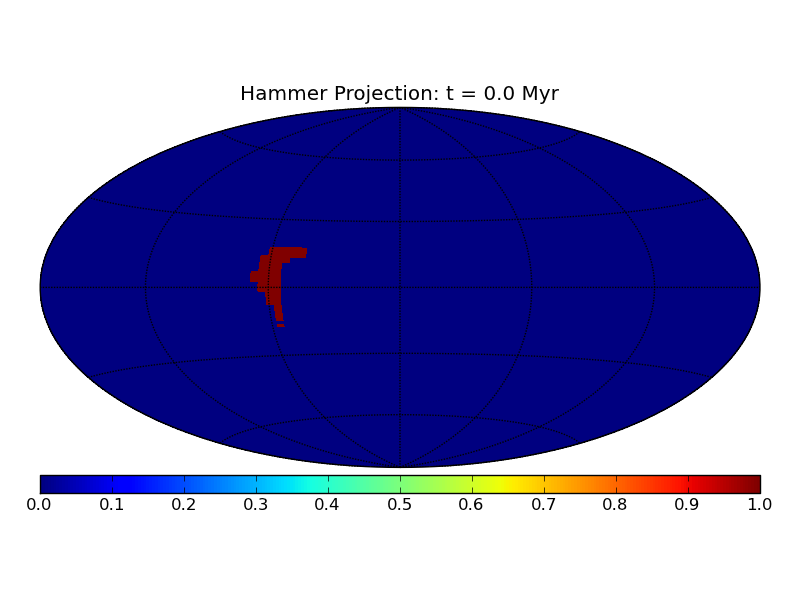} 
\includegraphics[width=0.32\textwidth]{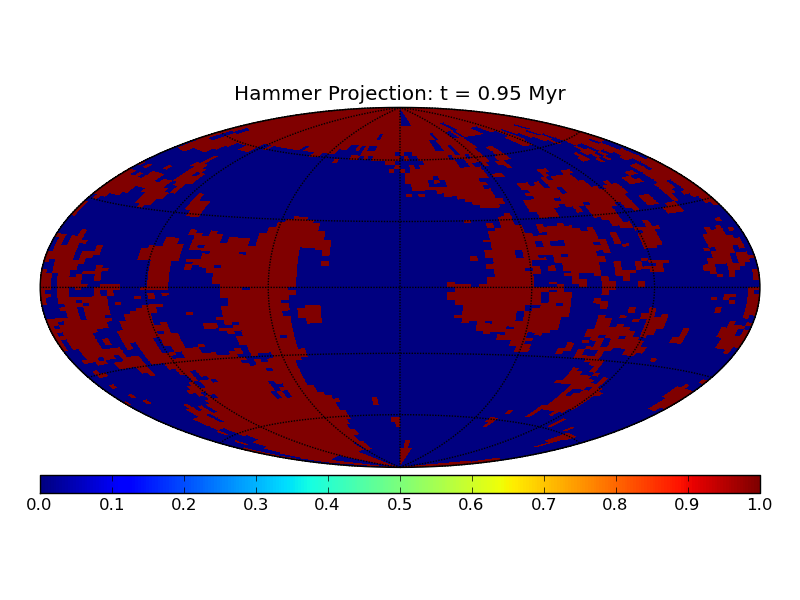} 
\includegraphics[width=0.32\textwidth]{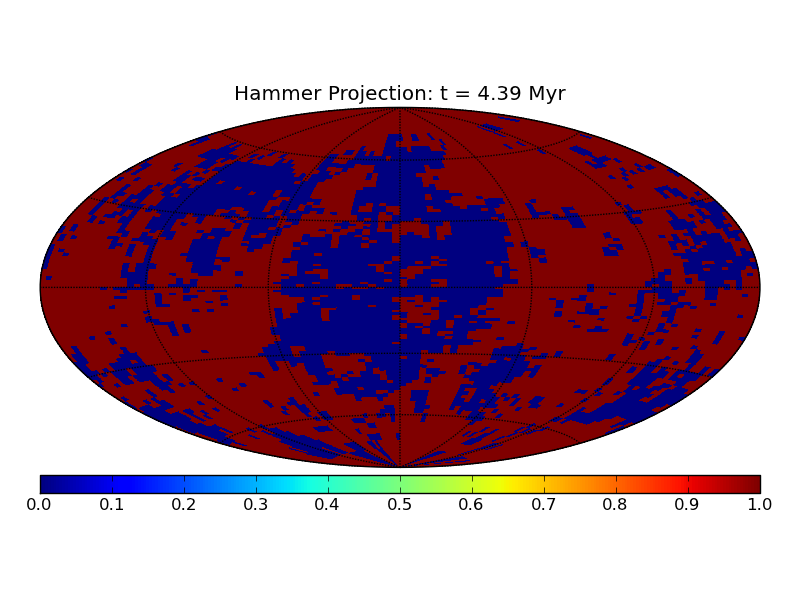} 
\caption{Hammer projections of the position of the ionization front at specific times. Red (blue) corresponds to the ionization front being within (outside) the grid in that direction, and identifies the directions where ionizing photons escape to large distances.\label{hammerplotfesc}}
\end{figure*}

Though we do not include photoionization in this work, it is interesting to estimate the fraction of ionizing photons which could escape to large distances. Along each sightline we evaluate whether the integral of $4 \pi r^{2} \alpha_{\rm B} n_{\rm e}^{2} dr$ exceeds $\dot{S}_{\rm cl}$, where $r$ is the radial distance to the stellar cluster, $n_{\rm e}$ is the electron number density, $\alpha_{\rm B}$ is the case B recombination coefficient and $\dot{S}_{\rm cl}$ is the rate of ionizing photons from the stellar cluster. If this condition is satisfied the ionization front is trapped on the grid in that direction.  Fig.~\ref{hammerplotfesc} shows the result of this calculation. Initially the ionization front is almost completely trapped within the GMC clump surrounding the stellar cluster, with less than 1 per cent of ionizing photons escaping (as indicated by the red colour). However, by $t=0.95$\,Myr, the stellar winds have sufficiently cleared away dense molecular and atomic material that about 40 per cent of the ionizing radiation escapes. This increases to nearly 60 per cent by $t=3.99$\,Myr. The RSG-enhanced cluster wind is then sufficiently dense to completely prevent any ionizing radiation escaping to large distances. Once the lower density WR-enhanced cluster wind clears the RSG dominated material off the grid the escape fraction increases once more, reaching nearly 65 per cent just prior to the first SN explosion and 75 per cent at $t=4.94$\,Myr.

Recently, \citet{Pellegrini12} determined that the mean, luminosity weighted, escape fraction of ionizing photons from the HII region population in the LMC and SMC is $\gtsimm 0.42$ and $\gtsimm 0.40$, respectively. Our results are consistent with these values, and similar values from \citet{Reines08}, but a more rigorous and detailed analysis of the photoionization beyond this work is necessary.
 
\subsection{Comparison of SimA and SimB and evolution of the molecular mass}
\label{sec:molecularmass}
\begin{figure}
\centering
\includegraphics[width=0.22\textwidth, height=0.22\textwidth]{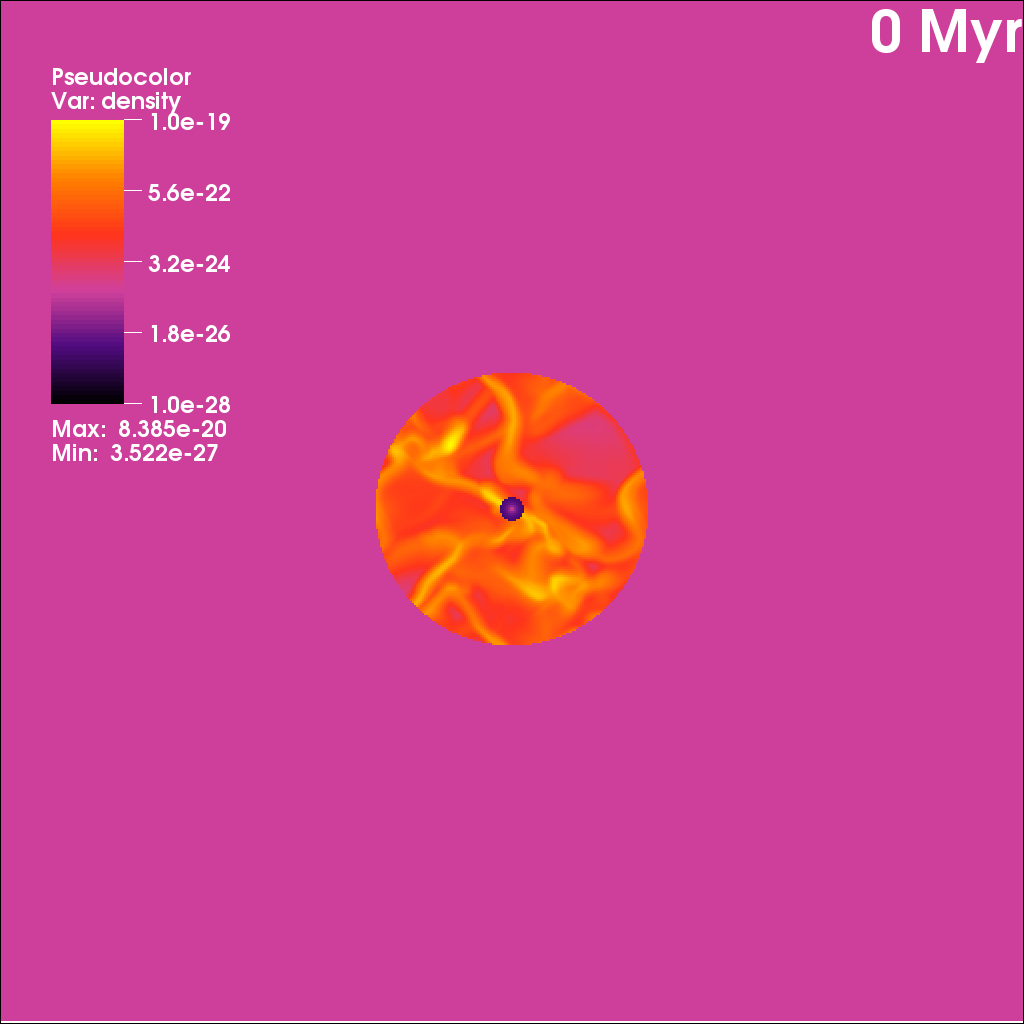}
\includegraphics[width=0.22\textwidth, height=0.22\textwidth]{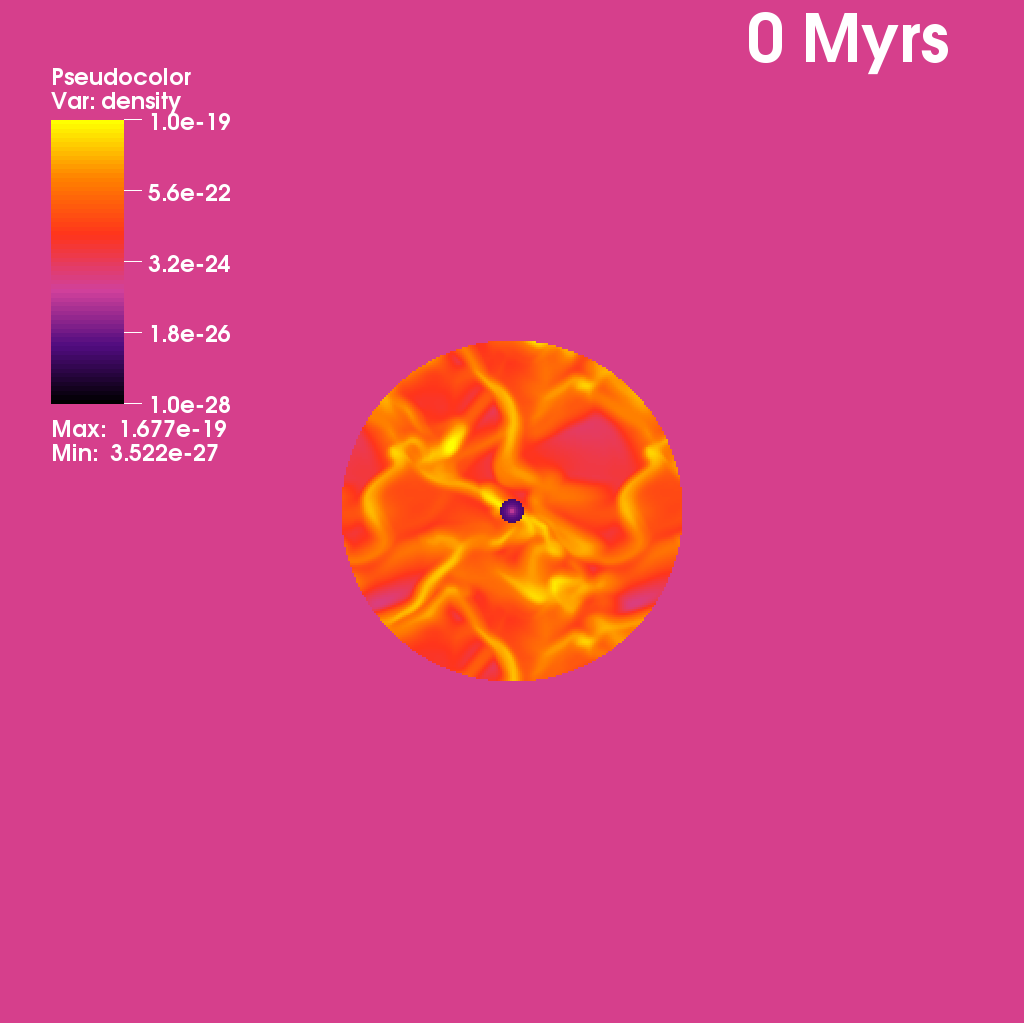}
\includegraphics[width=0.22\textwidth, height=0.22\textwidth]{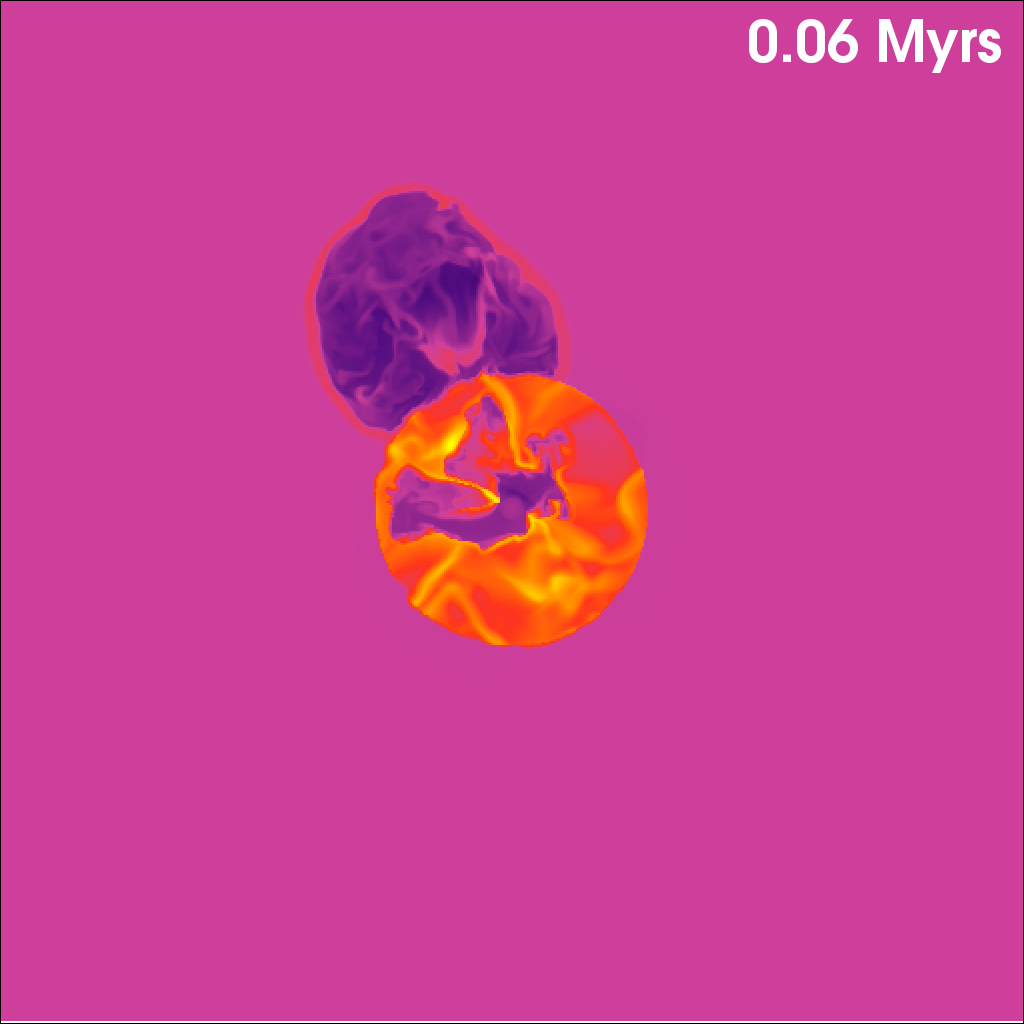}
\includegraphics[width=0.22\textwidth, height=0.22\textwidth]{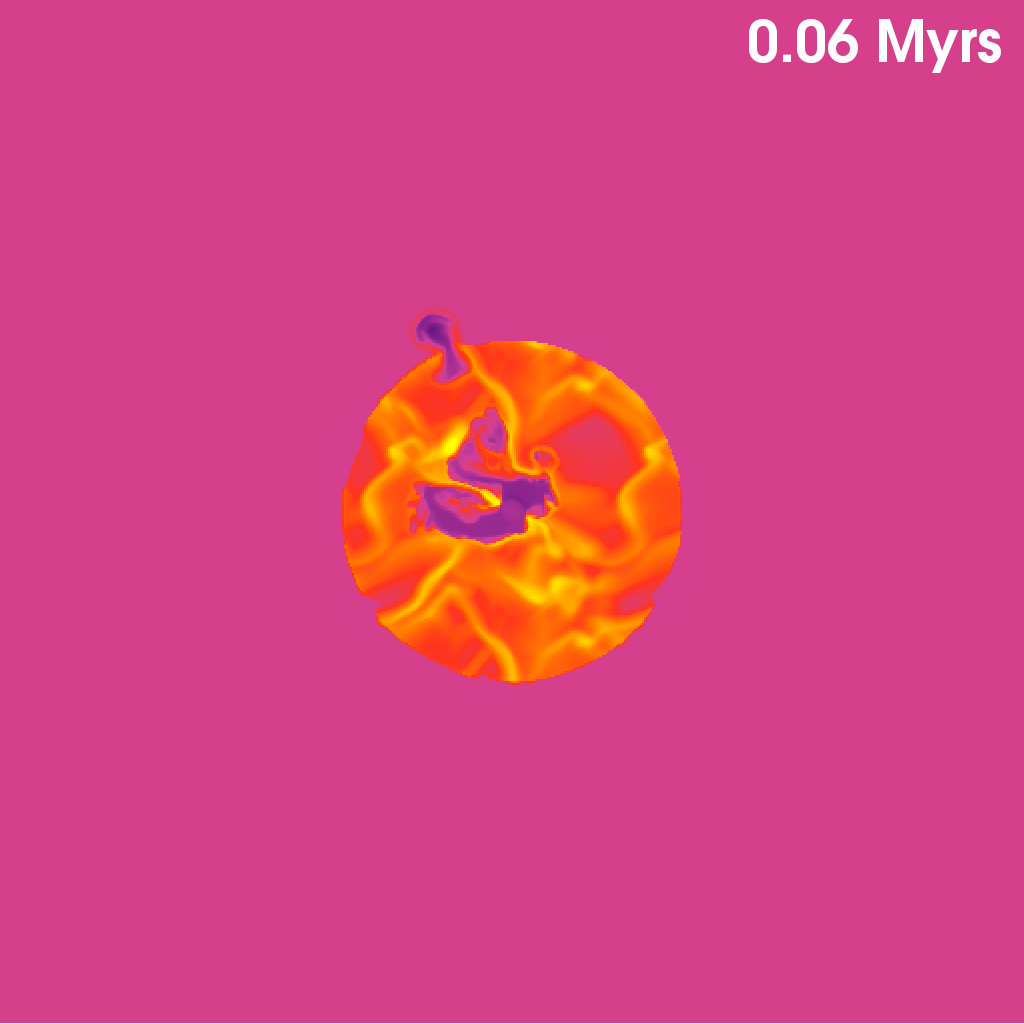}
\includegraphics[width=0.22\textwidth, height=0.22\textwidth]{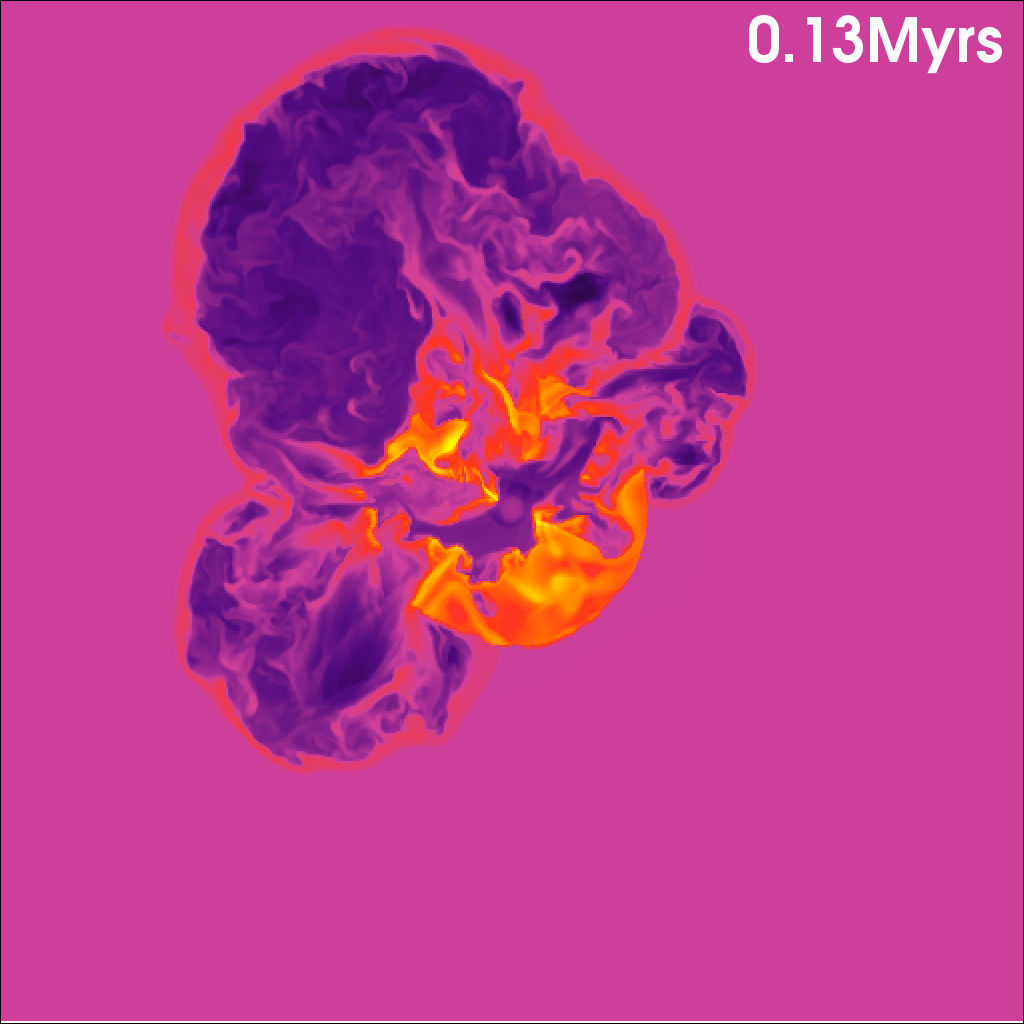}
\includegraphics[width=0.22\textwidth, height=0.22\textwidth]{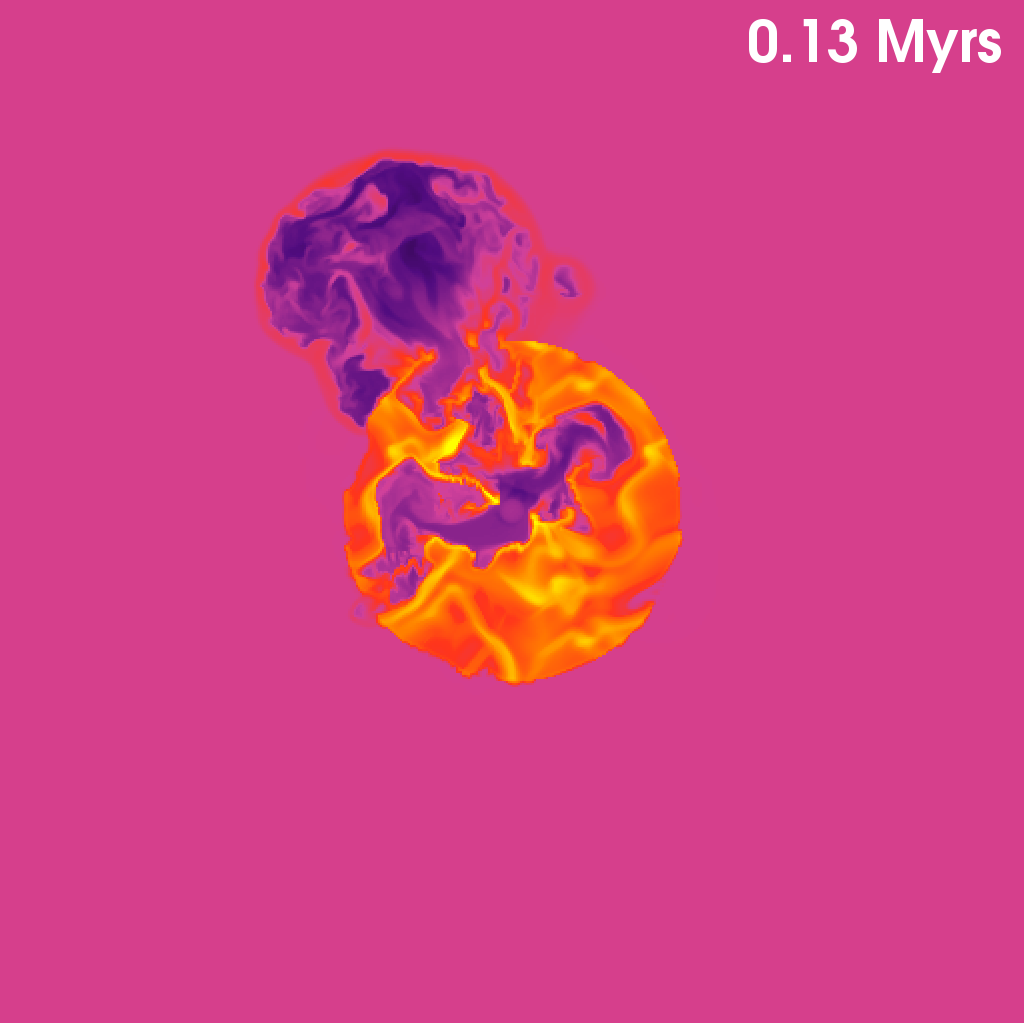}
\includegraphics[width=0.22\textwidth, height=0.22\textwidth]{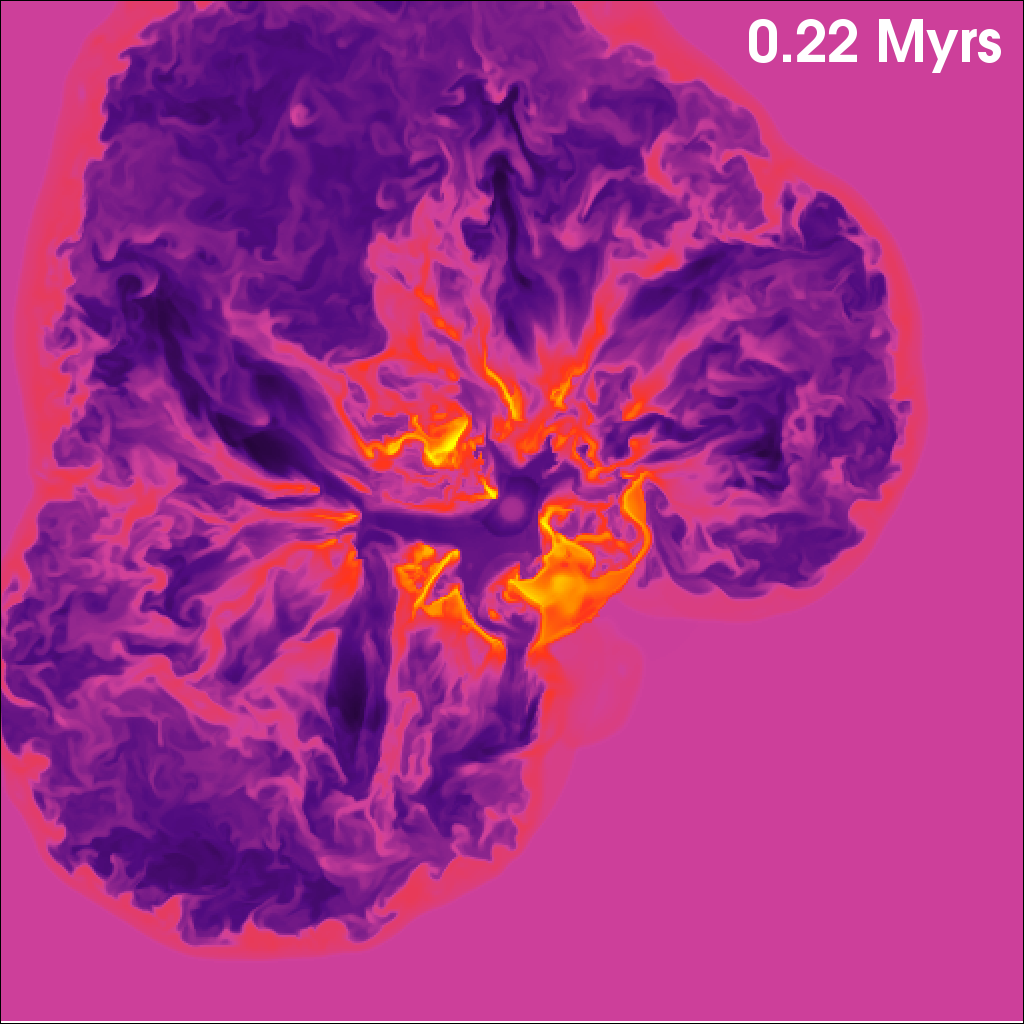}
\includegraphics[width=0.22\textwidth, height=0.22\textwidth]{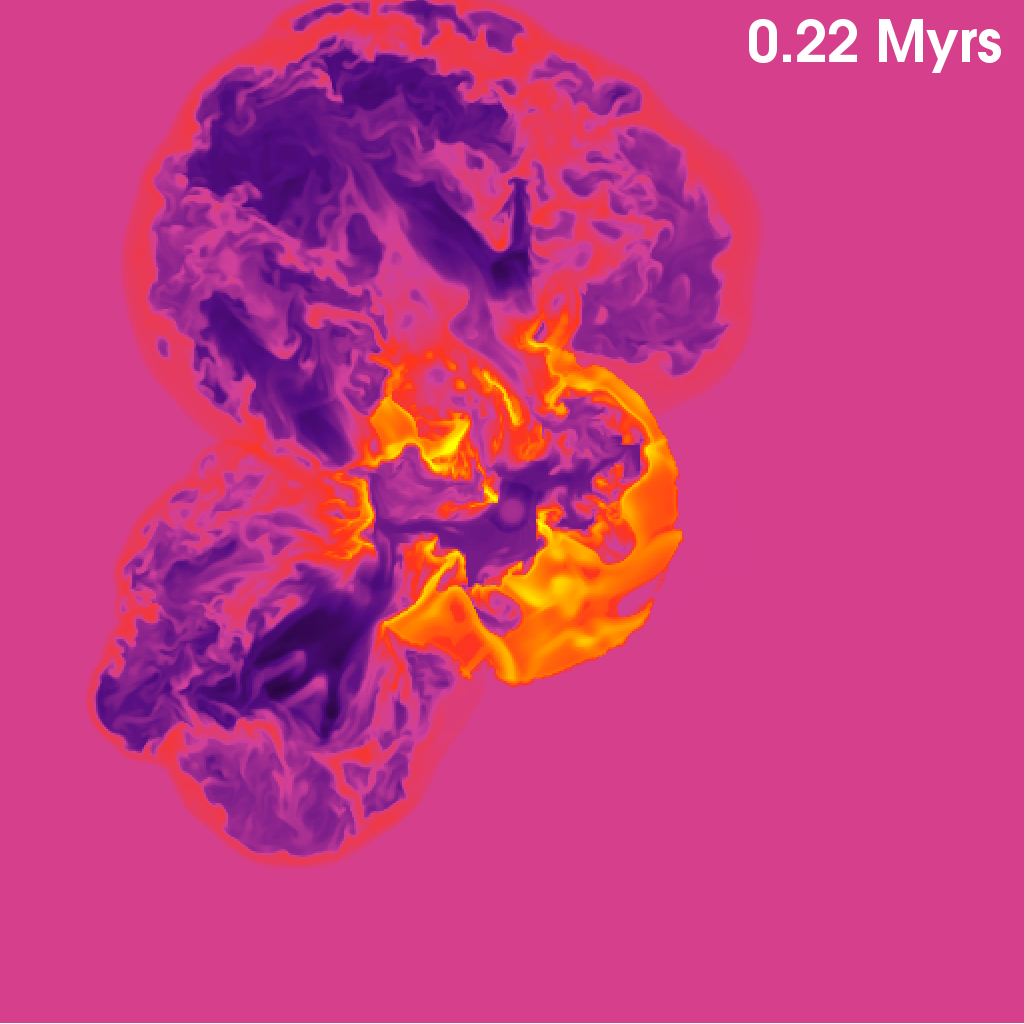}
\caption{Comparison of the initial expansion of the cluster wind in the xy-plane. [Left]: Sim A and [Right]: Sim B.  Sim B has a higher ambient density and a larger clump radius than Sim A.  The density scales for each simulation are shown in the top panels. \label{compsims}}
\end{figure}

Fig.~\ref{compsims} shows the initial destruction of the GMC clump for SimA and SimB.  The higher density clump and increased clump radius in SimB inevitably delays the blowout and subsequent expansion of the cluster wind.  However, the nature of the interaction is broadly similar, and the same general features are seen as the cluster wind percolates through the ``porous'' molecular environment, including the formation of fast-flowing low-density channels. 

The mass flux off the grid from SimB behaves qualitatively similarly to that from SimA, although the exact values are a little higher. There is again a broad peak caused by the initial encounter of the shell with the boundaries, in which the mass flux off the grid peaks at a rate of $4.6\times10^{-4} \rm\,M_{\odot}\,yr^{-1}$ at $t=0.67$\,Myr. The mass flux declines to a minimum of $2.9\times10^{-4} \rm\,M_{\odot}\,yr^{-1}$ at $t\approx 1.6$\,Myr, and then increases roughly linearly with time to reach a value of $7\times10^{-4} \rm\,M_{\odot}\,yr^{-1}$ at $t = 4$\,Myr. The latter indicates a ``mass-loading'' factor of nearly 1000.  The mass flux increases to $1.6\times10^{-3} \rm\,M_{\odot}\,yr^{-1}$ during the first WR stage, a mass-loading factor of nearly 80. A peak mass flux of $4.3\times10^{-3} \rm\,M_{\odot}\,yr^{-1}$ is attained following the first SN explosion, with peaks at $7$ and $8\times10^{-3} \rm\,M_{\odot}\,yr^{-1}$ following the second and third explosions respectively.

The energy flux off the grid from SimB peaks at $\approx 7.4 \times 10^{35}\,\rm{ergs\,s^{-1}}$ at $t=1.5$\,Myr, followed by a gradual decline to a roughly constant rate of $\approx 6.4 \times 10^{35}\,\rm{ergs\,s^{-1}}$ between $t=2.5-4$\,Myr. During this latter period, nearly half of the cluster wind power is radiated. The energy flux plateau's at about $1.45 \times 10^{37}\,\rm{ergs\,s^{-1}}$ during the first WR stage (indicating that again nearly half of the input energy is being radiated), but similar values are reached only towards the very end of the other two WR stages. 

\begin{figure}
\centering
\includegraphics[width=0.49\textwidth, height=0.25\textwidth]{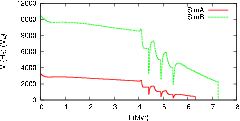}
\caption{Mass of H$_2$ contained within the full computational volume of SimA (solid red line) and SimB (dotted green line). \label{hmassstatstotal}}
\end{figure}

The evolution of the total H$_2$ mass throughout the simulation volume for both SimA and SimB is shown in Fig.~\ref{hmassstatstotal}.  Due to the larger clump radius and higher density, SimB has approximately three times the initial H$_2$ mass as SimA.  At very early times during the initial blowout the H$_2$ mass decreases by $\approx\,10\%$ as molecular gas at relatively low densities is cleared out of the GMC clump. Both simulations then show the same general trend of steady H$_2$ depletion during the MS phase of the cluster wind.  SimA loses H$_2$ mass at a rate of $1.74\times 10^{-4}\rm\,M_{\odot}\,yr^{-1}$ between $t=1-4$\,Myr. Since this is comparable to the mass flux off the grid during this time it further reinforces the point that most of the mass streaming into the wider environment was orginally molecular material. 

The evolution of the H$_2$ mass becomes more exciting when the stars undergo evolutionary transitions. There are slight rises during periods when the cluster wind is dominated by RSG mass-loss, and more significant decreases during WR dominated periods when the molecular material is ablated at a much faster rate. But most dramatic of all is the response of the H$_2$ gas to a supernova explosion. Immediately after an explosion, the H$_2$ mass undergoes a rapid and steep decline which is quickly followed by a recovery to a similar or higher value than the pre-SN H$_2$ mass.  The depletion of H$_2$ at this time is caused by its conversion to neutral and ionized hydrogen as it is heated by the SN shockwave which propagates relatively slowly through such dense molecular material. Note that it is not due to the 
H$_2$ gas being expelled from the grid.  Once the shockwave has passed the densest shocked (neutral and atomic) gas cools and reverts quickly to its previous molecular state, resulting in the replenishment of the H$_2$ mass seen in Fig.~\ref{hmassstatstotal}. The overshoot of the H$_2$ mass relative to its pre-SN value is likely a result of the higher gas pressure which now exists within the cluster environment, and which allows some of the originally atomic hydrodgen to become molecular.
The mass of molecular gas resumes its decline once the next most massive star in the cluster enters its WR stage. 

Since most of each SN's energy escapes along the low density channels through the remains of the GMC clump (see Sec.~\ref{sec:fluxes}), the SNe couple very weakly to the densest, low volume filling factor, gas. Hence the very densest regions are relatively immune to the effects of the supernovae (see also Figs.~\ref{evo1} and ~\ref{evotemp}). Indeed, Fig.~\ref{hmassstatstotal} shows that overall, the supernova shocks actually {\em increase} the amount of molecular material in the cluster environment. This rather unexpected result demonstrates that the inhomogeneity of the environment, the density range of the gas, and its effective porosity, are key considerations in understanding massive star feedback. This picture also differs substantially from the many models of mass-loaded supernova remnants in the literature which envisage the complete mixing of material injected into the remant from embedded clumps \citep[e.g.][]{White91,Dyson02,Pittard03}.

The last supernova occurs at $t = 5.4$\,Myr, after which there is $\approx 470\,\rm M_{\odot}$ (14.4\% of the original mass) of H$_2$ remaining in SimA and $\approx 3100\,\rm M_{\odot}$ (29.8\% of the original mass) in SimB. 

\begin{figure}
\centering
\includegraphics[width=0.49\textwidth, height=0.25\textwidth]{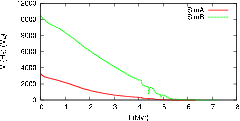}
\caption{Mass of H$_2$ contained within the initial GMC clump radius of SimA ($4$\,pc, solid red line) and SimB ($5$\,pc, dotted green line). \label{hmassstatsrclump}}
\end{figure}

Fig.~\ref{hmassstatsrclump} shows the mass of H$_2$ contained within the initial clump radius (4 and 5\,pc for SimA and SimB respectively).  The depletion of H$_2$ is much more rapid within this radius as the cluster wind not only ablates but also pushes molecular material away from the stars.  By the time the lowest mass star explodes there is very little H$_2$ present within the original volume of the GMC clump in either simulation ($< 0.1$ per cent of the original mass for SimA).  The features due to the evolution of the stars shown in Fig.~\ref{hmassstatstotal} are apparent here also, although to a lesser degree.

\section{Conclusion}
\label{sec:conclusions}

This paper investigates the effects of massive star feedback, via stellar winds and supernovae, on the inhomogeneous molecular environment left over from the formation of a stellar cluster from a GMC clump.  The remains of the GMC clump confines and shapes the initial structure of the expanding wind-blown bubble, which breaks out of the clump along paths of least resistance. Hot, high speed gas flows away from the cluster through low-density channels opened up by these blow-outs.  Mass is loaded into these flows from the ablation of dense clumps embedded within them and from material stripped from the dense gas which confines and directs the flows.  This complex interaction of the cluster wind with its environment is far removed from the results of simple spherically symmetric models. Increasing the density (by a factor of two) and the radius (by 25 per cent) of the cluster-forming GMC clump does not significantly affect this qualitative picture.

The density, temperature, pressure and velocity of gas in the cluster environment all span many orders of magnitude. The hottest gas typically occurs at the reverse shock of the cluster wind, and cools as it expands away from the cluster and mixes in with denser surrounding material. A multitude of weaker shocks exist around dense inhomogenities entrained into gas flowing at mildly supersonic speeds. In addition, gas within the cluster environment is subject to changes of several orders of magnitude in the dynamic pressure as the stars in the cluster evolve through their MS, RSG and WR stages, and explode as supernovae.

Our simulations show how molecular material is gradually ablated by the cluster wind and pushed away from the stellar cluster.  Despite mass-loading or entrainment factors of several hundreds during the MS phase of the cluster wind, and several tens during the later WR-dominated phase, the destruction and sweeping up of molecular material is a relatively slow process, and a substantial amount of molecular mass remains when the first star explodes as a SN. We find that the shocks resulting from SN explosions couple very weakly to the molecular gas, due to its small volume filling fraction, and the ease with which the energy from the SN can ``by-pass'' it. The high porosity of the GMC clump at this stage allows the SN blast to rip through the cluster in a largely unimpeded fashion, with the forward shock refracting around dense inhomogeneities. The early evolution of the remnant is markedly different from the standard spherically symmetric picture. Although the SN shock destroys molecular material which it overruns, we find that the cooling times of the densest regions are very short and allow molecular material to quickly reform. At least in our simulations, the stellar winds appear to be a more effective agent at removing molecular material from the cluster environment, despite injecting less energy than the SNe.

Examination of the energy flux off the hydrodynamic grid reveals that between one quarter and one half of the energy injected by the stellar winds is radiated away, with the remaining energy available to do work on the immediate surroundings. In comparison, more than 99 per cent of the energy from each SN explosion escapes into the wider environment. These fractions are clearly dependent on the initial conditions of our models. In particular, our limited study suggests that the fractions of energy radiated away will increase for denser cluster environments, and vice-versa. We are performing additional calculations to determine the conditions necessary for almost all of the injected energy to be radiated. This is relevant to some of the super star clusters in M82 where the thermalization effiency of the stellar feedback (winds plus supernovae) does not exceed a few percent \citep{Silich09}. 

We also estimate the fraction of ionizing photons which escape the cluster environment. This increases from less than 1 per cent at the start of the simulation, to 40 per cent when the cluster is 1\,Myr old, and to 60 per cent after 4\,Myr. The escape fraction momentarily reduces when the cluster wind is RSG dominated, but the overall trend is of an increasing escape fraction as dense material is pushed away from the cluster.  

In a future paper we will examine synthetic emission calculated from our simulations which we will compare against observations. We will also address some of the simplifications of our current model such as the neglect of radiative pressure and photoionization.

\section*{Acknowledgements}

HR acknowledges a Henry Ellison Scholarship from The University of Leeds and JMP acknowledges funding from the Royal Scoiety for a University Research Fellowship. We would also like to thank Tom Hartquist for discussions which have improved this paper and Enrique V\'{a}zquez-Semadeni for providing files of the turbulent background.



\label{lastpage}

\end{document}